\documentclass[journal=jctcce,manuscript=article]{achemso}
\setkeys{acs}{usetitle=true}

\usepackage[version=3]{mhchem} 

\usepackage{color, xcolor, colortbl}
\usepackage{graphicx,epstopdf}
\usepackage{geometry}
\usepackage{amsmath,amssymb,amsthm}
\usepackage{algorithm}
\usepackage{algorithmic}
\usepackage{bm}
\usepackage[caption=false]{subfig}
\usepackage{appendix}
\usepackage{multirow}
\usepackage{braket}
\usepackage[english]{babel}
\usepackage{hyperref}
\usepackage[capitalize]{cleveref}
\usepackage{adjustbox}
\usepackage{xspace}
\usepackage[roman]{complexity}
\usepackage{comment}
\usepackage{ulem}

\newcommand{\bvec}[1]{\mathbf{#1}}

\newcommand{\vk}{\bvec{k}}

\newcommand{\vq}{\bvec{q}}
\newcommand{\vr}{\bvec{r}}

\newcommand{\vG}{\bvec{G}}

\newcommand{\vR}{\bvec{R}}

\newcommand{\conj}[1]{#1^*}

\newcommand{\xsum}{\mathop{\sum\nolimits'}}

\newcommand{\I}{\mathrm{i}}

\newcommand{\mc}[1]{\mathcal{#1}}

\newcommand{\wt}[1]{\widetilde{#1}}

\newcommand{\abs}[1]{\left\lvert#1\right\rvert}

\newcommand{\ud}{\,\mathrm{d}}
\newcommand{\Or}{\mathcal{O}}

\newcommand{\REV}[1]{{#1}}

\title{Staggered mesh method for correlation energy
calculations of solids: Second order
M{\o}ller-Plesset perturbation theory}

\author{Xin Xing}
\affiliation{Department of Mathematics, University of California,
Berkeley, CA 94720, USA}

\author{Xiaoxu Li}
\affiliation{Department of Mathematics, University of California,
Berkeley, CA 94720, USA} 
\alsoaffiliation{School of Mathematical Sciences, Beijing Normal University, No.19, Xinjiekouwai St, Haidian District, Beijing, 100875, P.R.China}

\author{Lin Lin}
\email{linlin@math.berkeley.edu}
\affiliation{Department of Mathematics, University of California,
Berkeley, CA 94720, USA}
\alsoaffiliation{Computational Research Division, Lawrence Berkeley
National Laboratory, Berkeley, Berkeley, CA 94720, USA}
\alsoaffiliation{Challenge Institute for Quantum Computation, University of California, Berkeley, CA 94720, USA}

\begin{document}

\begin{abstract}
The calculation of the MP2 correlation energy for extended systems can be viewed as a multi-dimensional integral in the thermodynamic limit, and the standard method for evaluating the MP2 energy can be viewed as a trapezoidal quadrature scheme. We demonstrate that existing analysis neglects certain contributions due to the non-smoothness of the integrand, and may significantly underestimate finite-size errors. We propose a new staggered mesh method, which uses two staggered Monkhorst-Pack meshes for occupied and virtual orbitals, respectively, to compute the MP2 energy. 
The staggered mesh method circumvents a significant error source in the standard method, in which certain quadrature nodes are always placed on points where the integrand is discontinuous. 
\REV{One significant advantage of the proposed method is that there are no tunable parameters, and the additional numerical effort needed can be negligible compared to the standard MP2 calculation.}
Numerical results indicate that the staggered mesh method can be particularly advantageous for quasi-1D systems, as well as quasi-2D and 3D systems with certain  symmetries.
\end{abstract}

\section{Introduction}\label{sec:intro}
Correlated wavefunction based methods have long been the standard in quantum chemistry for accurate solution of the many-electron Schr\"odinger equation in molecular systems. In recent years, they are also increasingly  used for evaluating energies beyond the mean-field level in extended systems~\cite{MarsmanGruneisPaierKresse2009,GruneisMarsmanKresse2010, MuellerPaulus2012,SchaeferRambergerKresse2017,McClainSunChanEtAl2017,GruberLiaoTsatsoulisEtAl2018}. In contrast to the zero dimensional molecular systems, properties in bulk solids, surfaces and other low-dimensional extended systems need to be calculated properly in the thermodynamic limit (TDL). Due to the steep increase of the computational cost with respect to the system size, reaching convergence in a brute force fashion is often beyond reach, and finite-size corrections must be applied.
Common correction methods used to reduce the finite-size errors in correlation energy calculations include power-law extrapolation\cite{MarsmanGruneisPaierKresse2009,BoothGruneisKresseAlavi2013,LiaoGrueneis2016,MihmMcIsaacShepherd2019, MihmYangShepherd2020}, structure factor extrapolation\cite{ChiesaCeperleyMartinEtAl2006,LiaoGrueneis2016,GruberLiaoTsatsoulisEtAl2018}, and twist averaging\cite{LinZongCeperley2001,GruberLiaoTsatsoulisEtAl2018, MihmMcIsaacShepherd2019}. 

Unless otherwise stated, throughout the paper,  we assume the system extends along all three dimensions, and a standard Monkhorst-Pack (MP) mesh with $N_\vk$ points sampled in the first Brillouin zone (BZ) is used. The power law extrapolation typically assumes that the finite-size error is proportional to $N_\vk^{-1/3}$, $N_\vk^{-1}$, or their linear combinations. 
The $N_\vk^{-1/3}$ scaling is due to the fact that the correlation energy may inherit the $\mc{O}(N_\vk^{-1/3})$ finite-size error in HF orbital energies\cite{McClainSunChanEtAl2017}. 
The finite-size errors in the orbital energies can be reduced to $\mc{O}(N_\vk^{-1})$ via the Madelung-constant correction\cite{BroqvistAlkauskasPasquarello2009,ShepherdHendersonScuseria2014}. 
With this error removed, it has been argued based on structure factor analysis that the finite-size error in the correlation energy scales as $\mc{O}(N_\vk^{-1})$ due to the omission of certain terms in the structure factor \cite{LiaoGrueneis2016,GruberLiaoTsatsoulisEtAl2018}.
The structure factor extrapolation method, as its name suggests, computes the finite-size correction by extrapolating the omitted structure factor around the \REV{singular point of the Coulomb kernel in the reciprocal space}. The twist averaging technique calculates and averages the structure factors, and consequently the correlation energies using a set of shifted $\vk$-point meshes, and is often used as a pre-processing for power-law extrapolation and structure factor interpolation. 
The effectiveness of these correction methods can often be strongly system-dependent in practice\cite{LiaoGrueneis2016,GruberLiaoTsatsoulisEtAl2018}.

In this paper, we focus on the finite-size error of correlation energy calculations and its correction in the simplest scenario, namely the correlation energy from the second order M{\o}ller-Plesset perturbation theory (MP2) for insulating systems (the MP2 energies for metallic systems may diverge \cite{GruneisMarsmanKresse2010,GellMannBrueckner1957}).
In the TDL, the MP2 energy can be expressed as an integral in the BZ.
The numerical evaluation of the MP2 energy then uses a trapezoidal quadrature to replace the  integral by a finite sum over the MP mesh. Correspondingly, the finite-size error in MP2 energy arises from two sources: the error of the integrand, and the error of the numerical quadrature. 
The first error comes from the basis set incompleteness and finite-size errors in orbitals and orbital energies, and can be reduced by various existing techniques\cite{MarsmanGruneisPaierKresse2009,HattigKlopperKohnEtAl2012,GruneisAndreasShepherdEtAl2013}.

The integrand of the MP2 energy calculation generally has many discontinuous points. In this paper, we demonstrate that existing structure-factor based error analysis\cite{LiaoGrueneis2016, GruberLiaoTsatsoulisEtAl2018} neglects certain contributions due to the discontinuous behavior of the integrand, and underestimates the finite-size errors from the numerical quadrature. 
\REV{
We show that the error of the numerical quadrature comes from placing certain quadrature nodes at points of discontinuity, and also from the overall non-smoothness of the integrand.}
In particular, the standard MP2 calculation uses the same MP mesh for both occupied and virtual orbitals. This leads to the sampling of certain $\vq$ points (the difference between the $\vk$ points of an occupied-virtual orbital pair) on which the integrand is discontinuous. The error due to such improper placement of the quadrature nodes is $\Or(N_\vk^{-1})$. 


We propose a simple modification to address this problem \REV{with negligible additional costs}. Our staggered mesh method uses one MP mesh for occupied orbitals, and another MP mesh shifted by half mesh size for virtual orbitals. We show that the integrand is well defined on all $\vq$ points in the numerical calculation, thus circumventing the need of structure factor interpolation. \REV{The staggered mesh method has no tunable parameters, and the additional cost required can be negligible when compared to that of the standard MP2 calculations.} We show that the finite-size error of the staggered mesh method is mainly affected by the intrinsic non-smoothness of the integrand in the MP2 calculation. 

We compare the performance of the staggered mesh method, the standard method, \REV{and the structure factor interpolation method\cite{LiaoGrueneis2016,GruberLiaoTsatsoulisEtAl2018}} for a model system, where the mean-field orbital energies and wavefunctions are obtained accurately from a given effective potential.  We then demonstrate numerical tests on periodic hydrogen dimer, \REV{lithium hydride, silicon, and diamond systems} in the quasi-1D, 2D and 3D bulk settings using the PySCF\cite{SunBerkelbachEtAl2018} package. Our results indicate that the use of the staggered mesh can significantly accelerate the convergence towards the TDL in two scenarios: 1) quasi-1D systems, where the non-smoothness of the integrand is removable, 2) quasi-2D or 3D bulk systems with certain symmetries.

\section{Theory}\label{sec:theory}

Let $\Omega$ be the unit cell, $\abs{\Omega}$ be its volume, and $\Omega^*$ be the associated BZ. The Bravais lattice is denoted by $\mathbb{L}$ and its associated reciprocal lattice is denoted by $\mathbb{L}^*$. The MP mesh is used for $\vk$-point sampling in $\Omega^*$ and $N_\vk$ denotes the total number of $\vk$ points. When the MP mesh contains the $\Gamma$-point, the system can be identified with a periodic supercell $\Omega^{S}$ with volume $\abs{\Omega^{S}}=N_{\vk}\abs{\Omega}$. Each molecular orbital can be written as
\begin{equation*}
\psi_{n\vk}(\vr) = \dfrac{1}{\sqrt{N_\vk}} e^{\I \vk\cdot\vr} u_{n\vk}(\vr)=
 \frac{1}{\abs{\Omega}\sqrt{N_{\vk}}} \sum_{\mathbf{G}\in\mathbb{L}^*} \hat{u}_{n\vk}(\mathbf{G}) e^{\I (\vk+\mathbf{G}) \cdot \mathbf{r}},
\end{equation*}
where $n$ is a generic band index, and $u_{n\vk}$ is periodic with respect to the unit cell.
\REV{Although we choose to use the planewave basis set for convenience, our analysis is applicable to other periodic basis sets as well (e.g., the periodic Gaussian basis set \cite{DovesiCivalleriEtc2005,McClainSunChanEtAl2017}) for orbital representations, by expressing these basis functions as a linear combination of planewaves. Our analysis mainly concerns the low-frequency modes (in particular, around $\vG=\mathbf{0}$) and is thus insensitive to the choice of basis sets.}
 We also define the pair product (of the periodic components) as
\begin{equation*}
\varrho_{n'\vk',n\vk}(\vr)=\conj{u}_{n'\vk'}(\vr)  u_{n\vk}(\vr):=\frac{1}{\abs{\Omega}} \sum_{\mathbf{G}\in\mathbb{L}^*} \hat{\varrho}_{n'\vk',n\vk}(\mathbf{G}) e^{\I \mathbf{G} \cdot \mathbf{r}}.
\end{equation*}

Throughout the paper, $n\in\{i,j\}$ refers to the occupied orbital and $n\in\{a,b\}$ refers to the unoccupied orbital. The two-electron repulsion integral (ERI) tensor in the molecular orbital basis can be written as
\begin{equation}
\braket{i\vk_i,j\vk_j|a\vk_a,b\vk_b}= \frac{1}{\abs{\Omega^{S}}} \xsum_{\vG\in\mathbb{L}^*}
\frac{4\pi}{\abs{\vq+\vG}^2}  \hat{\varrho}_{i\vk_i,a\vk_a}(\mathbf{G}) \hat{\varrho}_{j\vk_j,b\vk_b}(\vG_{\vk_i,\vk_j}^{\vk_a,\vk_b}-\mathbf{G}), 
\label{eqn:ERI_element}
\end{equation}
where  $\vk_a-\vk_i=:\vq$ and we have
\begin{equation*}
\REV{\vG_{\vk_i,\vk_j}^{\vk_a,\vk_b}:=\vk_i+\vk_j-\vk_a-\vk_b\in\mathbb{L}^*,}
\end{equation*}
by crystal momentum conservation.
The notation $\xsum_{\vG\in\mathbb{L}^*}$ means that the possible term with $\vq+\vG=\bm{0}$ is excluded. 

\REV{According to Nesbet's theorem,}
the correlation energy per unit cell 
in general is given by
\begin{equation}
E_c=\frac{1}{N_\vk}\sum_{ijab}\sum_{\vk_i \vk_j \vk_a \vk_b} (2\braket{i\vk_i,j \vk_j |a\vk_a, b\vk_b}-\braket{i\vk_i,j \vk_j |b\vk_b,a\vk_a})T_{i\vk_i,j\vk_j}^{a\vk_a,b\vk_b},
\label{eqn:correlation_energy}
\end{equation}
where $\vk_i, \vk_j, \vk_a, \vk_b\in\Omega^*$. Here $T_{i\vk_i,j\vk_j}^{a\vk_a,b\vk_b}=t_{i\vk_i,j\vk_j}^{a\vk_a,b\vk_b}+t_{i\vk_i}^{a\vk_a} t_{j\vk_j}^{b\vk_b}$, and $t_{i\vk_i}^{a\vk_a}$ and $t_{i\vk_i,j\vk_j}^{a\vk_a,b\vk_b}$ are
singles and doubles amplitudes obtained from solution of related amplitude equations. In the coupled cluster doubles (CCD) theory, we have $t_{i\vk_i}^{a\vk_a}=0$, and the MP2 energy is further given by setting the doubles amplitude to
\begin{equation}
t_{i\vk_i,j\vk_j}^{a\vk_a,b\vk_b}=\frac{\braket{ a\vk_a, b\vk_b|i\vk_i,j \vk_j}}{\varepsilon_{i\vk_i} + \varepsilon_{j\vk_j} - \varepsilon_{a\vk_a} - \varepsilon_{b\vk_b}}.
\label{eqn:mp2_amplitude}
\end{equation}


Note that \cref{eqn:correlation_energy} can be rewritten as
\begin{equation}
E_c=\frac{1}{N_\vk \abs{\Omega^S}}\sum_{ijab}\sum_{\vk_i \vk_j \vk_a \vk_b} \braket{i\vk_i,j \vk_j |a\vk_a, b\vk_b}\wt{T}_{i\vk_i,j\vk_j}^{a\vk_a,b\vk_b},
\label{eqn:correlation_redefine}
\end{equation}
where we have absorbed the exchange term into the redefined amplitude
\[
\wt{T}_{i\vk_i,j\vk_j}^{a\vk_a,b\vk_b}=\abs{\Omega^{S}}\left(2T_{i\vk_i,j\vk_j}^{a\vk_a,b\vk_b}-T_{i\vk_i,j\vk_j}^{b\vk_b,a\vk_a}\right),
\]
and the scaling factor $\abs{\Omega^{S}}$ ensures that each entry $\wt{T}_{i\vk_i,j\vk_j}^{a\vk_a,b\vk_b}$ does not vanish in the TDL.  

In order to write down the correlation energy in the TDL, we use the fact that both the ERI tensor and the $T$ amplitude do not change if we replace any $\vk$ by $\vk+\vG$ for some $\vG\in\mathbb{L}^*$. Then fixing $\vk_i\in\Omega^*$, we may shift $\vk_a$ by some $\vG$ vector so that the difference $\vq=\vk_a-\vk_i\in\Omega^*$. Similarly further fixing $\vk_j\in \Omega^*$, we may shift $\vk_b$ so that $\vG_{\vk_i,\vk_j}^{\vk_a,\vk_b}=\bm{0}$, i.e. $\vk_b=\vk_j-\vq$. 
Note that this requires redefining $\hat{\varrho}_{n'\vk',n\vk}$ to accommodate the case where $\vk$ is outside $\Omega^*$. More importantly, such manipulation is only formal and is introduced to simplify the theoretical analysis. In practical calculations, we may still keep $\vk_i, \vk_j, \vk_a, \vk_b\in\Omega^*$ as in standard implementations. After such modifications, $E_c$ in the TDL as $N_\vk\to \infty$ can be concisely written as a triple integral over BZ (which is a $9$-dimensional integral for 3D bulk systems):
\begin{equation}
\begin{split}
E_c^{\text{TDL}}=\int_{\Omega^*} \ud \vq &  \int_{\Omega^*} \ud \vk_i \int_{\Omega^*} \ud \vk_j 
\\
& \frac{\abs{\Omega}}{(2\pi)^9} \sum_{ijab}\xsum_{\vG\in\mathbb{L}^*}
\frac{4\pi}{\abs{\vq+\vG}^2}  \hat{\varrho}_{i\vk_i,a(\vk_i+\vq)}(\mathbf{G}) \hat{\varrho}_{j\vk_j,b(\vk_j-\vq)}(-\mathbf{G}) \wt{T}_{i\vk_i,j\vk_j}^{a(\vk_i+\vq),b(\vk_j-\vq)}.
\end{split}
\label{eqn:correlation_TDL}
\end{equation}
\REV{Using the fact that the intersection of $\Omega^*$ and $\mathbb{L}^*$ only includes the $\Gamma$-point, the singularity set $\{\vq+\vG=\bm{0},\vq\in\Omega^*,\vG\in\mathbb{L}^*\}=\{\vq=\bm{0},\vG=\bm{0}\}$ is only an isolated point. Hence in}
this continuous formulation, we may also write $\xsum_{\vG\in\mathbb{L}^*}$ simply as the regular summation $\sum_{\vG\in\mathbb{L}^*}$.

\subsection{Error analysis}\label{sec:error_analysis}

All numerical schemes for evaluating the correlation energy in the TDL amounts to approximating the triple integral \cref{eqn:correlation_TDL}. The quality of the numerical approximation can be affected by the following error sources:
1) The error introduced by replacing the integral \cref{eqn:correlation_TDL} by a numerical quadrature \cref{eqn:correlation_redefine}, 2) The mean-field orbital energies $\{\varepsilon_{n\vk}\}$ and orbitals $\{u_{n\vk}(\vr)\}$ are not evaluated in the TDL, 3) Basis set incompleteness error, 
4) Error in evaluating the $T$-amplitudes.
The last three sources contribute to the errors of the  integrand values used in the numerical quadrature \cref{eqn:correlation_redefine}.

This paper only concerns the first error, i.e. the quadrature error. We assume that mean-field calculations are less expensive than correlation energy calculations, and the finite-size error of the orbitals and orbital energies could be reduced by using other correction methods and/or a large enough MP mesh if needed. Even when the same MP mesh is used to evaluate mean-field energies and orbitals, after the Madelung-constant correction to the occupied orbital energies, the contribution of the finite-size from the orbital energies becomes $\mc{O}(N_\vk^{-1})$ \cite{McClainSunChanEtAl2017}.
The error due to the incompleteness of the basis set  
is more difficult to assess. Though such error can be reduced via power-law extrapolation~\cite{MarsmanGruneisPaierKresse2009} or explicit correlation methods \cite{HattigKlopperKohnEtAl2012,GruneisAndreasShepherdEtAl2013}, we will not consider such improvements in this paper.
We will also only consider the evaluation of the MP2 energy, where the $T$-amplitudes are given explicitly by orbital energies and ERIs. We will demonstrate below that even under such assumptions, the finite-size effect due to the quadrature error remains significant.



To connect to the commonly used argument in the literature \cite{ChiesaCeperleyMartinEtAl2006,LiaoGrueneis2016, GruberLiaoTsatsoulisEtAl2018} to analyze the quadrature error using structure factors, we note that the structure factor $S_{\vq}(\vG)$ corresponds to a part of the integrand in \cref{eqn:correlation_TDL} as
\begin{equation}\label{eqn:structure_factor}
S_{\vq}(\vG)=\int_{\Omega^*} \ud \vk_i \int_{\Omega^*} \ud \vk_j 
 \frac{\abs{\Omega}}{(2\pi)^9}\sum_{ijab}
  \hat{\varrho}_{i\vk_i,a(\vk_i+\vq)}(\mathbf{G}) \hat{\varrho}_{j\vk_j,b(\vk_j-\vq)}(-\mathbf{G}) \wt{T}_{i\vk_i,j\vk_j}^{a(\vk_i+\vq),b(\vk_j-\vq)}.
\end{equation}
The correlation energy is then 
\begin{equation}
E_c^{\text{TDL}}=\int_{\Omega^*} \ud \vq \xsum_{\vG\in\mathbb{L}^*}
\frac{4\pi}{\abs{\vq+\vG}^2} S_{\vq}(\vG).
\label{eqn:correlation_sf}
\end{equation} 
We may also combine the information from the structure factors and define the integrand of \cref{eqn:correlation_sf} as
\begin{equation}
h(\vq) = \xsum_{\vG\in\mathbb{L}^*}
\frac{4\pi}{\abs{\vq+\vG}^2} S_{\vq}(\vG).
\label{eqn:hq}
\end{equation}

The standard MP2 calculation \cref{eqn:correlation_redefine} can be interpreted as two quadrature steps in estimating each $S_\vq(\vG)$ at a finite set of $\vq$ points and $E_c^\text{TDL}$ as,
\begin{align}
S_{\vq}(\vG) 
&\approx
 \dfrac{|\Omega^*|^2}{N_\vk^2} \sum_{\vk_i, \vk_j \in \mathcal{K}} 
  \left(
 \frac{\abs{\Omega}}{(2\pi)^9}\sum_{ijab}
\hat{\varrho}_{i\vk_i,a(\vk_i+\vq)}(\mathbf{G}) \hat{\varrho}_{j\vk_j,b(\vk_j-\vq)}(-\mathbf{G}) \wt{T}_{i\vk_i,j\vk_j}^{a(\vk_i+\vq),b(\vk_j-\vq)}
\right)
\nonumber\\
&=: \wt{S}_\vq(\vG), \quad \vq \in \mathcal{K}_\vq,\vG \in\mathbb{L}^*,
\label{eqn:quadrature_SqG}
\\
E_c^{\text{TDL}}
&
 \approx
 \dfrac{|\Omega^*|}{N_\vk}
  \sum_{\vq \in \mathcal{K}_\vq}
   \left(
 \xsum_{\vG\in\mathbb{L}^*}
 \frac{4\pi}{\abs{\vq+\vG}^2} \wt{S}_{\vq}(\vG)
 \right),
 \label{eqn:quadrature_Ec}
\end{align}
where $\mathcal{K}$ denotes the  MP mesh and $\mathcal{K}_\vq$ is a same-sized MP mesh containing all $\vq \in \Omega^*$ 
defined as the minimum image of $\vk_a - \vk_i$ with $\vk_i,\vk_a \in \mathcal{K}$.
Furthermore, $\mathcal{K}_\vq$ always includes the $\Gamma$-point. These two steps apply the trapezoidal rules with uniform meshes $\mathcal{K}\times\mathcal{K}$ and $\mathcal{K}_\vq$ for \cref{eqn:structure_factor} and \cref{eqn:correlation_sf}, respectively.

Note that the integrand in \cref{eqn:correlation_sf} is discontinuous \REV{in the presence of zero momentum transfer (i.e., at $\vq = \bm{0}$)}, and its value at this point is indeterminate due to the term $(4\pi/|\vq|^2) S_\vq(\bm{0})$.
It has been argued that for $\vq+\vG\ne \bm{0}$, $S_{\vq}(\vG)$ converges quickly~\cite{LiaoGrueneis2016}, and hence the error is mainly due to the neglect of this discontinuous term from the primed summation in \cref{eqn:quadrature_Ec}, which scales as $N_{\vk}^{-1}\sim \abs{\Omega^{S}}^{-1}$. 
\REV{However, such an analysis neglects two other sources of discontinuity. }

1) Fixing $\vq$ and $\vG$, the amplitude $\wt{T}_{i\vk_i,j\vk_j}^{a(\vk_i+\vq),b(\vk_j-\vq)}$ in the integrand for $S_\vq(\vG)$ in \cref{eqn:structure_factor} is discontinuous as a function of $(\vk_i, \vk_j)$ when $\vk_j - \vk_i - \vq \in \mathbb{L}^*$ due to its exchange part, i.e., 
\[
|\Omega^S|T_{i\vk_i,j\vk_j}^{b(\vk_j - \vq),a(\vk_i + \vq)}
= 
\dfrac{
        \xsum_{\vG' \in \mathbb{L}^*}  \frac{4\pi}{|\vk_j - \vk_i - \vq + \vG'|^2} \hat{\varrho}_{i\vk_i, b(\vk_j-\vq)}^*(\vG') 
        \hat{\varrho}_{j\vk_j, a(\vk_i+\vq)}^*(-\vG') 
}{\varepsilon_{i\vk_i} + \varepsilon_{j\vk_j} - \varepsilon_{b(\vk_j - \vq)} - \varepsilon_{a(\vk_i + \vq)}}.
\]
For each pair $(\vk_i, \vk_j)$ satisfying the relation $\vk_j - \vk_i - \vq \in \mathbb{L}^*$, the exchange term above neglects the summation term associated with $\vk_j - \vk_i - \vq + \vG' = \bm{0}$, leading to $N_\vk^{-2}\sim |\Omega^S|^{-2}$ error in the associated volume element corresponding to the multi-index $(\vk_i, \vk_j)$. For each $\vq \in\mathcal{K}_\vq$, there are $\Or(N_\vk)$ such pairs $(\vk_i, \vk_j) \in \mathcal{K} \times \mathcal{K}$. 
Overall, neglecting the discontinuous terms when evaluating  $\wt{T}_{i\vk_i,j\vk_j}^{a(\vk_i+\vq),b(\vk_j-\vq)}$  at these quadrature nodes leads to  $\mc{O}(N_\vk^{-1})$ error in computing each $S_\vq(\vG)$. 
\REV{This leads to $\Or(N_\vk^{-1})$ error in computing the sum $\xsum_{\vG\in\mathbb{L}^*}
\frac{4\pi}{\abs{\vq+\vG}^2} S_{\vq}(\vG)$ at each $\vq\in \mathcal{K}_\vq$ in \cref{eqn:quadrature_Ec},} and hence additional $\mc{O}(N_\vk^{-1})$ error in computing $E_c^\text{TDL}$.

\REV{
2) For $\vq = \bm{0}$ and $\vG \neq \bm{0}$, the amplitude $\wt{T}_{i\vk_i,j\vk_j}^{a(\vk_i+\vq),b(\vk_j-\vq)}$ in the integrand for $S_{\vq}(\vG)$ also neglects another discontinuous term in its direct part, i.e., 
\[
|\Omega^S|T_{i\vk_i,j\vk_j}^{a(\vk_i + \vq), b(\vk_j - \vq)}
= 
\dfrac{
        \xsum_{\vG' \in \mathbb{L}^*}  \frac{4\pi}{|\vq + \vG'|^2} \hat{\varrho}_{i\vk_i, a(\vk_i+\vq)}^*(\vG') 
        \hat{\varrho}_{j\vk_j, b(\vk_j - \vq)}^*(-\vG') 
}{\varepsilon_{i\vk_i} + \varepsilon_{j\vk_j} - \varepsilon_{b(\vk_j - \vq)} - \varepsilon_{a(\vk_i + \vq)}}.
\]
The  terms $\frac{4\pi}{|\vq|^2} \hat{\varrho}_{i\vk_i, a(\vk_i+\vq)}^*(\bm{0}) 
\hat{\varrho}_{j\vk_j, b(\vk_j - \vq)}^*(\bm{0})$ are neglected at $\vq=\bm{0}$ for any $\vk_i,\vk_j$,
leading to $\Or(1)$ error in computing $S_{\vq}(\vG)$ at $\vq = \bm{0}, \vG\neq \bm{0}$. This leads to $\Or(1)$ error in computing the sum 
$\xsum_{\vG\in\mathbb{L}^*} \frac{4\pi}{\abs{\vq+\vG}^2} S_{\vq}(\vG)$ at $\vq = \bm{0}$ in \cref{eqn:quadrature_Ec}. 
Taking the prefactor $N_{\vk}^{-1}$ into account, neglecting these discontinuous terms leads to $\Or(N_\vk^{-1})$ error in computing $E_c^\text{TDL}$. 
}

\REV{
        To summarize, there is $\Or(N_\vk^{-1})$ error in the evaluation of each $S_\vq(\vG)$ at  $\vq\in\mathcal{K}_\vq, \vG\in\mathbb{L}^*$ 
        due to neglecting discontinuous terms in the exchange part of the amplitude, 
        and there is $\Or(1)$ error in the evaluation of $S_\vq(\vG)$ at $\vq=\bm{0}, \vG\neq \bm{0}$  due to neglecting discontinuous terms in the direct part. 
        The contribution from both error sources is $\Or(N_\vk^{-1})$ in computing $E_c^\text{TDL}$. This is in addition to the $\Or(N_\vk^{-1})$ error due to the neglect of $4\pi/|\vq|^2S_\vq(\bm{0})$ at $\vq=\bm{0}$.
                As a result, correction schemes only aiming at recovering the value of $4\pi/|\vq|^2S_\vq(\bm{0})$ at $\vq=\bm{0}$ cannot  lead to asymptotic improvement of accuracy in general.
}


Our analysis above is also applicable to quasi-1D and quasi-2D systems, 
which samples $\vk$ points on the corresponding 1D axis 
and 2D plane in $\Omega^*$, respectively.  Without loss of generality we may assume the MP mesh includes $\vk$ points of the form $\vk=(0,0,k_z)$ for quasi-1D systems, and $\vk=(0,k_y,k_z)$ for quasi-2D systems.
The correlation energies of this model in the TDL can be written in an integral form similar to \cref{eqn:correlation_TDL}, while only changing the integration domains for $\vk_i, \vk_j$, and $\vq$ from $\Omega^*$ to the 
corresponding axis/plane in $\Omega^*$.
The discontinuity of the integrands in \cref{eqn:structure_factor} and \cref{eqn:correlation_sf} described for 3D systems earlier
\REV{is also present in low-dimensional systems, and 
neglecting discontinuous terms also leads to $\Or(N_\vk^{-1})$ quadrature error in the MP2 energy.}

\subsection{Staggered mesh method}
\begin{figure}
        \centering
        \includegraphics[width=0.4\textwidth]{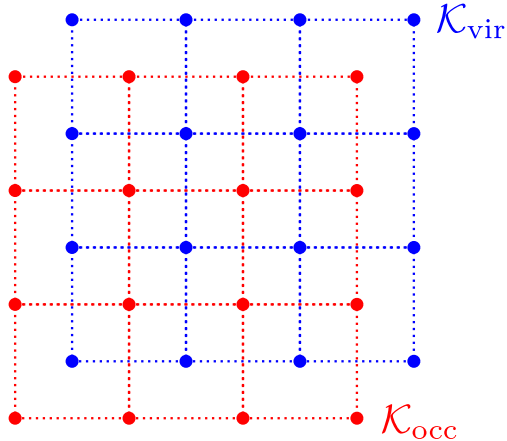}
        \caption{Illustration of the staggered meshes $\mathcal{K}_\text{occ}$ and $\mathcal{K}_\text{vir}$ for a quasi-2D system.}
        \label{fig:staggered_mesh}
\end{figure}

Based on the analysis above, the standard method for MP2 calculations places certain quadrature nodes on points of discontinuity of the integrand, which leads to finite-size errors of size $\Or(N_{\vk}^{-1})$. We propose a simple modification of the procedure to evaluate the MP2 energy, called \textit{the staggered mesh method}. 
The main idea is to use an MP mesh $\mathcal{K}_\text{occ}$ for occupied momentum vectors $\vk_i, \vk_j$, but a different, same-sized MP mesh $\mathcal{K}_\text{vir}$ for virtual momentum vectors $\vk_a, \vk_b$, where $\mathcal{K}_\text{vir}$ is obtained by shifting $\mathcal{K}_\text{occ}$ with half mesh size in all extended directions to create a staggered mesh (see \Cref{fig:staggered_mesh}). The MP2 energy is then computed as 
\begin{equation}\label{eqn:staggere_mp2}
E_c^\text{staggered}=\frac{1}{N_\vk \abs{\Omega^S}}\sum_{ijab}\sum_{\vk_i, \vk_j \in \mathcal{K}_\text{occ}} \sum_{\vk_a, \vk_b\in\mathcal{K}_\text{vir}} \braket{i\vk_i,j \vk_j |a\vk_a, b\vk_b}\wt{T}_{i\vk_i,j\vk_j}^{a\vk_a,b\vk_b},
\end{equation}
with $N_\vk = |\mathcal{K}_\text{occ}| =  |\mathcal{K}_\text{vir}|$ and $|\Omega^S| = N_\vk |\Omega|$.

This calculation \cref{eqn:staggere_mp2} can still be interpreted as a two-step numerical quadrature scheme in \cref{eqn:quadrature_SqG} and \cref{eqn:quadrature_Ec}, but with a different set of quadrature nodes. 
The induced mesh $\mathcal{K}_\vq$ in \cref{eqn:quadrature_Ec} shifts the $\Gamma$-centered MP mesh by half mesh size (recall that $\mathcal{K}_\vq$ is the set of all possible minimum images of  $\vk_a - \vk_i$ with $\vk_a \in \mathcal{K}_\text{vir}, \vk_i\in\mathcal{K}_\text{occ}$) and does not contain $\vq = \bm{0}$. Recall that in \cref{eqn:quadrature_SqG}  for computing $S_\vq(\vG)$, the integrand becomes discontinuous when $\vk_j - \vk_i - \vq \in \mathbb{L}^*$. In the staggered mesh method, for each $\vq\in\mathcal{K}_\vq$, all possible values of $ \vk_j - \vk_i - \vq$ (for any $\vk_i, \vk_j \in \mathcal{K}_\text{occ}$) belong to $\mathcal{K}_\vq$ and are always outside $\mathbb{L}^*$.
As a result, all the defined quadrature nodes in the staggered mesh method do not overlap with any points of discontinuity of the integrand for computing $S_\vq(\vG)$, $h(\vq)$,  or $E_c^\text{TDL}$. 
\REV{
        This completely eliminates the error due to the neglect of discontinuous terms when evaluating the integrand at those points of discontinuity.}

\REV{
In order to implement the staggered mesh method, we need to obtain the orbitals and orbital energies on the shifted MP mesh. 
Once the self-consistent Hartree-Fock equations are solved, 
these quantities can be evaluated by solving the Hartree-Fock equations non-self-consistently on the shifted mesh, and such additional cost  calculations can be negligible compared to the cost of MP2 calculations. 
The remaining cost of the staggered mesh method is exactly the same as that of the standard method. 
}


\section{Numerical results}\label{sec:numer}

According to the discussion in \cref{sec:error_analysis}, there are multiple factors contributing to the finite-size errors of the MP2 correlation energy. In order to focus on the contribution from the quadrature error, we first compare the performance of the standard and the staggered mesh methods for MP2 calculations for a series of model systems with given effective potentials in \cref{sec:numer_model}. We then compare the performance of the two methods for periodic \REV{hydrogen dimer, lithium hydride, silicon, and diamond systems} in \cref{sec:numer_pyscf}, using the PySCF software package\cite{SunBerkelbachEtAl2018}.
 
In all the following tests, the MP mesh for virtual orbitals includes the $\Gamma$ point. The standard method uses the same MP mesh for occupied orbitals.
 The staggered mesh method shifts the MP mesh by half mesh size for occupied orbitals. For quasi-1D, quasi-2D, and 3D systems, the MP meshes are  of size $1\times 1\times N_\vk$, $1\times N_\vk^{1/2} \times N_\vk^{1/2}$, and 
$N_\vk^{1/3}\times N_\vk^{1/3} \times N_\vk^{1/3}$, respectively.
Atomic units are used in all the tests.

\subsection{Model systems}\label{sec:numer_model}
\REV{We first study} a model system with a (possibly anisotropic) Gaussian effective potential field. 
\REV{In this model, no finite-size error correction is needed for orbitals and orbital energies.}
More specifically, let the unit cell be $[0,1]^3$, and use $14\times 14 \times 14$ planewave basis functions to discretize functions in the unit cell.  The Gaussian effective potential takes the form
\begin{equation}\label{eqn:gaussian_model}
V(\vr) = \sum_{\vR \in \mathbb{L}}C \exp\left(-\dfrac{1}{2}(\vr + \vR- \vr_0)^{\top}\Sigma^{-1}(\vr +\vR - \vr_0)\right),
\end{equation}
with $\vr_0 = (0.5, 0.5, 0.5)$.
For each momentum vector $\vk$ in $\Omega^*$, we solve the corresponding effective Kohn-Sham equation to obtain $n_\text{occ}$ occupied orbitals and $n_\text{vir}$ virtual orbitals. The covariance matrix $\Sigma$ controls the isotropicity of system. For the isotropic case, we choose $\Sigma = \text{diag}(0.2^2, 0.2^2, 0.2^2)$, $C = -200$, $n_\text{occ} = 1$, $n_\text{vir} = 3$. For  the anisotropic case, we choose $\Sigma = \text{diag}(0.1^2, 0.2^2, 0.3^2)$, $C = -200$, $n_\text{occ} = 1$, $n_\text{vir} = 1$. For such model problems, the selected $n_\text{vir}$ virtual bands are separated from the remaining virtual bands, which ensures that the MP2 correlation energy with a fixed number of virtual bands is a well-defined problem.
There is also a direct gap between the occupied and virtual bands in all cases.

\begin{figure}[htbp]
        \centering
        \subfloat[$|\wt{T}_{i\vk_i,j\vk_j}^{a(\vk_i+\vq),b(\vk_j-\vq)}|$ with  $\vq= (0,0,\frac{\pi}{2})$ ]{
                \includegraphics[width=0.47\textwidth]{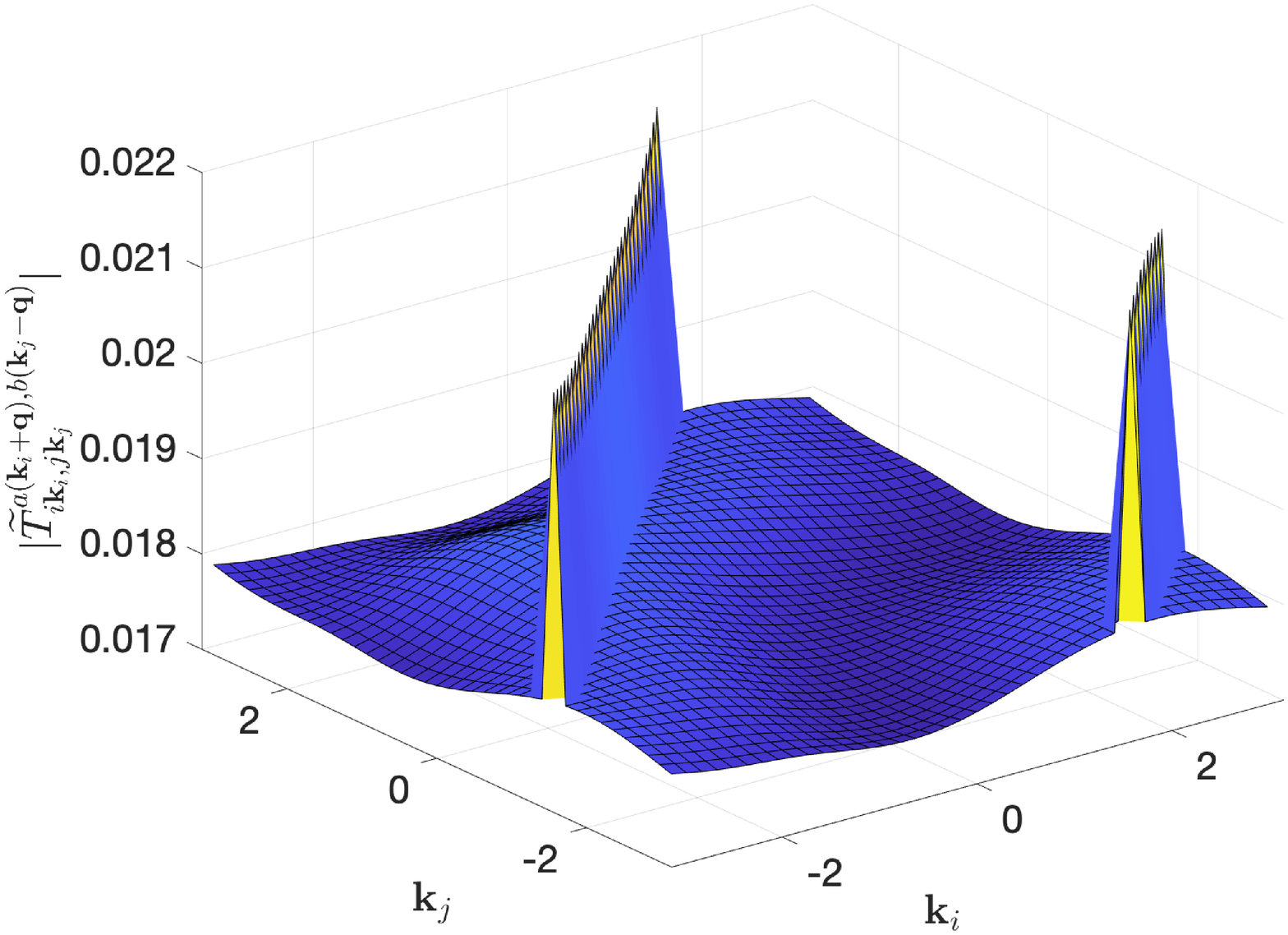}
        }
        \subfloat[$h(\vq)$]{
                \includegraphics[width=0.47\textwidth]{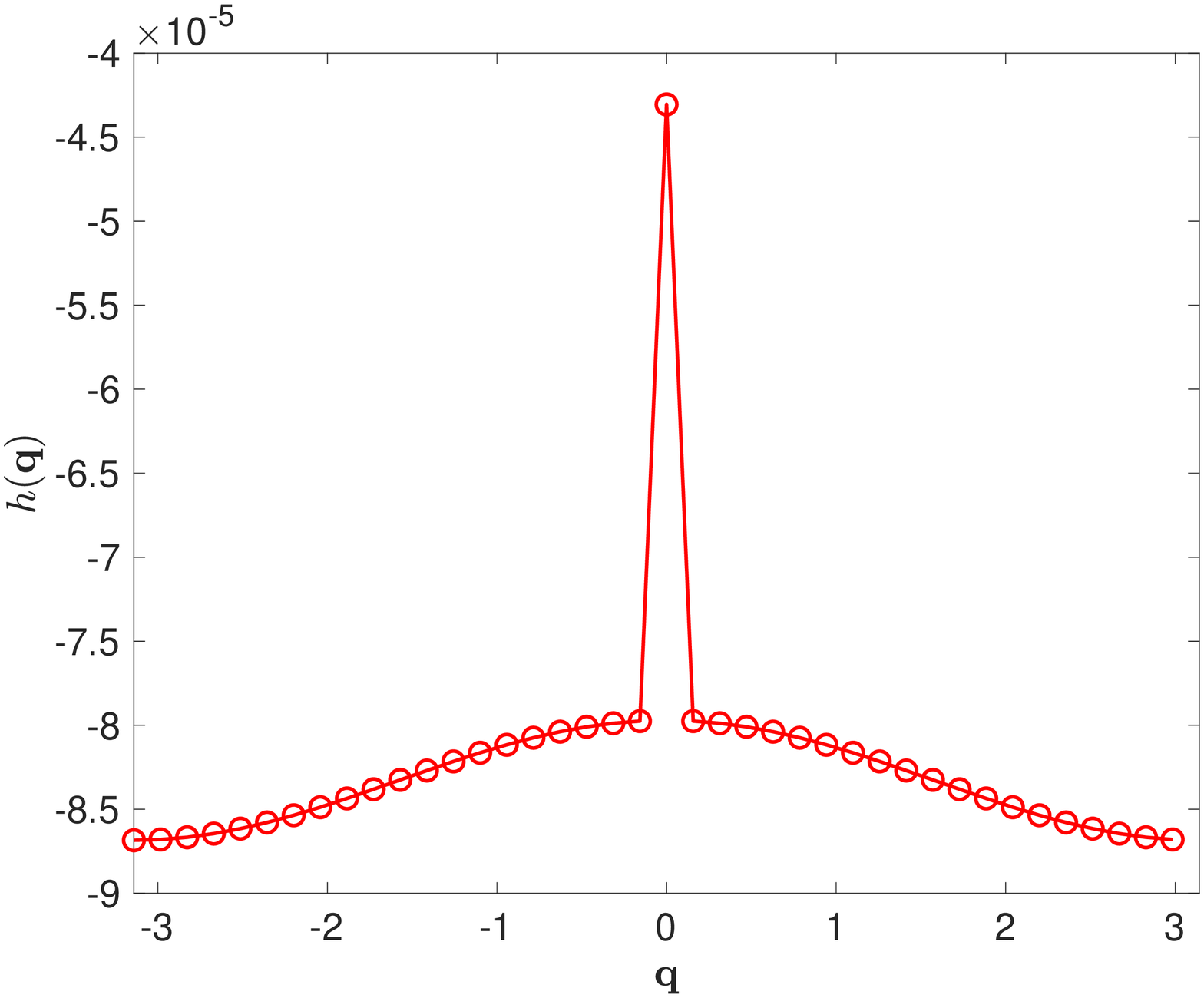}
        }
        
        \subfloat[$S_\vq(\vG)$]{
                \includegraphics[width=\textwidth]{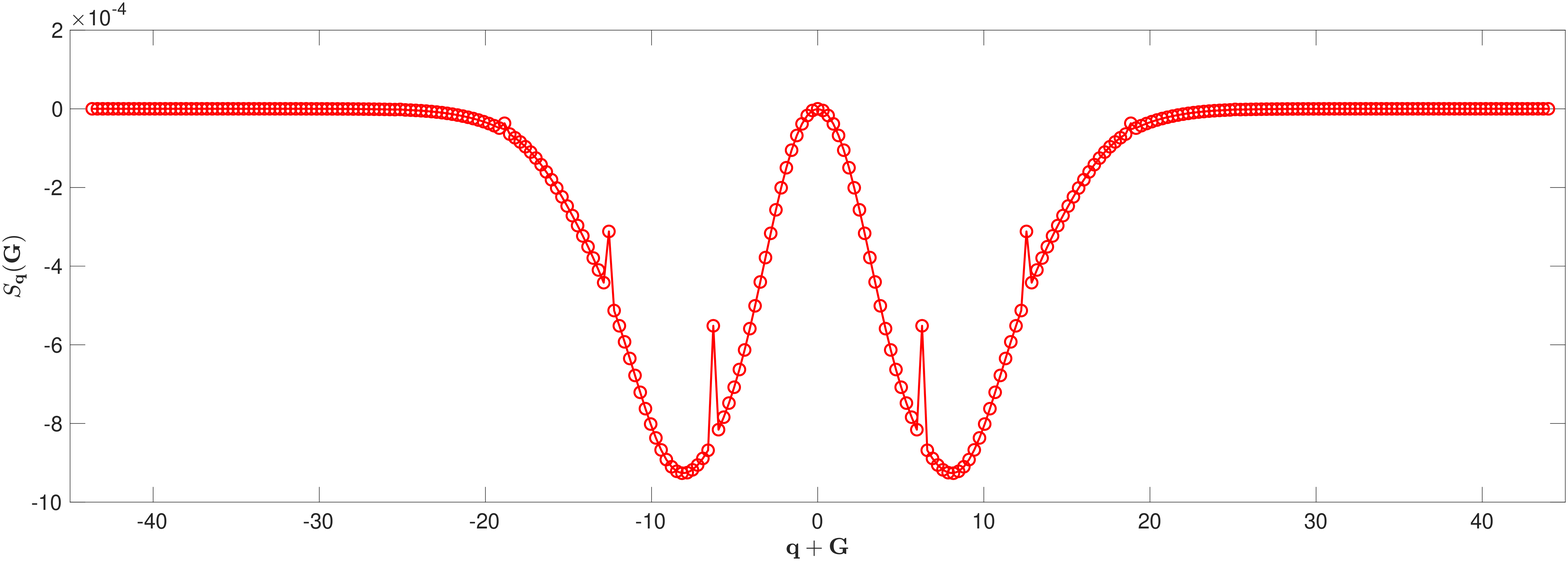}
        }
        
        \caption{
                Illustration of discontinuities in $\wt{T}_{i\vk_i,j\vk_j}^{a(\vk_i+\vq),b(\vk_j-\vq)}$, $h(\vq)$, and \REV{$S_\vq(\vG)$} for a quasi-1D model system with the anisotropic Gaussian effective potential field.         
                All sampled $\vk$ points are of the form $(0, 0, k)$ with $k \in [-\pi, \pi]$.
                \REV{
                        The structure factor $S_\vq(\vG)$ with $\vG = (0, 0, G_z)$, $G_z \in \{0, \pm 2\pi, \pm 4\pi, \ldots\}$ is plotted.  
                        The six notable discontinuous points in $S_{\vq}(\vG)$ correspond to $\vq = 0$ and $G_z = \pm 2\pi, \pm 4\pi, \pm 6\pi$.} 
                        The two lines of discontinuities in $\wt{T}_{i\vk_i,j\vk_j}^{a(\vk_i+\vq),b(\vk_j-\vq)}$ are $k_j - k_i - \frac{\pi}{2} = 0$ and $k_j - k_i - \frac{\pi}{2} = -2\pi$.        
                \label{fig:integrand_discontinuity}
        }
\end{figure}

\REV{\Cref{fig:integrand_discontinuity} first  illustrates the discontinuities of $\wt{T}_{i\vk_i,j\vk_j}^{a(\vk_i+\vq),b(\vk_j-\vq)}$, $S_\vq(\vG)$, and $h(\vq)$ for a quasi-1D model system. 
        According to the discussion in \cref{sec:error_analysis}, such discontinuous behaviors are generic in MP2 calculations. The standard MP2 calculation with any $\vk$-point mesh $\mathcal{K}$ always places some of its quadrature nodes at such points of discontinuity.}

\begin{figure}[H]
        \centering
        \includegraphics[width=0.6\textwidth]{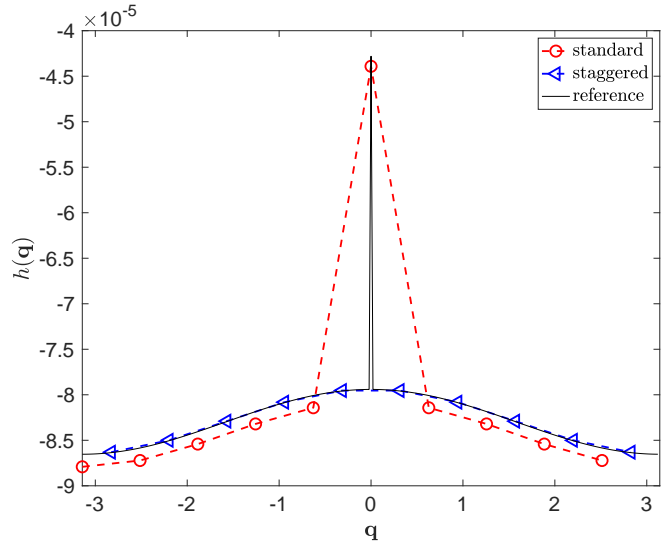}
        \caption{Illustration of $h(\vq)$ computed by the standard and the staggered mesh methods with mesh size  $1\times 1 \times 10$ for a quasi-1D model system with the anisotropic Gaussian effective potential field. 
                All sampled $\vq$ points are of the form $(0, 0, \vq_z)$ with $\vq_z \in [-\pi, \pi]$.
                The reference curve for $h(\vq)$ is computed based on the standard method with mesh size $1\times 1 \times 300$. The discontinuity of the reference value $h(\vq=0)$ is removable.
                \label{fig:hq_1d}
        }        
\end{figure}

\REV{\Cref{fig:hq_1d} illustrates the $\vq$-point mesh $\mathcal{K}_\vq$ and the computed $h(\vq)$ in the standard and the staggered mesh methods for a quasi-1D model system.
        We note that the staggered mesh method successfully avoids sampling $h(\vq)$ at $\vq = \bm{0}$. It also avoids sampling discontinuous points of the integrand in \cref{eqn:structure_factor}, and the computed values of $h(\vq)$ are more accurate than those computed by the standard method at every sampled point.}

We further consider the error for estimating the integrand $h(\vq)$ in \cref{eqn:hq} \REV{with different mesh sizes}.
For quasi-1D systems, we consider the evaluation of $h(\vq)$ at $\vq_1 = (0, 0, \pi)$. 
This particular point is selected because $h(\vq_1)$ can be directly evaluated by the standard method when $N_\vk$ is even, and by the staggered mesh method when $N_\vk$ is odd.
Similarly, for quasi-2D and 3D systems, we consider the evaluation of $h(\vq)$ at $\vq_2 = (0, \pi, \pi)$ and 
$\vq_3 = (\pi, \pi, \pi)$, respectively.

\begin{figure}[htbp]
    \centering
    \subfloat[Quasi-1D, isotropic]{
        \includegraphics[width=0.33\textwidth]{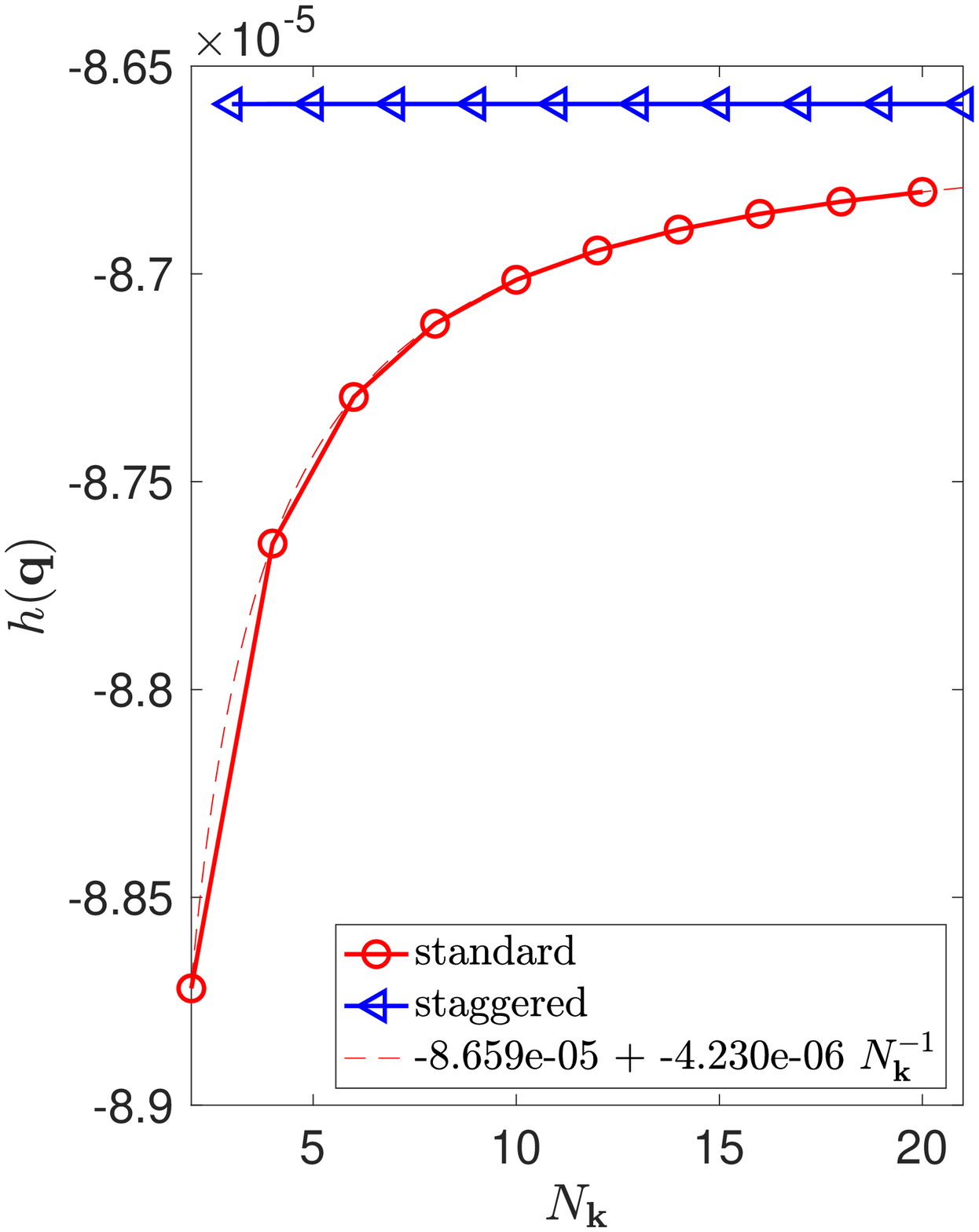}
    }
    \subfloat[Quasi-2D, isotropic]{
        \includegraphics[width=0.33\textwidth]{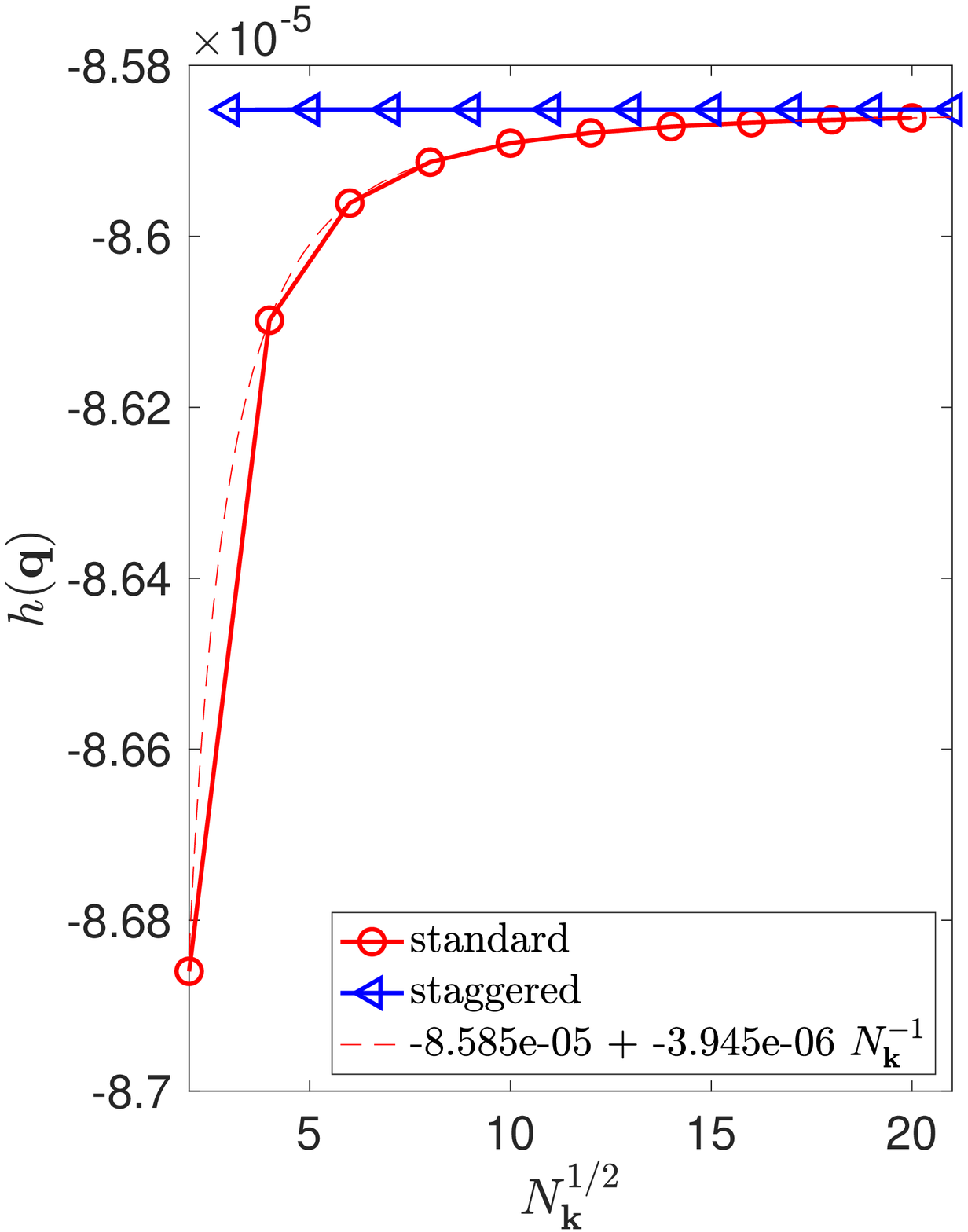}
    }
    \subfloat[3D, isotropic]{
        \includegraphics[width=0.33\textwidth]{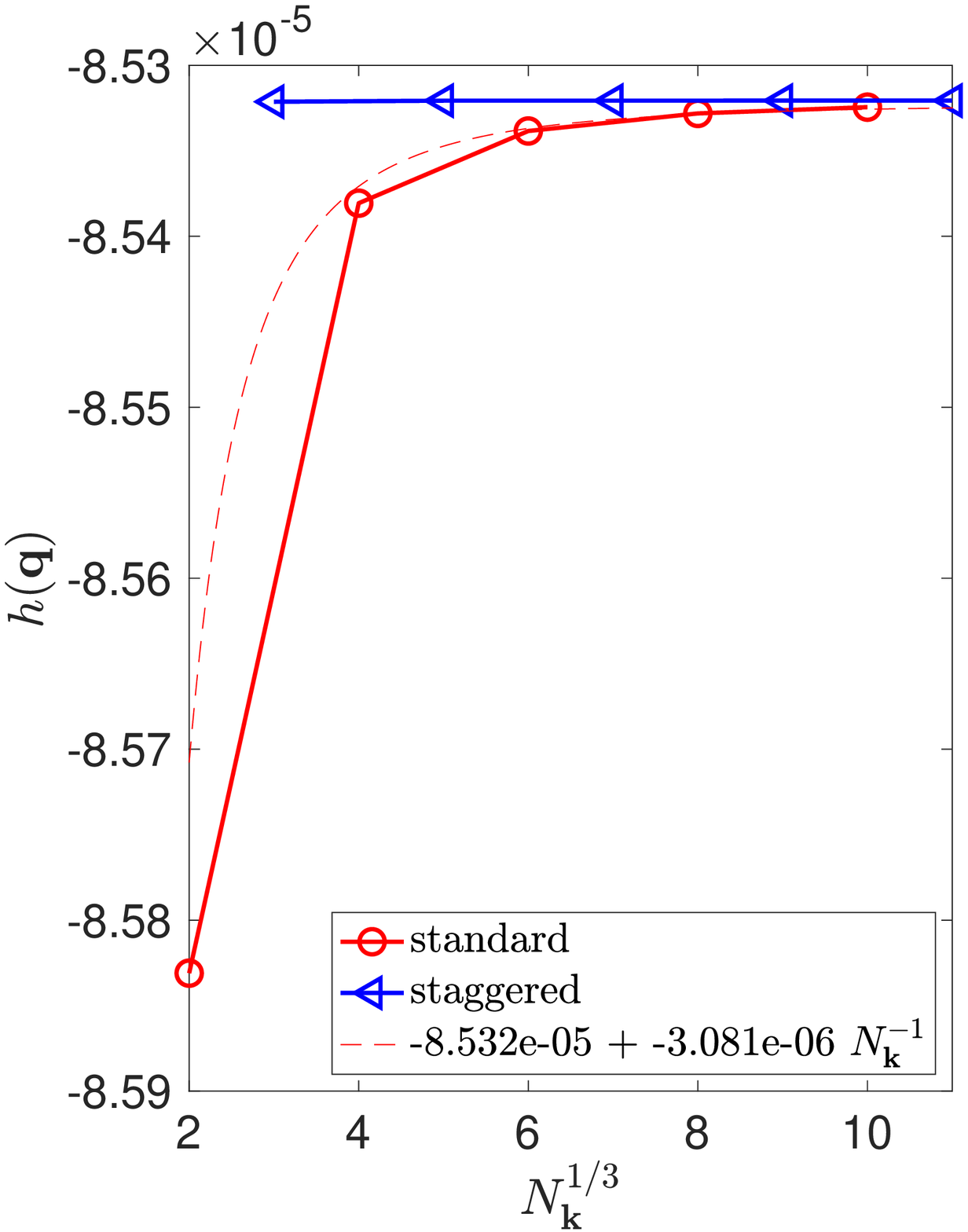}
    }
    
     \subfloat[Quasi-1D, anisotropic]{
        \includegraphics[width=0.33\textwidth]{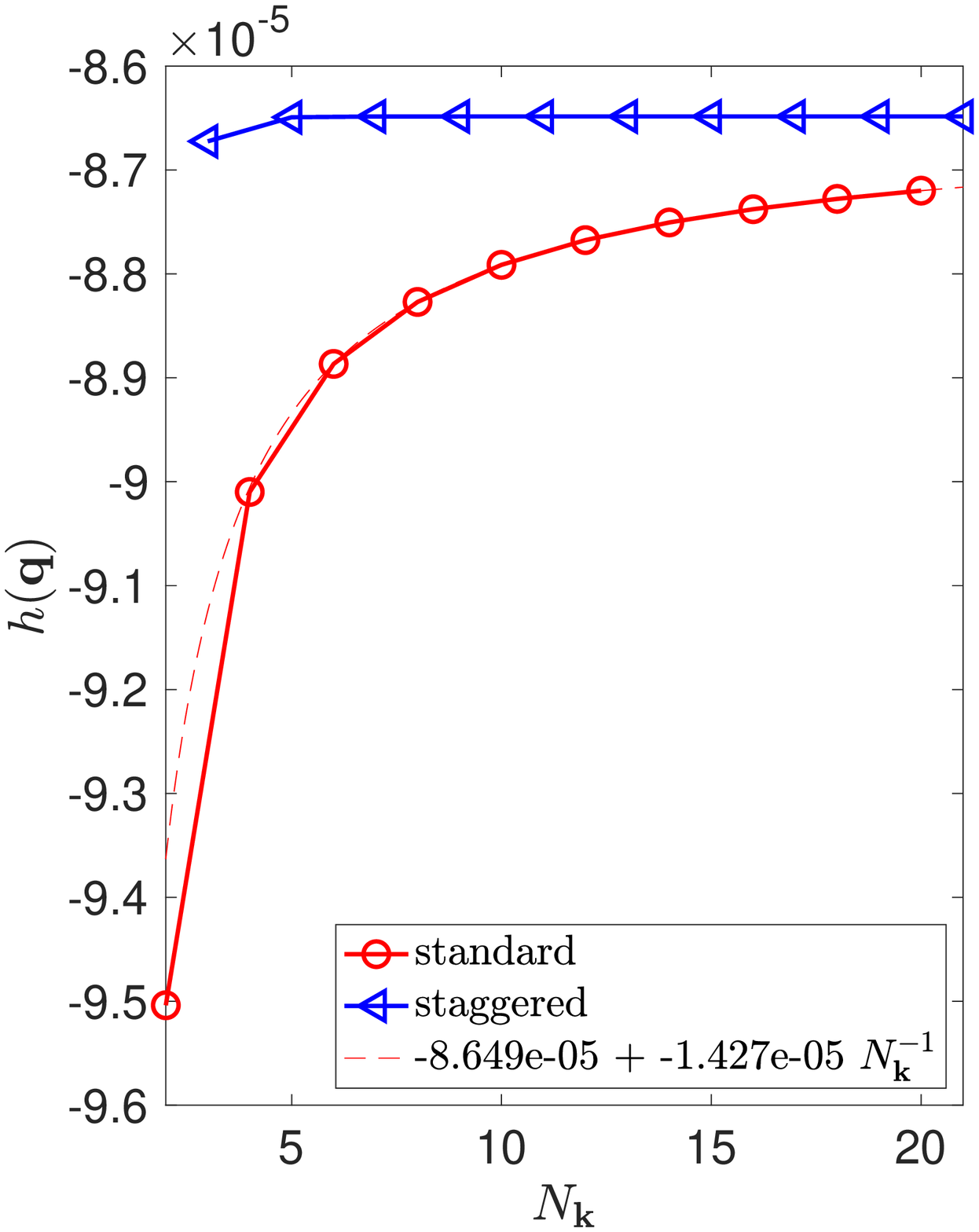}
    }
    \subfloat[Quasi-2D, anisotropic]{
        \includegraphics[width=0.33\textwidth]{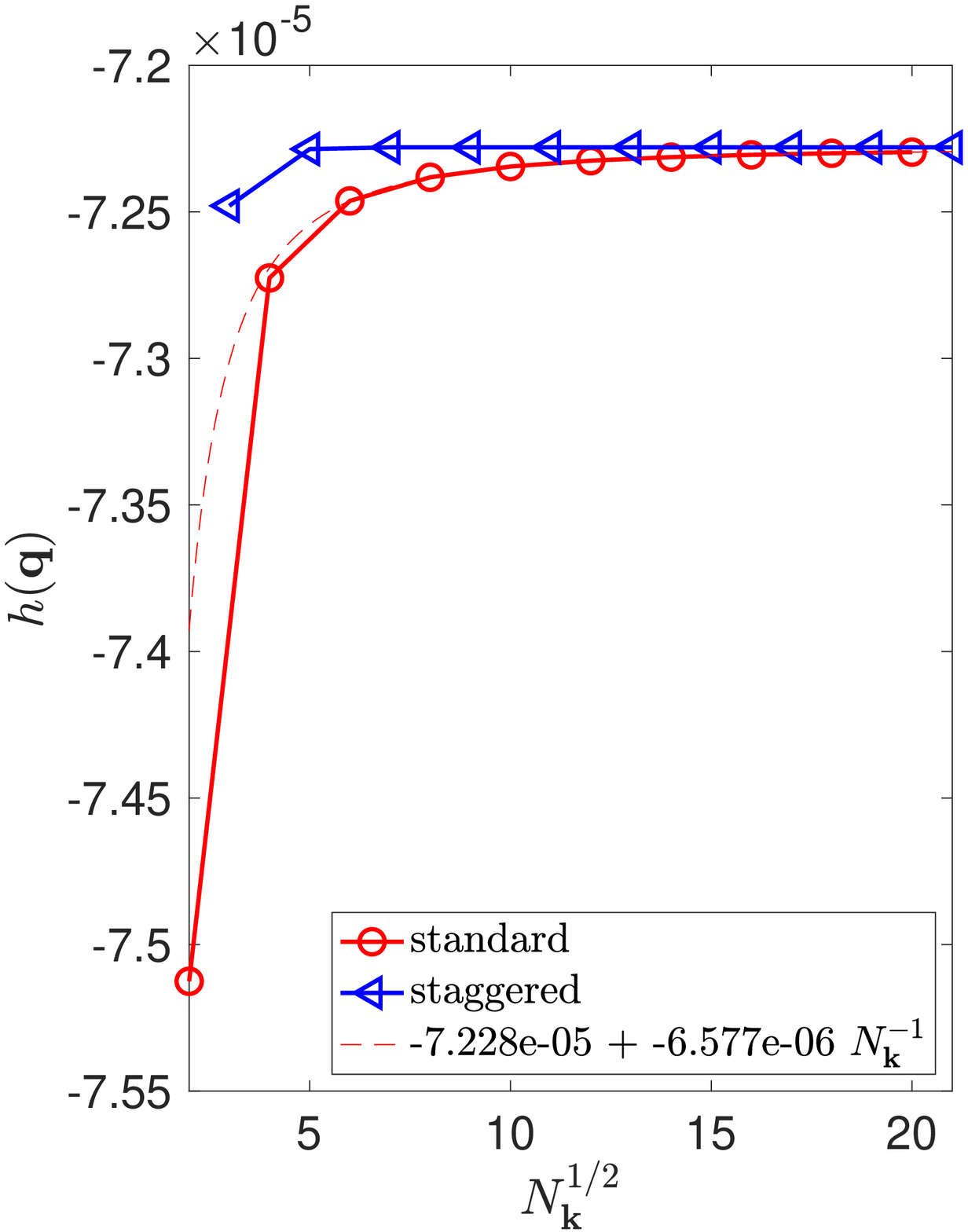}
    }
    \subfloat[3D, anisotropic]{
        \includegraphics[width=0.33\textwidth]{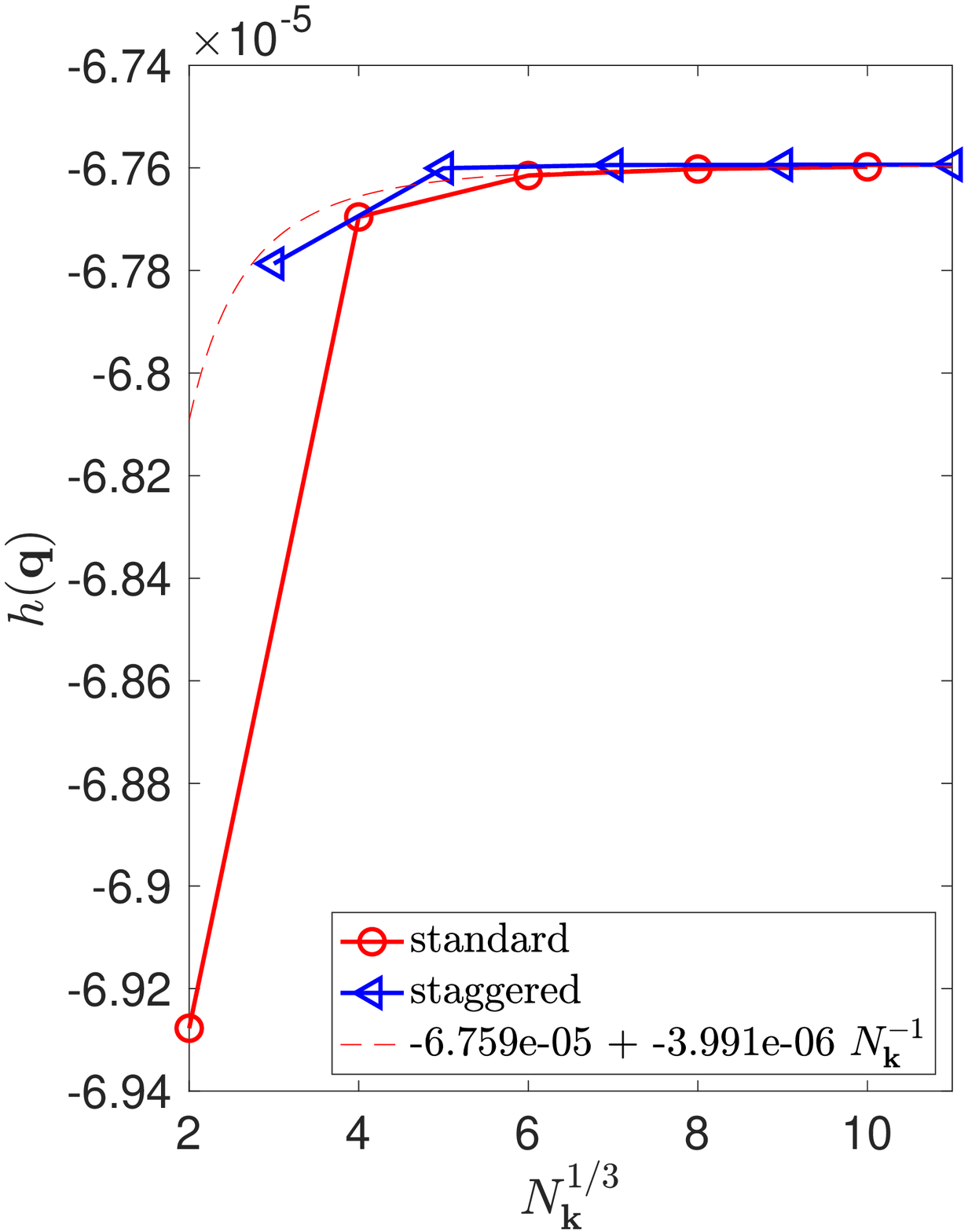}
    }
    
    \caption{Estimate of $h(\vq)$ at $\vq_1$/$\vq_2$/$\vq_3$ using the standard and the staggered mesh methods for quasi-1D/quasi-2D/3D model systems with isotropic and anisotropic Gaussian effective potential fields. 
    Each of these curve fittings omits the first two data points.}
    \label{fig:estimate_hq}
\end{figure}

\Cref{fig:estimate_hq} demonstrates the convergence of $h(\vq)$ with respect to $N_{\vk}$ using the standard and the staggered mesh methods. For all the systems, we find that the finite-size error of the staggered mesh method in estimating $h(\vq)$ at $\vq \neq \bm{0}$ is much smaller than that of the standard method, regardless of the dimension or the anisotropicity of the system. 

\begin{figure}[htbp]
    \centering
    \subfloat[Quasi-1D, isotropic]{
        \includegraphics[width=0.33\textwidth]{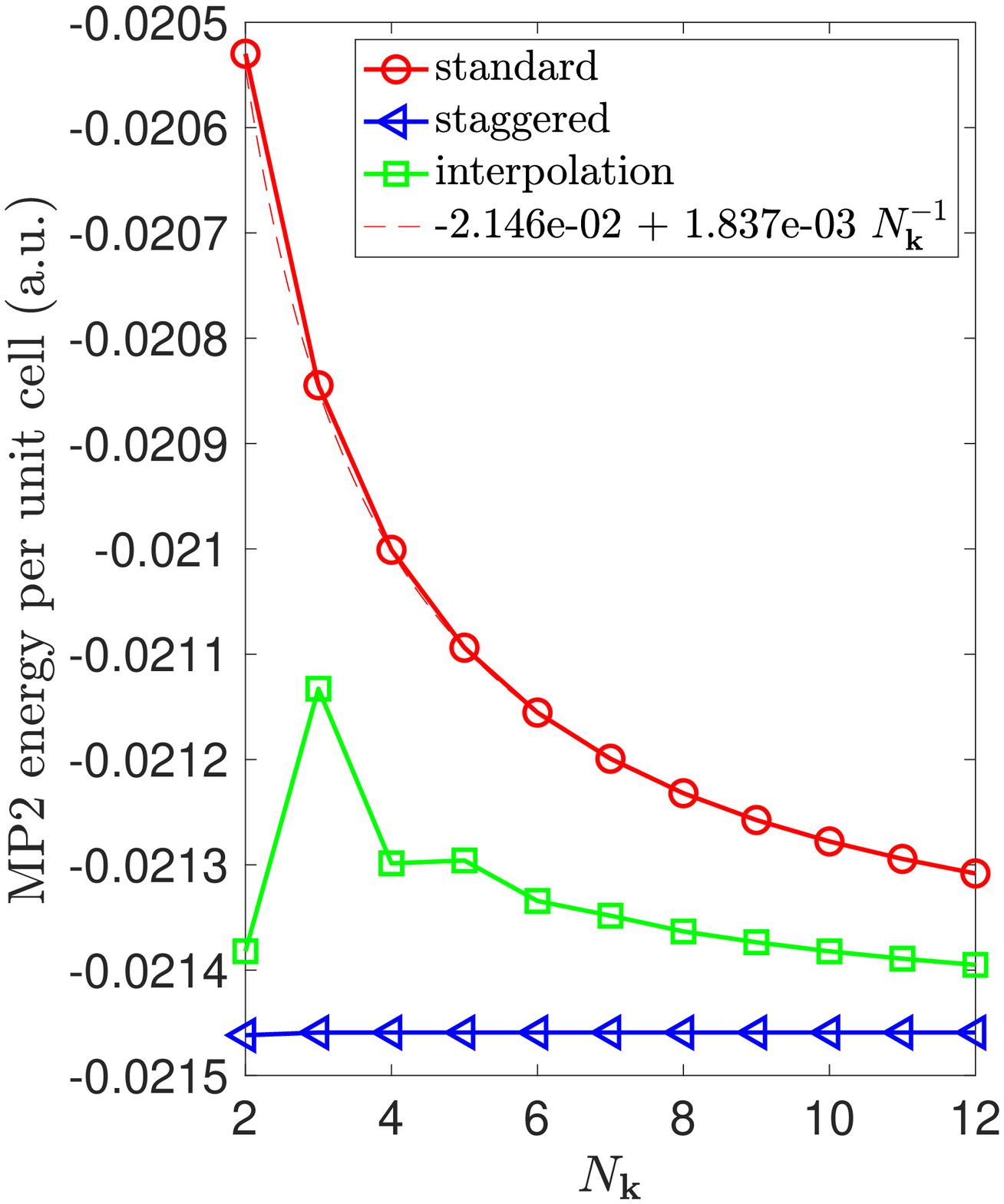}
        }
        \subfloat[Quasi-2D, isotropic]{
                \includegraphics[width=0.33\textwidth]{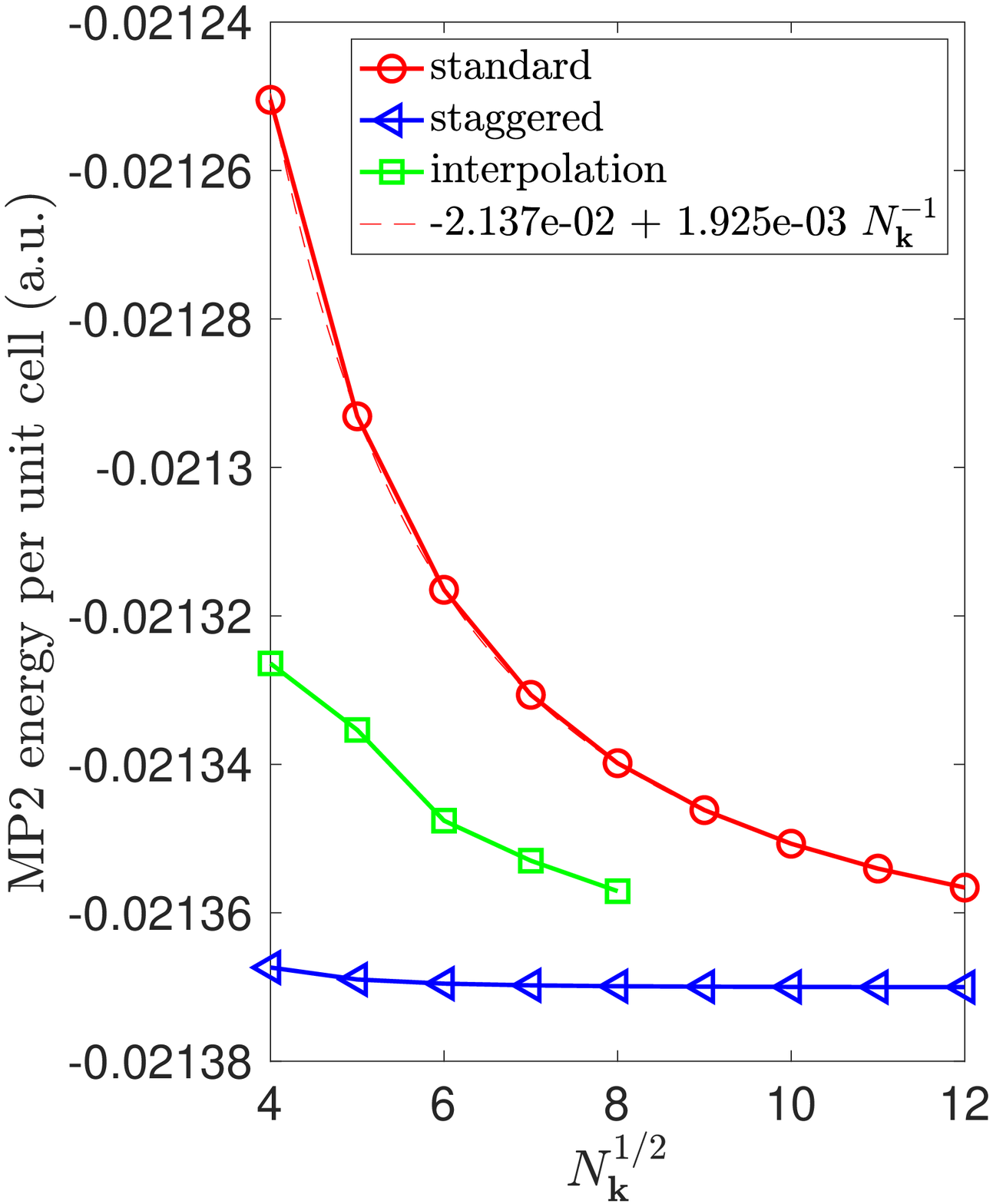}
        }
        \subfloat[3D, isotropic]{
                \includegraphics[width=0.33\textwidth]{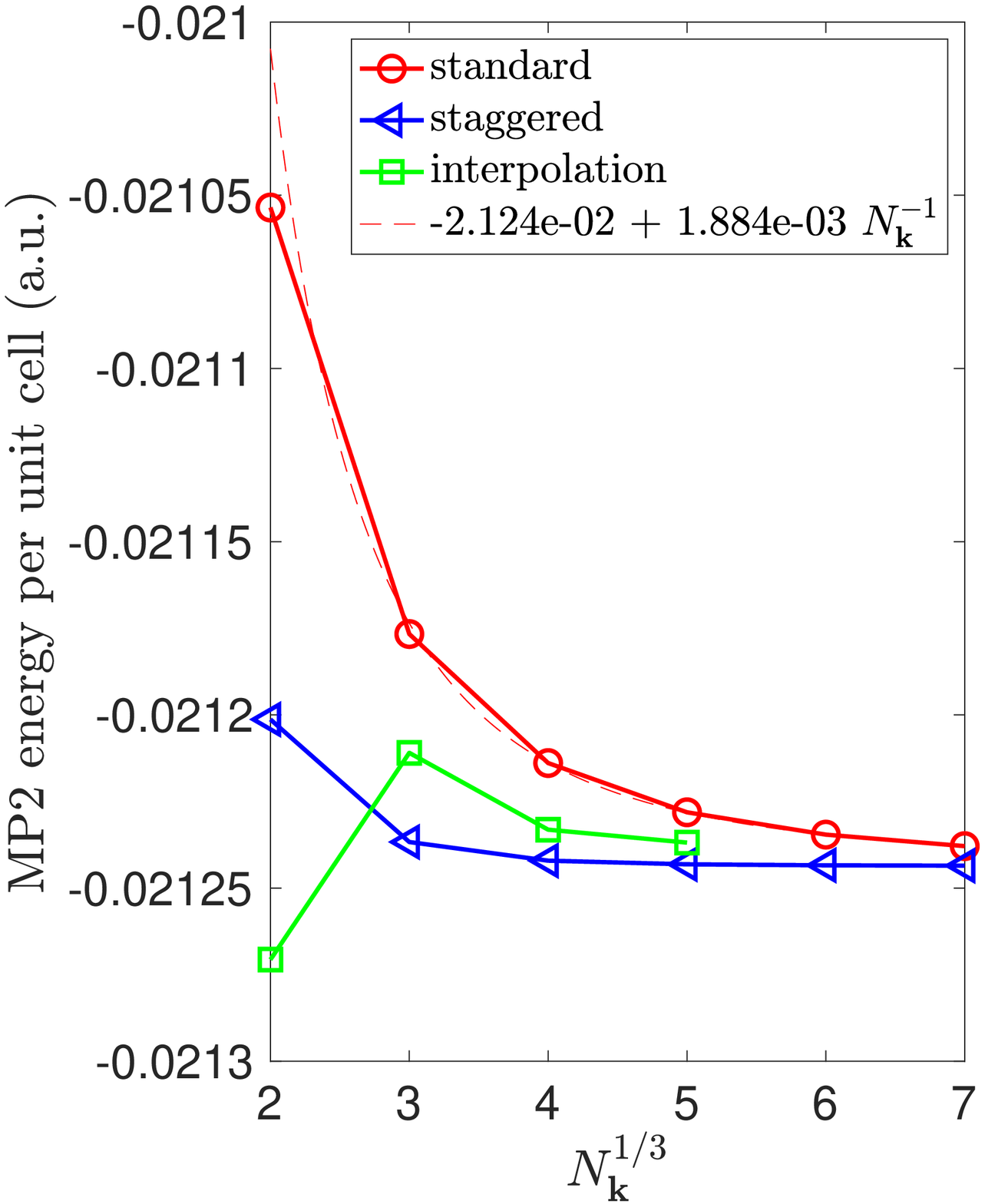}
        }
        
        \subfloat[Quasi-1D, anisotropic]{
                \includegraphics[width=0.33\textwidth]{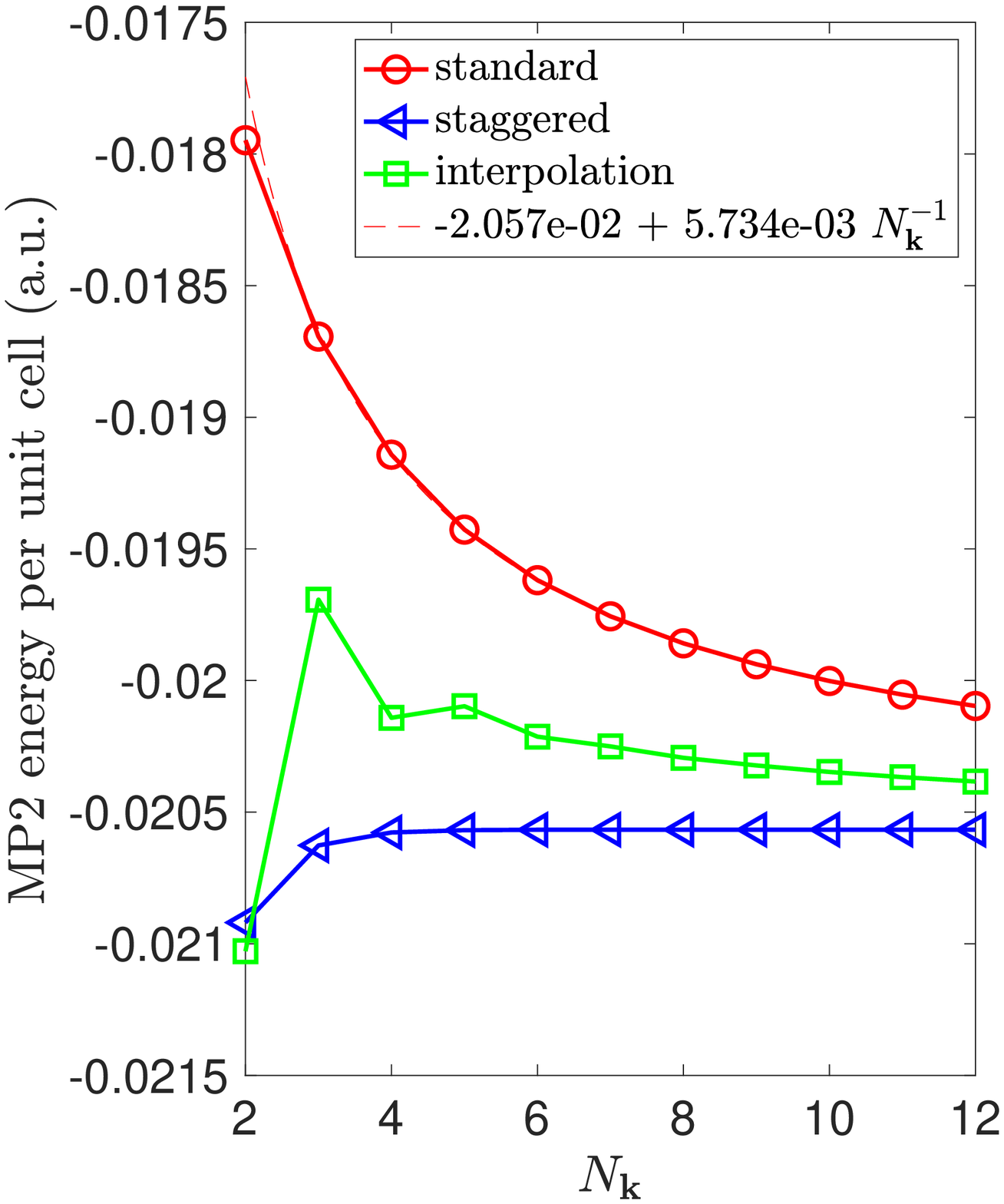}
        }
        \subfloat[Quasi-2D, anisotropic]{
                \includegraphics[width=0.33\textwidth]{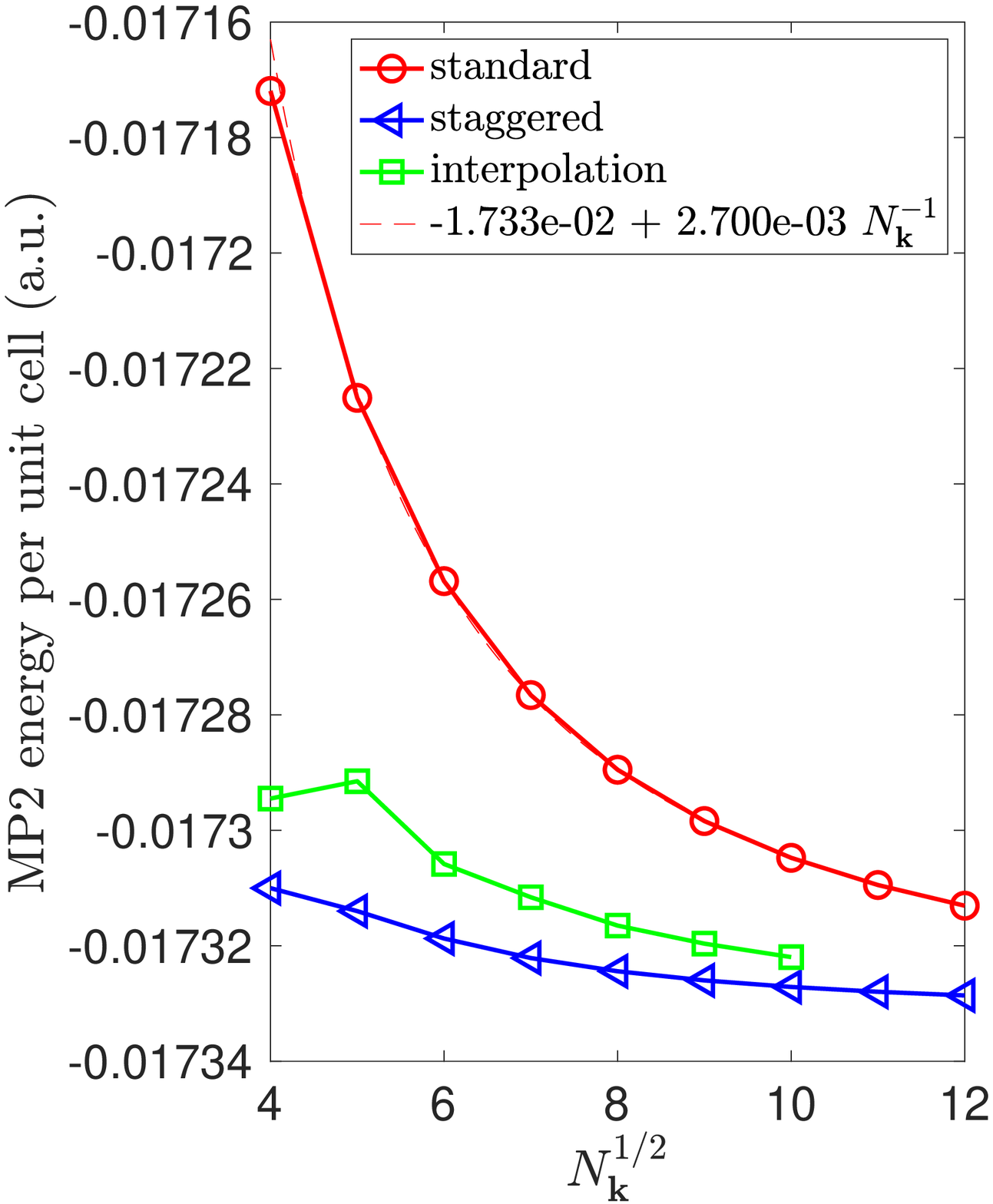}
        }
        \subfloat[3D, anisotropic]{
                \includegraphics[width=0.33\textwidth]{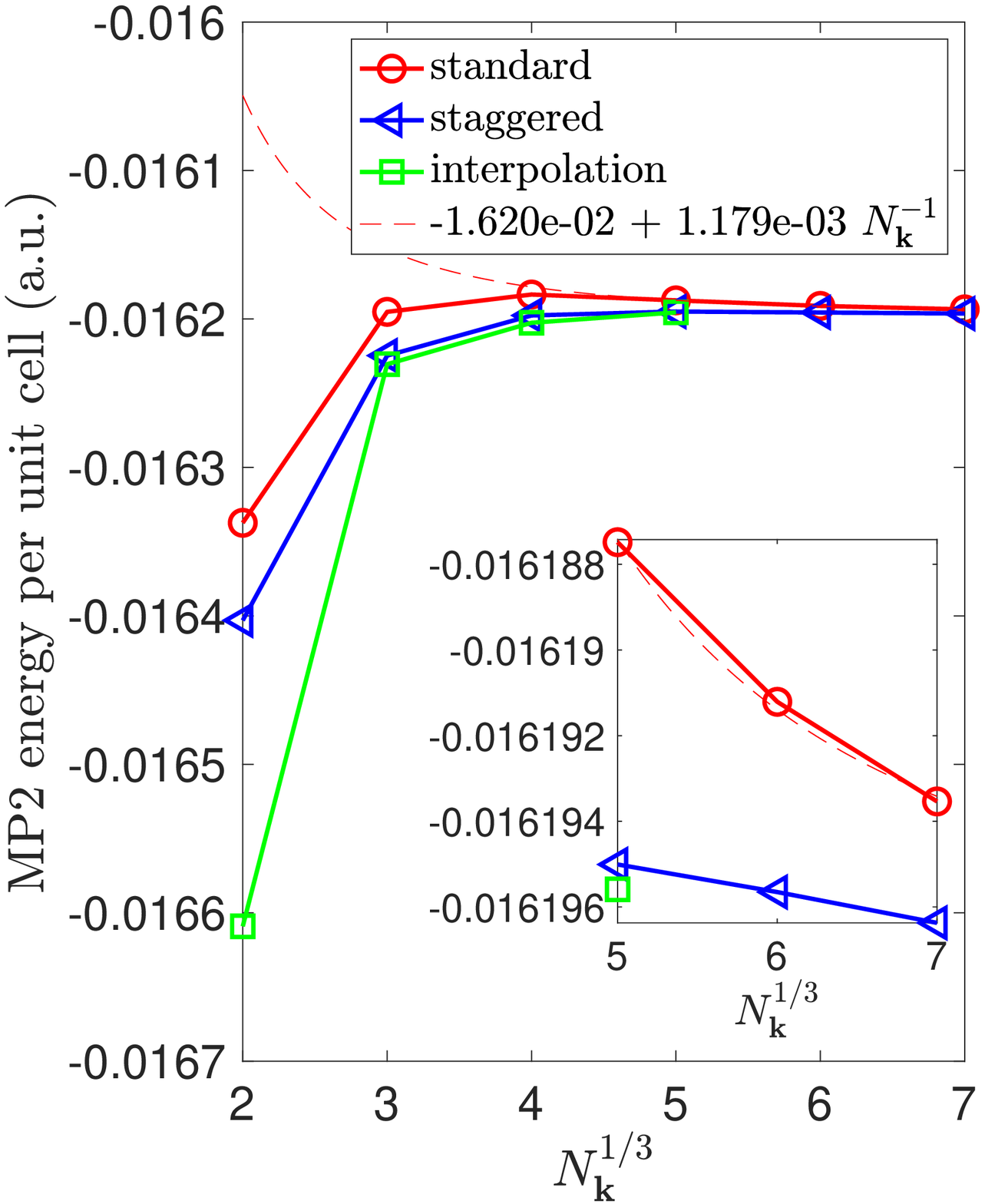}
        }
    \caption{MP2 energy per unit cell computed by the standard method,  the staggered mesh method, and \REV{the structure factor interpolation method} for quasi-1D, quasi-2D, and 3D model systems with isotropic and anisotropic Gaussian effective potential fields.
    Each of these curve fittings omits the first two or three data points.
    \REV{
        In most cases, the decay rate of finite-size error in the staggered mesh method is unclear, and thus no curve fitting is plotted for the method in all the figures.
        Due to excessive computational cost, results of the structure factor interpolation method for some large systems are not provided. Separate plots of the direct and the exchange parts of the MP2 energies are given in the Appendix.
    }
        }
    \label{fig:model_mp2}
\end{figure}

\Cref{fig:model_mp2} demonstrates the convergence of the MP2 correlation energy per unit cell computed by the standard method, the staggered mesh method, and \REV{the structure factor interpolation method}\cite{LiaoGrueneis2016,GruberLiaoTsatsoulisEtAl2018} for quasi-1D, quasi-2D, and 3D model systems. 
\REV{For each system, the structure factor $S_\vq(\vG)$ is computed by the standard method, and then extrapolated by cubic interpolation to a mesh that is $50$ times finer along each extended dimension compared to the original mesh for $\vq + \vG$.}
For quasi-1D systems, we find that the finite-size errors in the staggered mesh method decay very rapidly with respect to $N_\vk$, and the curve is nearly flat.
For quasi-2D and 3D model systems, the finite-size errors of the staggered mesh method are also much smaller for the isotropic systems. However, for the anisotropic systems, the convergence rates of the two methods are comparable \REV{and both numerically close to $\Or(N_\vk^{-1})$,} though the error of the staggered mesh method still exhibits a smaller preconstant.
\REV{
The varying performance of the staggered mesh method in different systems, and the remaining $\Or(N_\vk^{-1})$ quadrature error in the staggered mesh method for anisotropic quasi-2D and 3D systems are both closely related to the lack of overall smoothness in the integrand of MP2 calculation  \cref{eqn:correlation_TDL}, which will be discussed in more details in \cref{sec:discussion}. 
}

\REV{
We also observe that the performance of the structure factor interpolation lies between that of the standard and the staggered mesh methods. 
This is because the quality of the interpolation still suffers from the inherent discontinuity (see \cref{fig:integrand_discontinuity} (c)) and the quadrature error in the structure factor computed from the standard MP2 method.}

\subsection{Real systems}\label{sec:numer_pyscf}
We have implemented the staggered mesh method in the PySCF \cite{SunBerkelbachEtAl2018} software package. In order to focus on the quadrature error, we perform our comparisons between the standard and the staggered mesh methods as follows.
\REV{For each system, we first perform a self-consistent HF calculation with a fixed $\vk$-point mesh, and employ the spherical cutoff method \cite{SpencerAlavi2008} (given by the option exxdiv=`vcut\_sph' in PySCF) to reduce the finite size error due to the Fock exchange operator.
All orbitals and orbital energies used in MP2 calculations are then evaluated via non-self-consistent HF\ calculations at any required $\vk$ points and mesh sizes.
Therefore the orbitals and orbital energies are generated from an effective (non-local) potential field, and do not require further correction to the finite-size errors.} 
\REV{We employ the gth-szv basis set and the gth-pade pseudopotential in all tests. 
Results with the larger gth-dzvp basis set are given in Appendix. 
The kinetic energy cutoff for plane-wave calculations is set to $100$ a.u.\ in all tests.}

\begin{figure}[htbp]
\centering
\subfloat[Quasi-1D]{
        \includegraphics[width=0.33\textwidth]{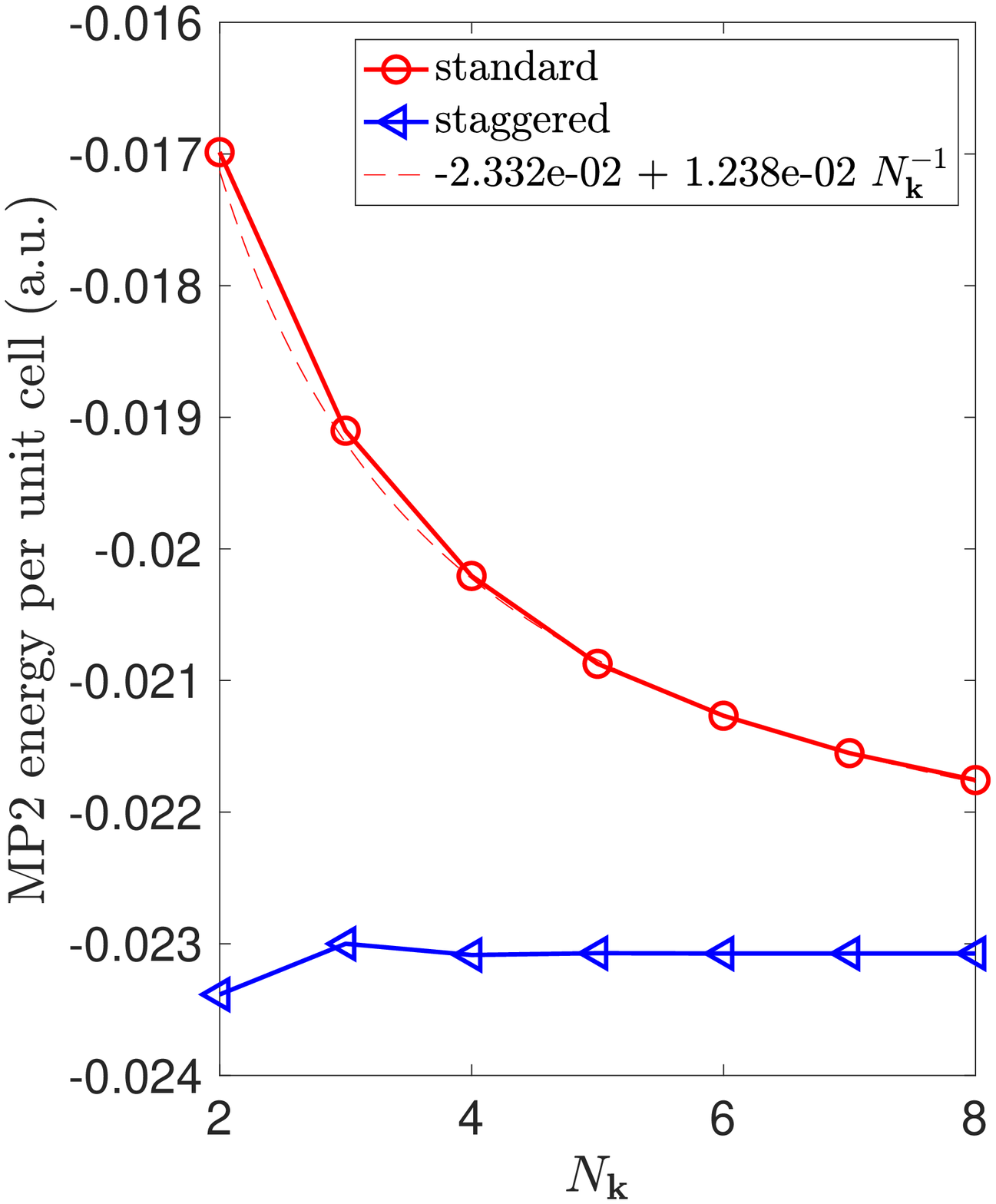}
}
\subfloat[Quasi-2D]{
        \includegraphics[width=0.33\textwidth]{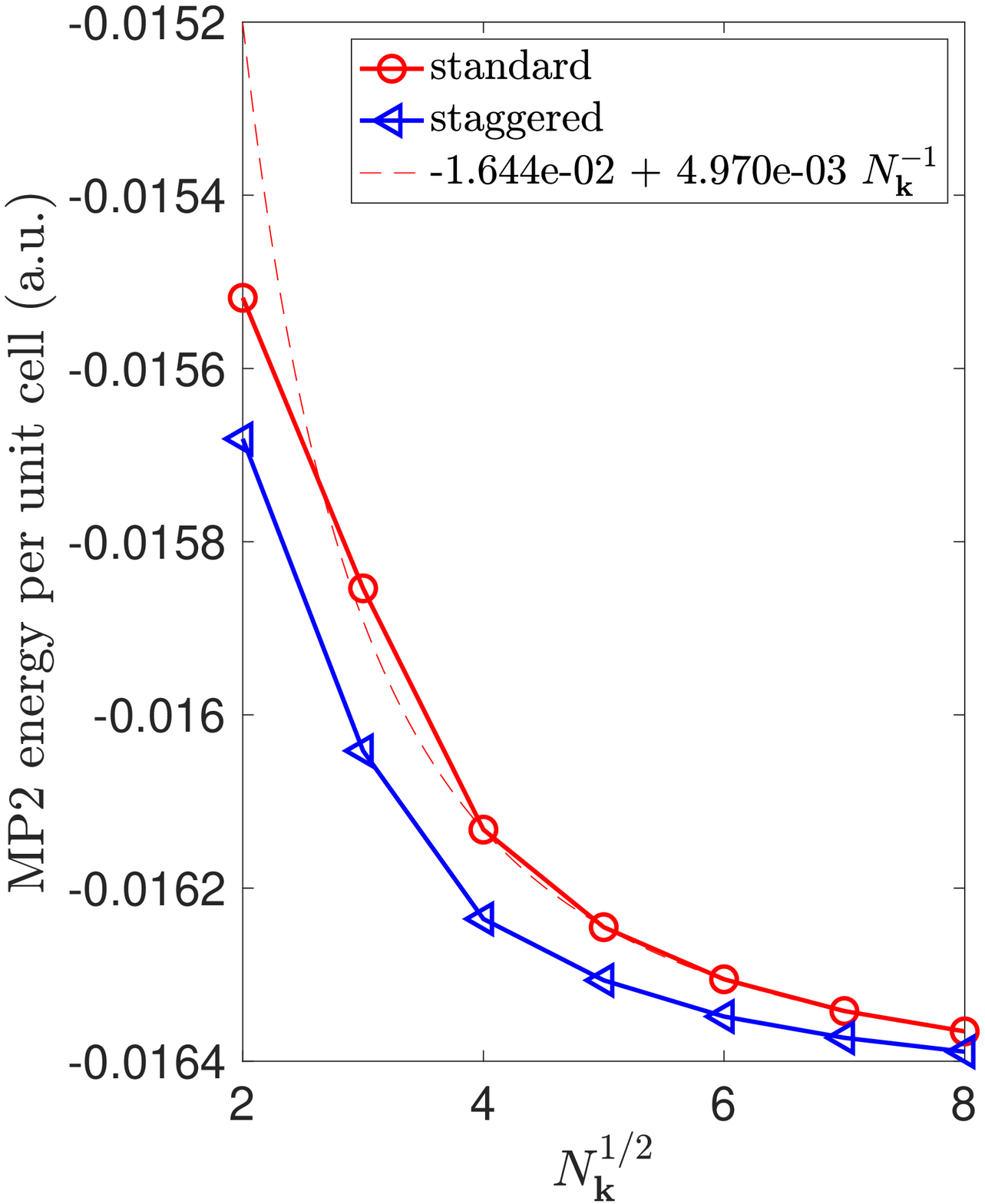}
}
\subfloat[3D]{
        \includegraphics[width=0.33\textwidth]{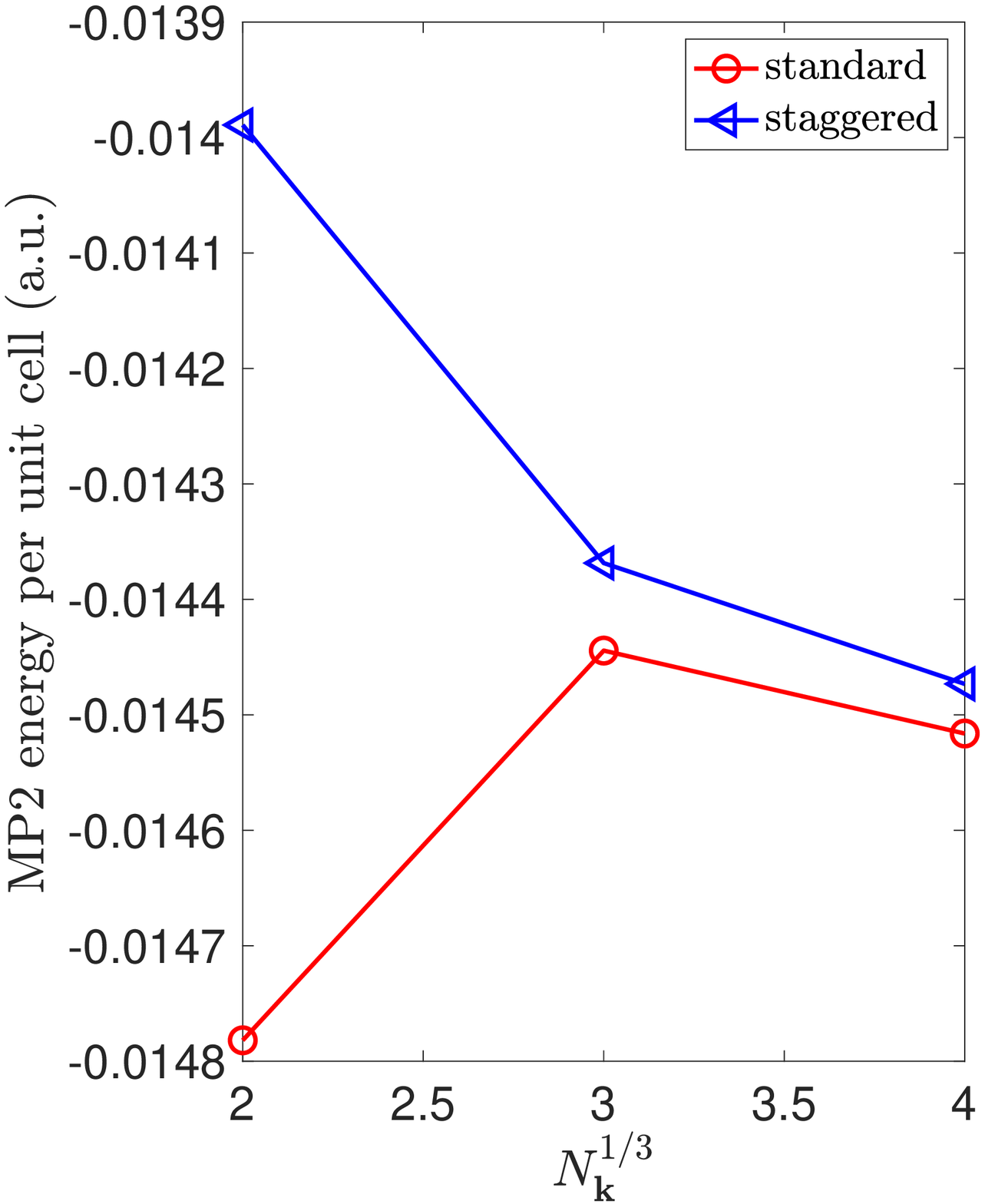}
}

\caption{MP2 energy per unit cell computed by the standard and the staggered mesh methods for periodic hydrogen dimer systems.
}
\label{fig:H2_mp2_energy}
\end{figure}

\begin{figure}[htbp]
        \centering
        \subfloat[Quasi-1D]{
                \includegraphics[width=0.33\textwidth]{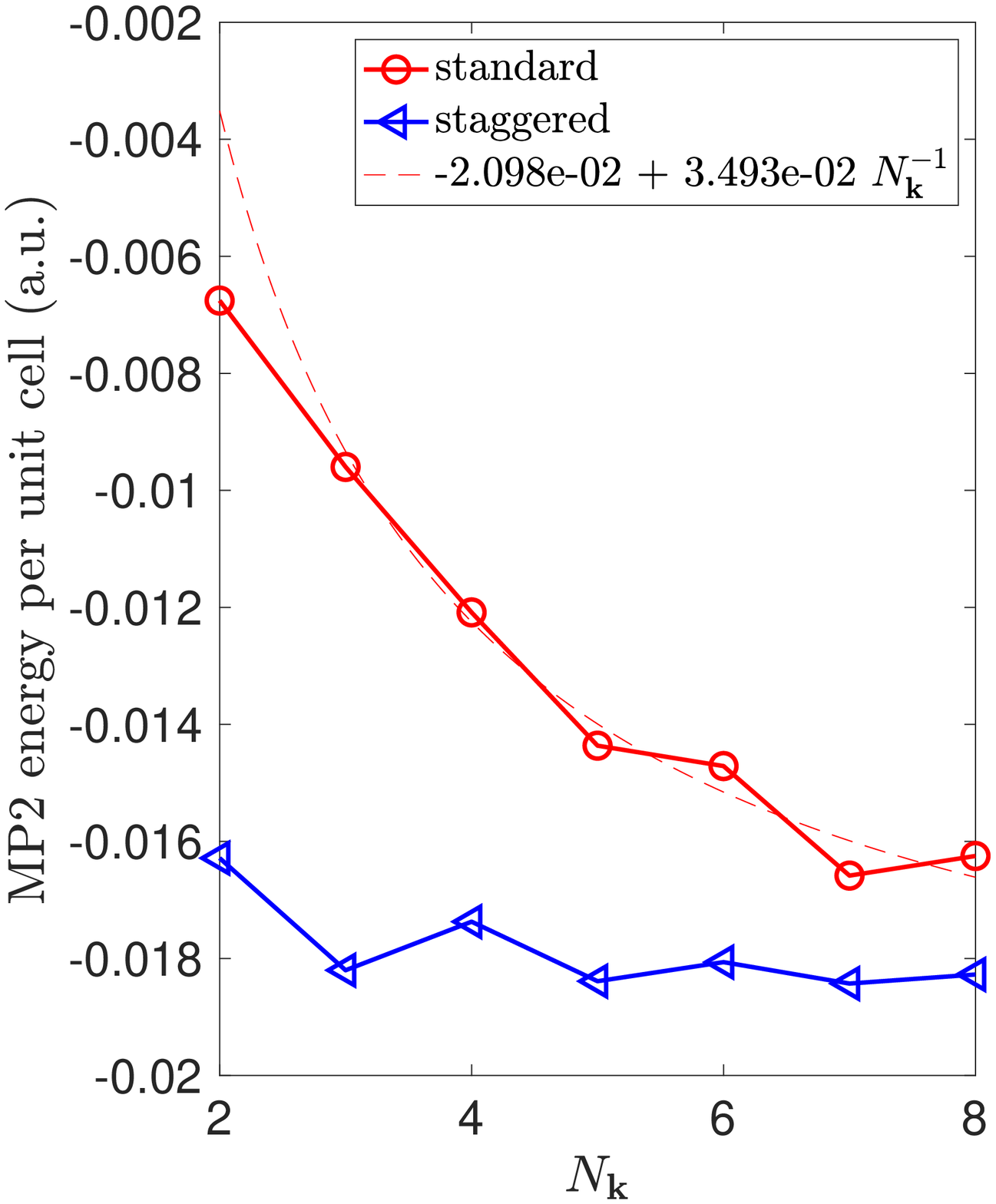}
        }
        \subfloat[Quasi-2D]{
                \includegraphics[width=0.33\textwidth]{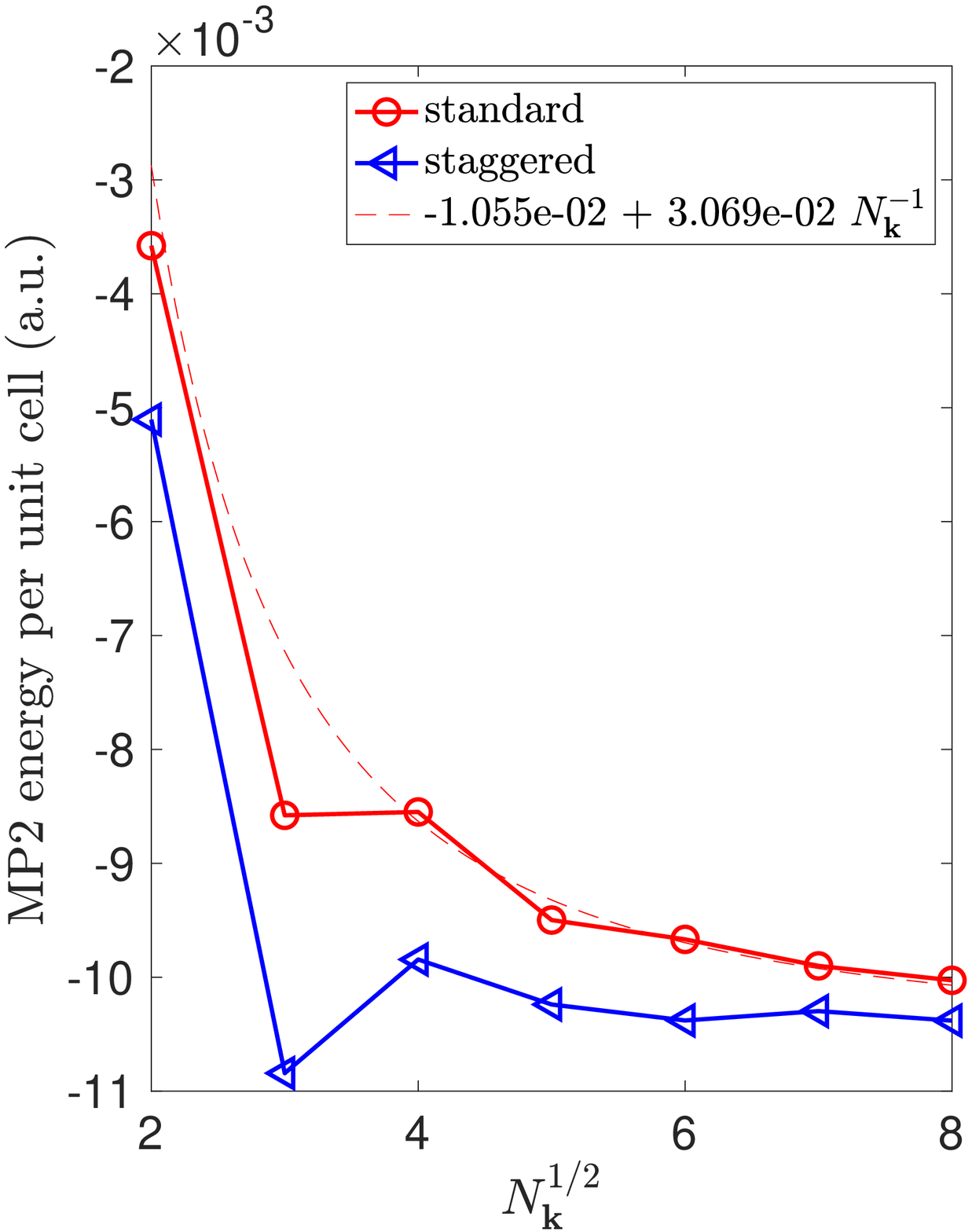}
        }
        \subfloat[3D]{
                \includegraphics[width=0.33\textwidth]{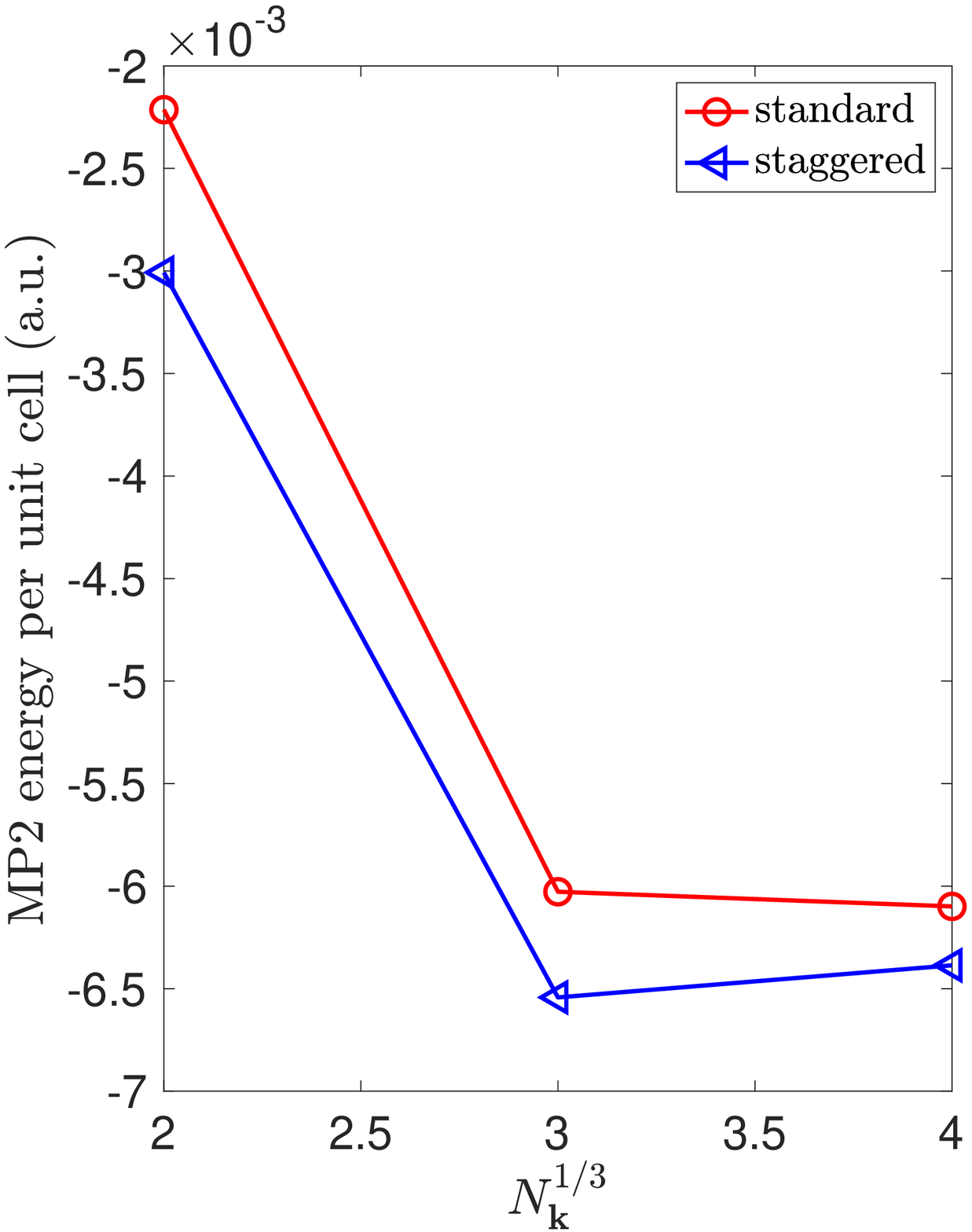}
        }
        
        \caption{MP2 energy per unit cell computed by the standard and the staggered mesh methods for periodic LiH systems. The fluctuation might be due to the small size of the basis set, as the amount of fluctuation is reduced when using the larger gth-dzvp basis set in \cref{fig:dzvp_mp2}.
        }
        \label{fig:LiH_mp2_energy}
\end{figure}

\begin{figure}[htbp]
        \centering
        \subfloat[Quasi-1D]{
                        \includegraphics[width=0.33\textwidth]{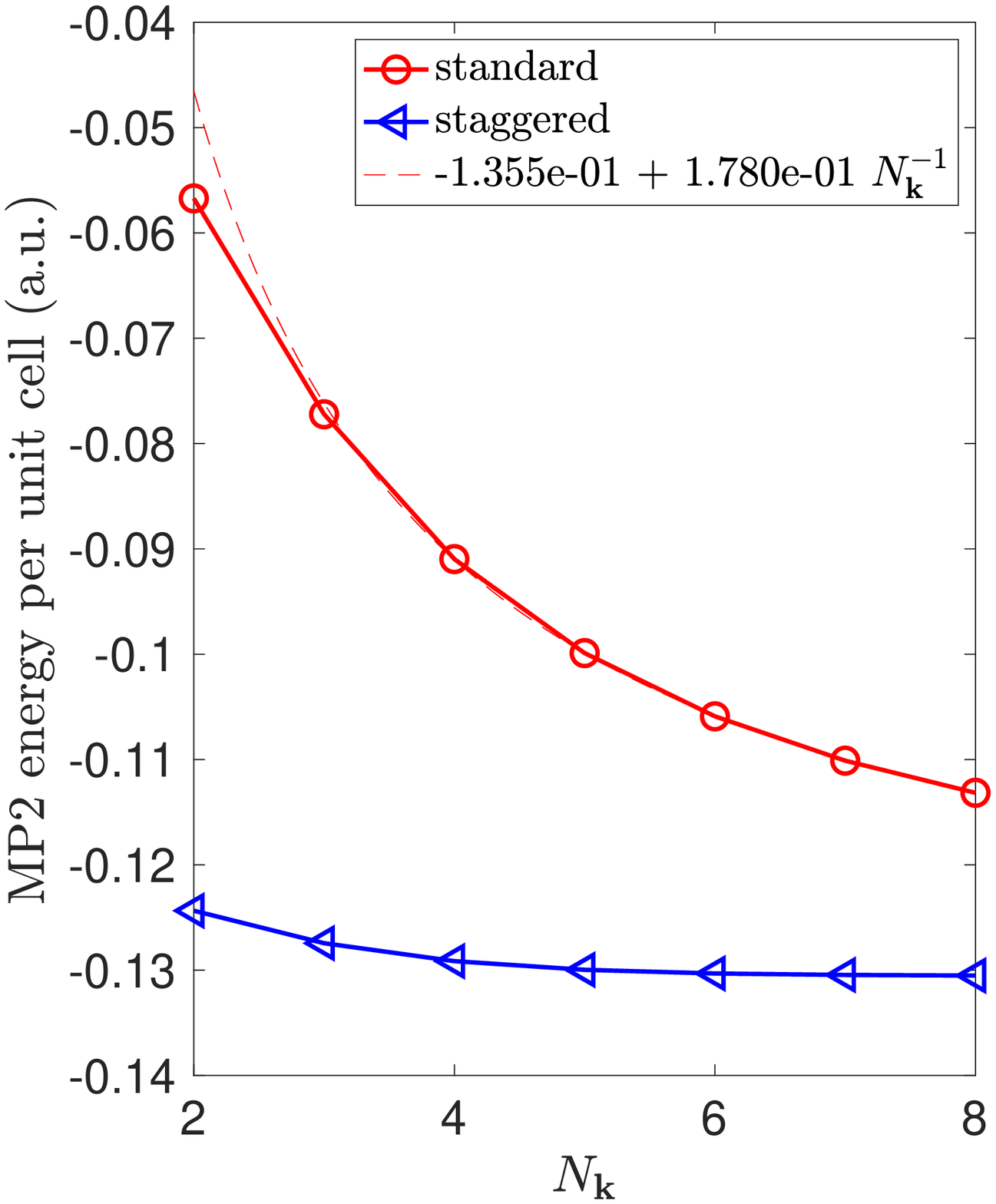}
        }
        \subfloat[Quasi-2D]{
                        \includegraphics[width=0.33\textwidth]{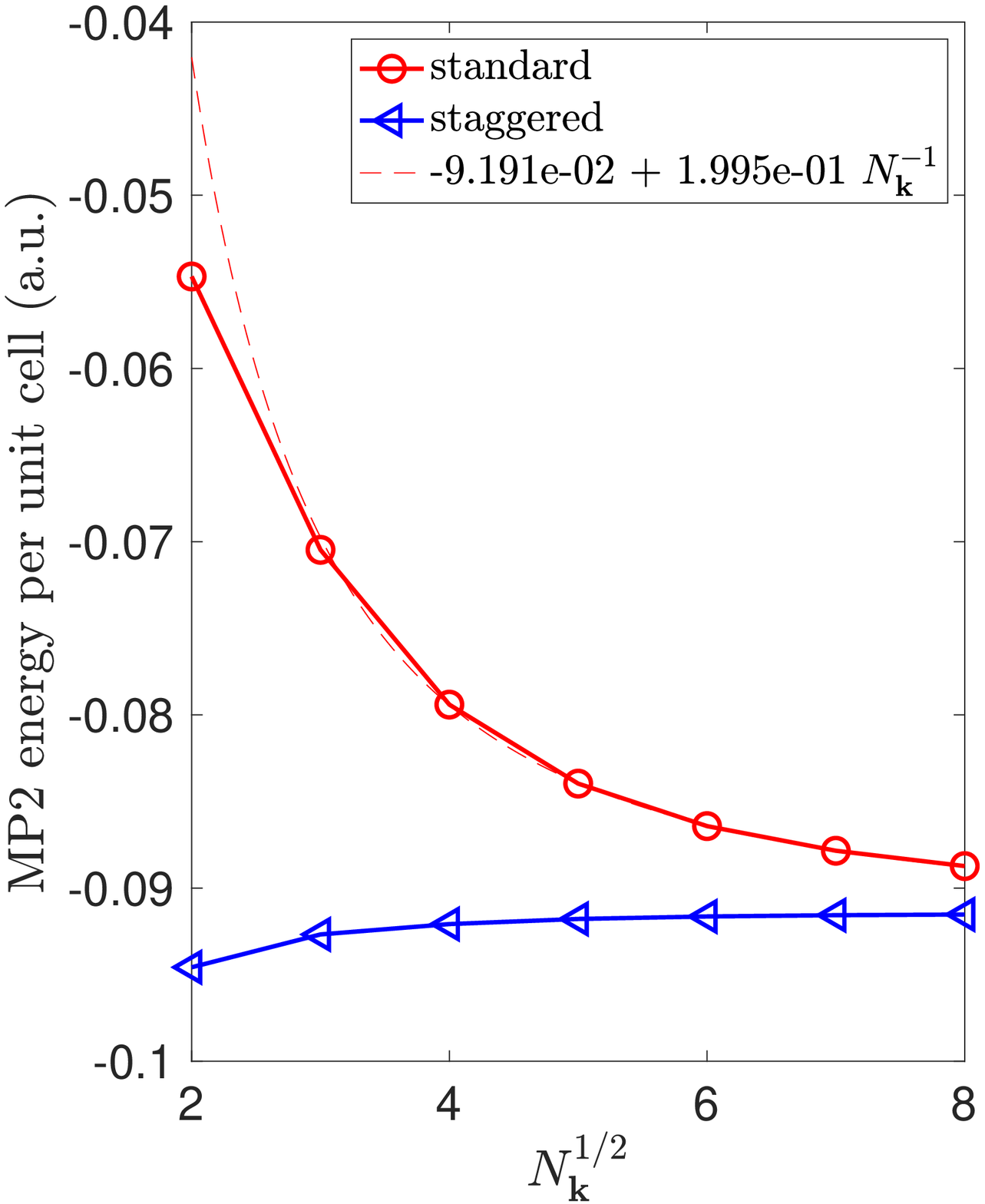}
        }
        \subfloat[3D]{
                        \includegraphics[width=0.33\textwidth]{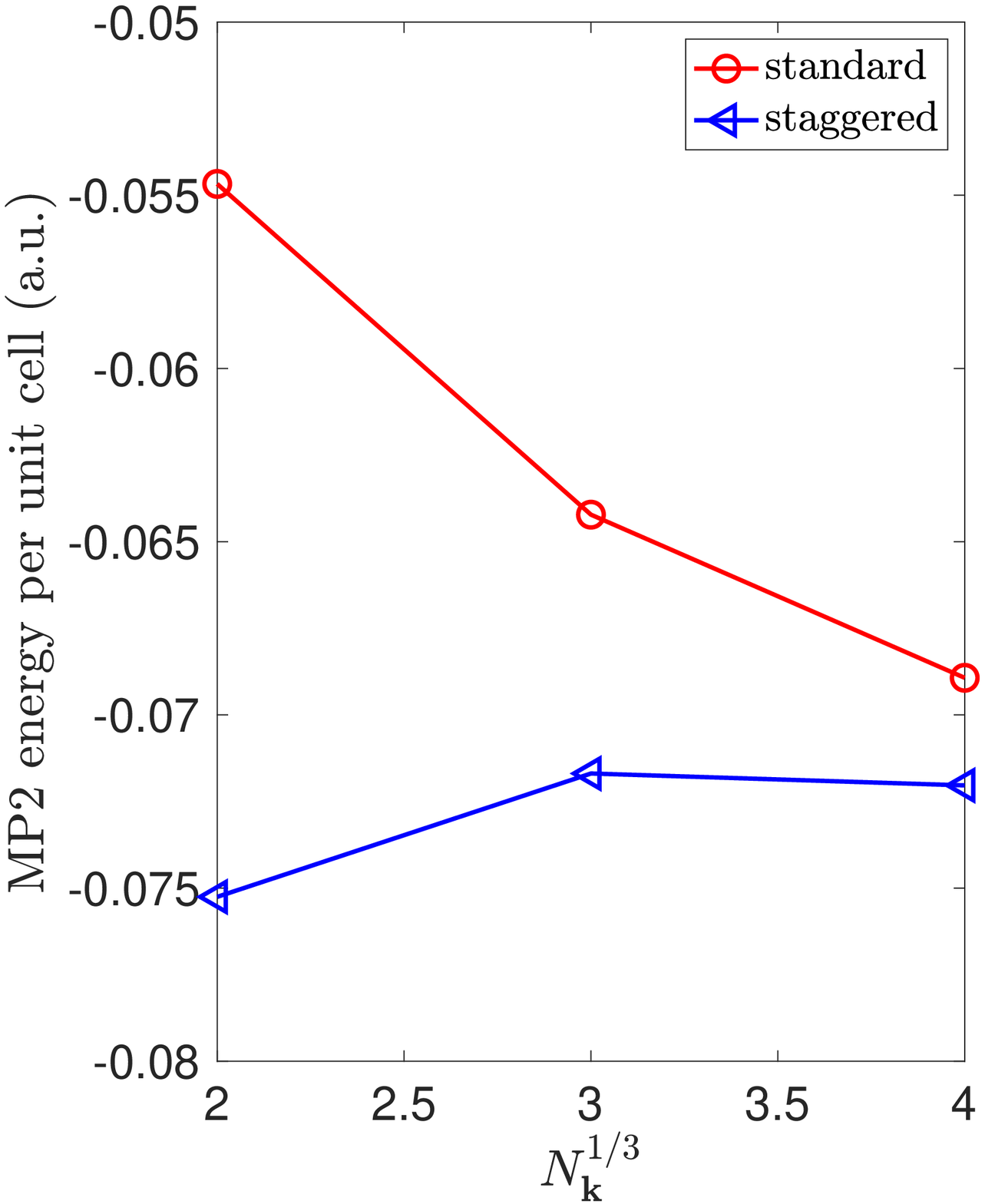}
        }
        
        \caption{MP2 energy per unit cell computed by the standard and the staggered mesh methods for periodic silicon systems.
        }
        \label{fig:Si_mp2_energy}
\end{figure}

\begin{figure}[htbp]
        \centering
        \subfloat[Quasi-1D]{
                \includegraphics[width=0.33\textwidth]{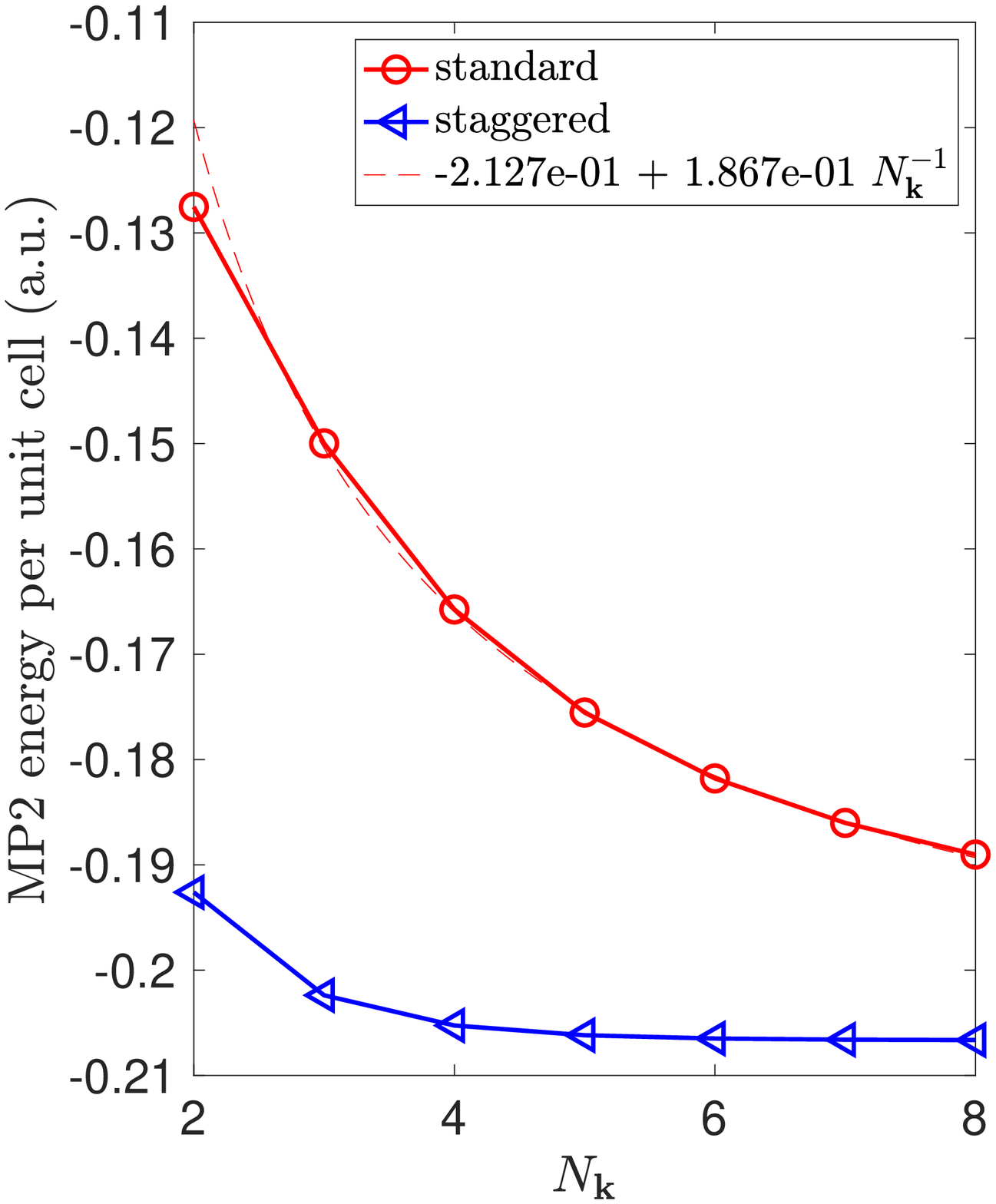}
        }
        \subfloat[Quasi-2D]{
                \includegraphics[width=0.33\textwidth]{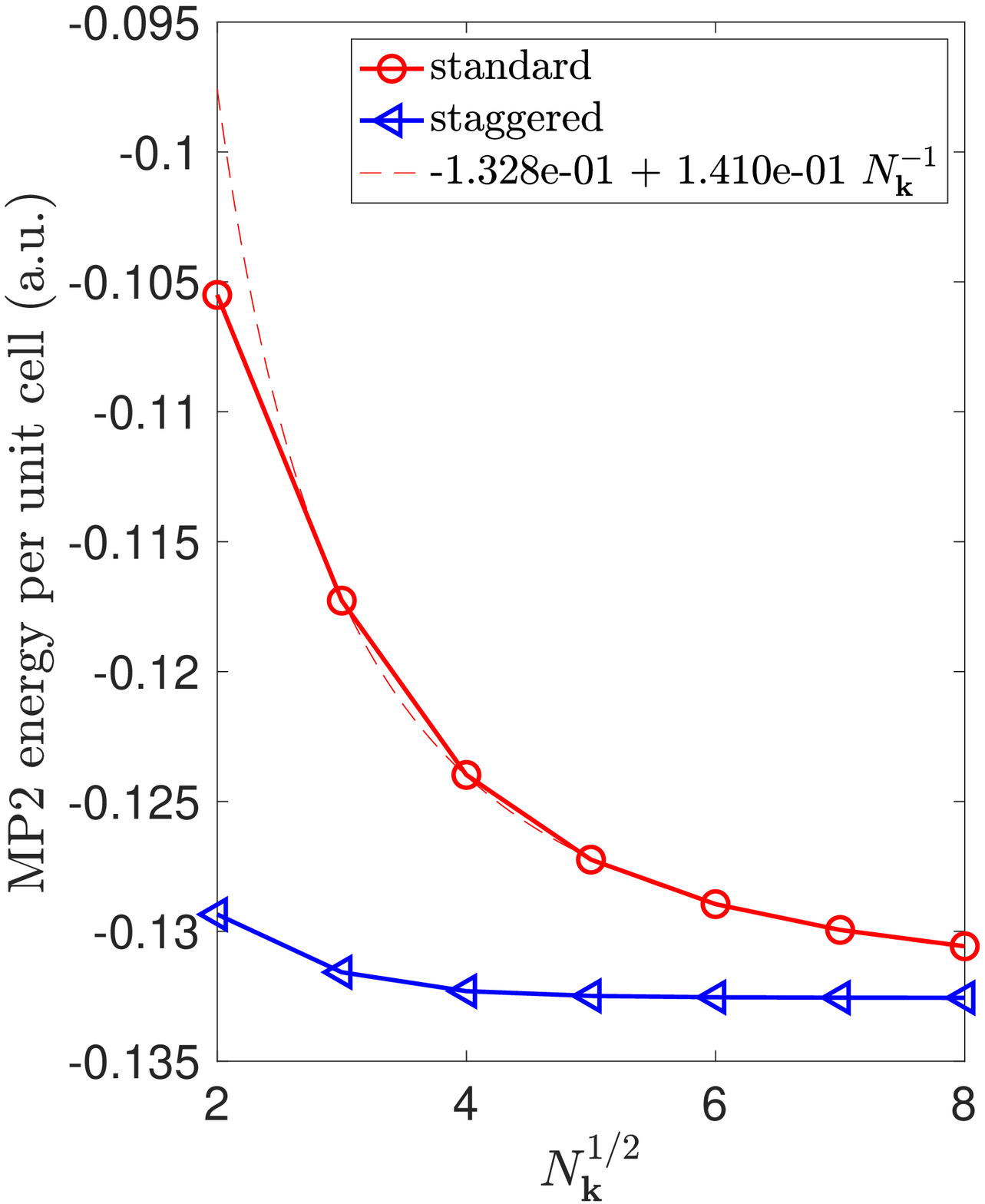}
        }
        \subfloat[3D]{
                \includegraphics[width=0.33\textwidth]{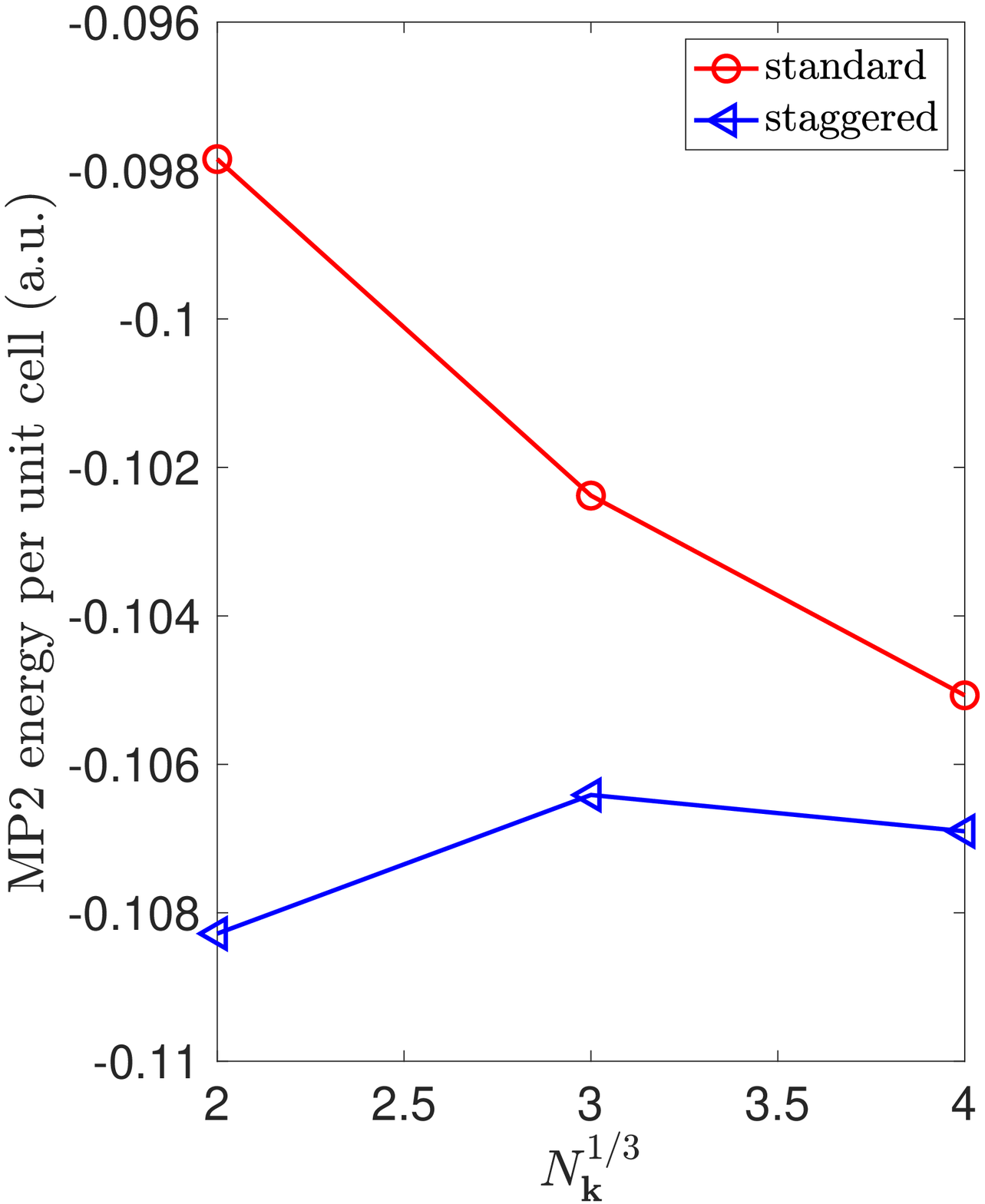}
        }
        
        \caption{MP2 energy per unit cell computed by the standard and the staggered mesh methods for periodic diamond systems.
        }
        \label{fig:diamond_mp2_energy}
\end{figure}

We consider four sets of periodic systems: \REV{hydrogen dimer, lithium hydride, silicon, and diamond.}
The hydrogen dimer is placed at the center of a cubic unit cell of edge length $6$ Bohr pointing in the $x$-direction and has separating distance $1.8$ Bohr. 
\REV{Lithium hydride has a cubic crystal structure, and silicon and diamond have a diamond cubic crystal structure. 
For these three systems, we use primitive unit cells containing 2 atoms.
Note that lithium hydride, silicon, and diamond systems have higher degrees of symmetry than the hydrogen dimer system.
The reference HF calculations for all the tests are based on a $3\times 3\times 3$ $\vk$-point mesh.}
\Cref{fig:H2_mp2_energy,fig:LiH_mp2_energy,fig:Si_mp2_energy,fig:diamond_mp2_energy} show the MP2 energy results for  quasi-1D, quasi-2D, and 3D systems for the four periodic systems. 

The staggered mesh method outperforms the standard one in quasi-1D case for all the systems. 
For quasi-2D and 3D cases, the staggered mesh method performs significant better than the standard one for  lithium hydride, silicon, and diamond. 
In comparison, the performance of the two methods becomes similar for the quasi-2D and 3D hydrogen dimer systems. 
\REV{
These observations are consistent with those over model systems, and the staggered mesh method can significantly outperform the standard method over all quasi-1D systems and over certain quasi-2D and 3D systems with high symmetries. }

\section{Further discussions}\label{sec:discussion}
\REV{Numerical results in \cref{sec:numer} indicate that for anisotropic systems (model and real systems), the finite-size errors in the staggered mesh method can still be $\Or(N_\vk^{-1})$. 
The staggered mesh method significantly reduces the error in the evaluation of the integrand for $E_c^\text{TDL}$. However, in the presence of discontinuity, the remaining quadrature error to the trapezoidal rule may still be significant due to the overall non-smoothness of the integrand, even when the integrand can be evaluated exactly on each well-defined point.}

\REV{More specifically,} the integrand of \cref{eqn:correlation_sf} in MP2 calculation, i.e., $h(\vq)$, is \REV{periodic but not smooth}. The error of a trapezoidal rule can be generally analyzed using the well-known Euler-Maclaurin formula. Let $\delta k$ denote the mesh size along each direction (i.e., $N_{\vk}\sim (\delta k)^{-d}$ for systems that extend along $d$ dimensions). 
For a periodic function with continuous derivatives up to $m$-th order, the quadrature error can be as small as $\Or(\delta k^m)$. However, the integrand for $E_c^{\text{TDL}}$ already has unbounded second order derivatives. Therefore standard error analysis predicts that the quadrature error can be $\Or(\delta k^2)=\Or(N_{\vk}^{-2/3})$, or even worse, for three-dimensional systems. If so, the finite-size errors would in general be dominated by such quadrature errors. Fortunately, the points of discontinuity are isolated, and we find that the quadrature error should be $\Or(\delta k^3)=\Or(N_{\vk}^{-1})$  for 3D systems \REV{and $\Or(\delta k^2)=\Or(N_{\vk}^{-1})$ for quasi-2D systems in the worst case (in addition to the $\Or(N_\vk^{-1})$ error from possible neglect of discontinuous terms in integrand evaluation).} However, the analysis is much more involved than the direct application of the Euler-Maclaurin expansion. Instead it generalizes the result of Lyness \cite{Lyness1976} for a class of punctured trapezoidal rules, and we will report the full numerical analysis in a future publication. Furthermore, for systems with certain symmetries (for instance, three-dimensional systems with cubic symmetries), the smoothness condition of the integrand can be improved, which leads to quadrature error that decays faster than $\Or(N_{\vk}^{-1})$, and such faster decay agrees with the observations in the literature~\cite{ChiesaCeperleyMartinEtAl2006,DrummondNeedsSorouriEtAl2008} \REV{and our numerical results in \cref{sec:numer}.}

The situation for quasi-1D system is qualitatively different. This is because all the discontinuous points in quasi-1D systems \REV{turn out to be  removable, i.e. by properly redefining the integrand values at these isolated points, $h(\vq)$ can become a smooth function (see the numerical examples in \cref{fig:integrand_discontinuity,fig:hq_1d})}. 
Therefore with a properly defined integrand, the quadrature error for  quasi-1D systems decays super-algebraically (i.e., the quadrature error  decays asymptotically faster than $\Or(\delta k^m)$ for any $m>0$) according to the Euler-Maclaurin formula.
Note that in practice, there is no need to find the proper integrand values at discontinuous points if no quadrature node overlaps with such points, \REV{which is the case for the staggered mesh method.}

\begin{figure}[htbp]
        \centering
        \subfloat[isotropoic]{
                \includegraphics[width=0.47\textwidth]{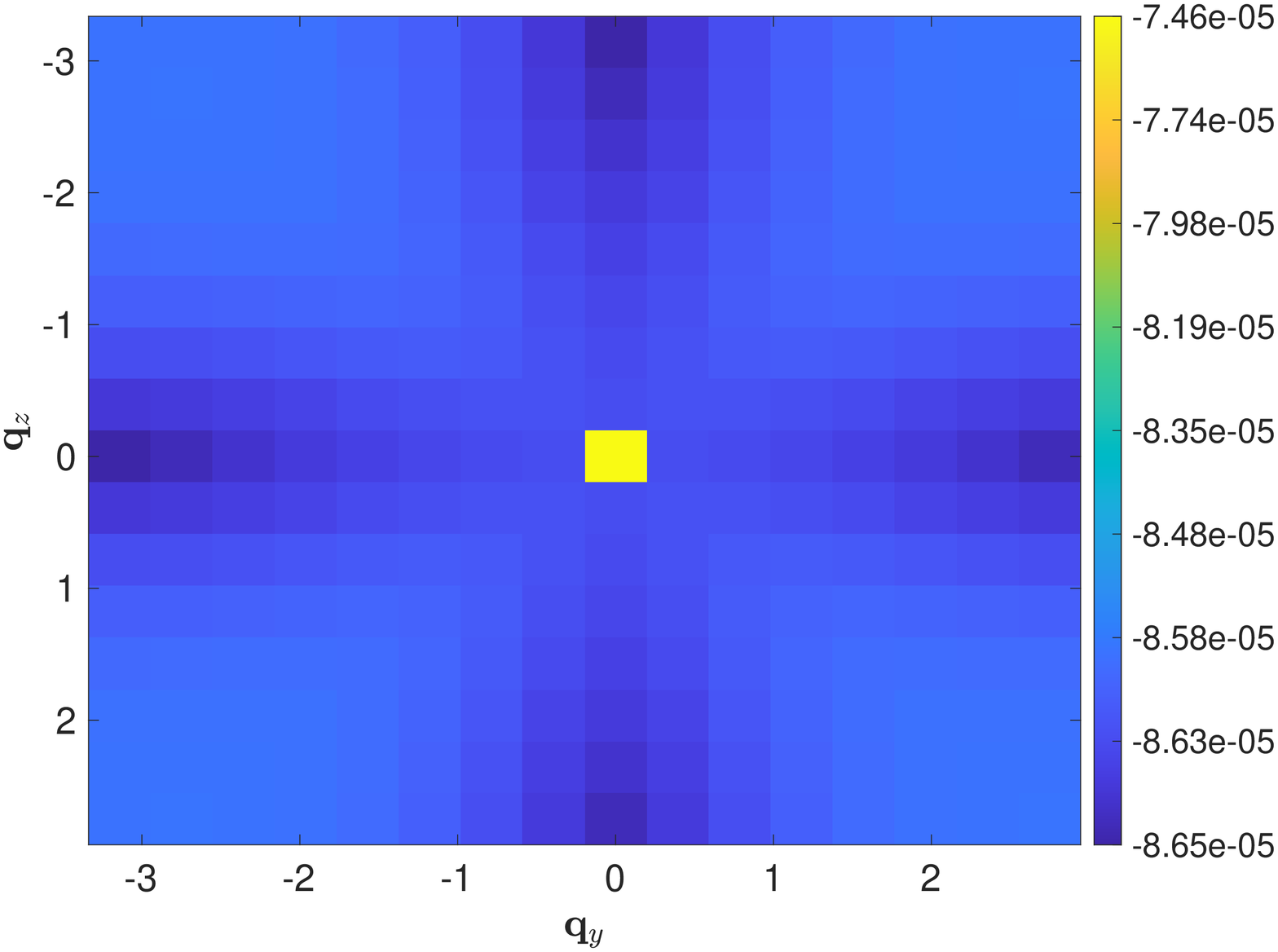}
        }
        \subfloat[anisotropic]{
                \includegraphics[width=0.47\textwidth]{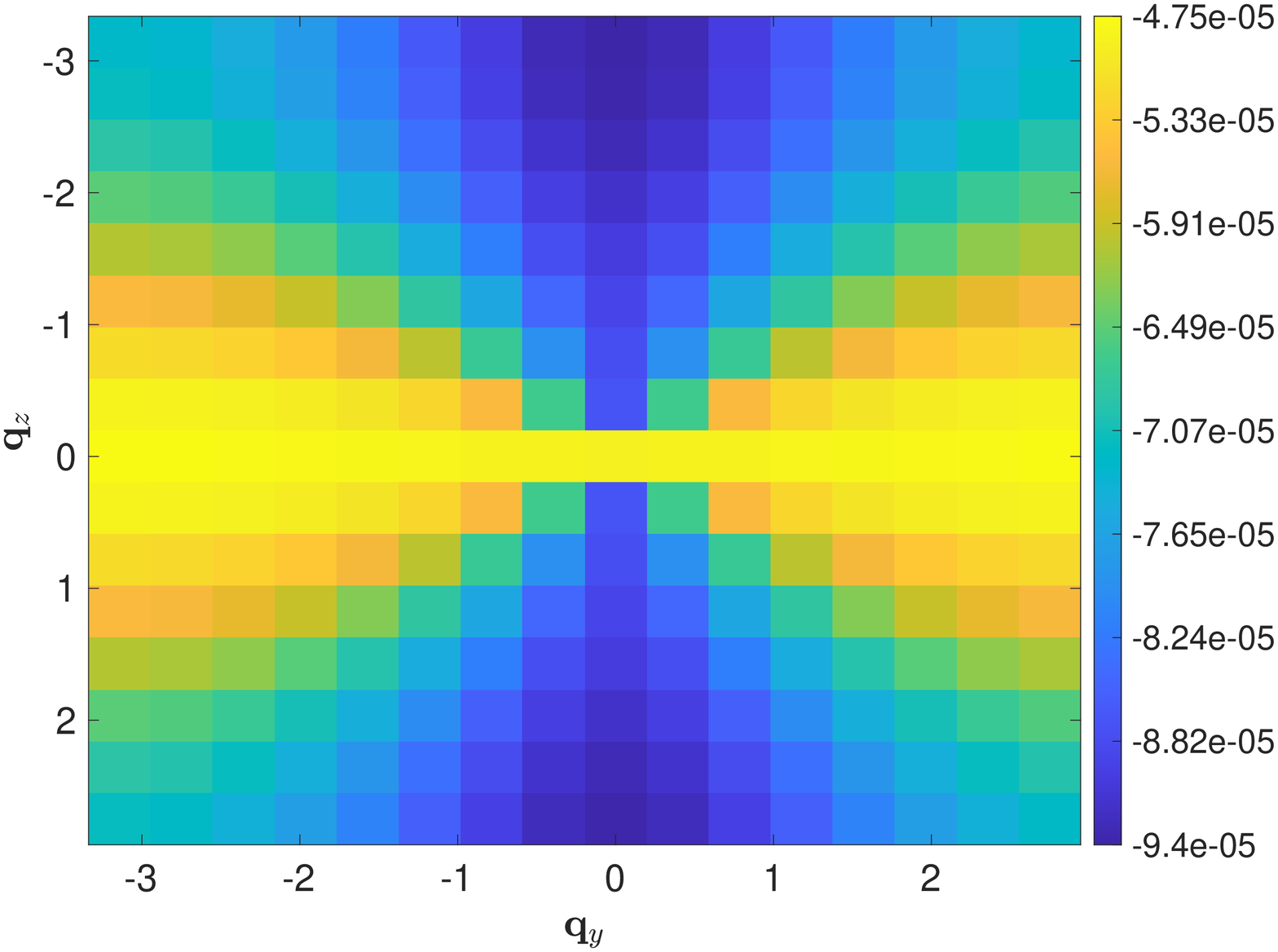}
        }
        \caption{
                Illustration of discontinuities in  $h(\vq)$ from two quasi-2D model systems with an isotropic and an anisotropic Gaussian effective potential fields, respectively. 
                All sampled $\vq$ points are of the form $(0, \vq_y, \vq_z)$ with $\vq_y,\vq_z \in [-\pi, \pi]$.
                \label{fig:hq_discontinuity}
        }
\end{figure}

The discontinuity of $h(\vq)$ at $\vq = \bm{0}$ is generally not removable in quasi-2D and 3D systems (similarly for the discontinuity of the integrand in \cref{eqn:structure_factor} for computing $S_\vq(\vG)$ and $h(\vq)$). 
For systems with certain symmetries, $\lim_{\vq\to\bm{0}}h(\vq)$ may exist. 
\REV{Redefining $h(\bm{0})$ as this limit improves the integrand smoothness and can lead to quadrature error smaller than $\Or (N_\vk^{-1})$ for a general trapezoidal rule.} 
In this scenario, the overall quadrature error is dominated by \REV{placing the quadrature nodes at those discontinuous points while not properly defining their integrand values, which is the case in the standard MP2 calculation.}
As an example, \Cref{fig:hq_discontinuity} illustrates the discontinuity of $h(\vq)$ obtained from two quasi-2D model systems which have an isotropic and an anisotropic Gaussian effective potential fields, respectively.
The additional symmetry from the isotropic potential leads to the removable discontinuity at $\vq = \bm{0}$ for $h(\vq)$, while in the anisotropic case, the values of $h(\vq)$ along the $x,y$ axes are very different near $\vq=\bm{0}$, and hence $\lim_{\vq\to\bm{0}}h(\vq)$ is not well defined.

\REV{
        To summarize, the remaining quadrature error in the staggered mesh method is closely related to the non-smoothness of the integrand for $E_c^\text{TDL}$. For quasi-1D systems and certain quasi-2D and 3D systems with certain symmetries, the integrand can have improved smoothness condition and the staggered mesh method  can have quadrature error smaller than $\Or(N_{\vk}^{-1})$. 
}
%

\section{Conclusion}\label{sec:conclusion}

The convergence of the MP2 correlation energy towards the TDL is a fundamental question in materials science. Existing analysis in the literature focuses on the missing contribution of the structure factor $S_{\vq}(\vG)$ at $\vq+\vG=\mathbf{0}$, but neglects contributions from 1) certain quadrature nodes coincide with points of discontinuity of the integrand  2) the quadrature error due to the intrinsic non-smoothness of the integrand. We demonstrate that such contributions can be at least equally important and scale as $\Or(N_\vk^{-1})$. We propose the staggered mesh method that uses a different set of quadrature nodes for the trapezoidal quadrature, which allows us to completely avoid the first source of the error \REV{with negligible additional costs}. Numerical evidence shows that the staggered mesh method is particularly advantageous over the standard method for quasi-1D systems and systems with symmetries, which reduces the contribution from the second error source. We expect that the new approach can also be useful for correlation energy calculations beyond the MP2 level, such as higher levels of perturbation theories and \REV{coupled cluster} theories. 

\appendix
\section*{Appendix}\label{appendix}
\begin{figure}[htbp]
        \centering
        \subfloat[Quasi-1D, anisotropic, direct]{
                \includegraphics[width=0.33\textwidth]{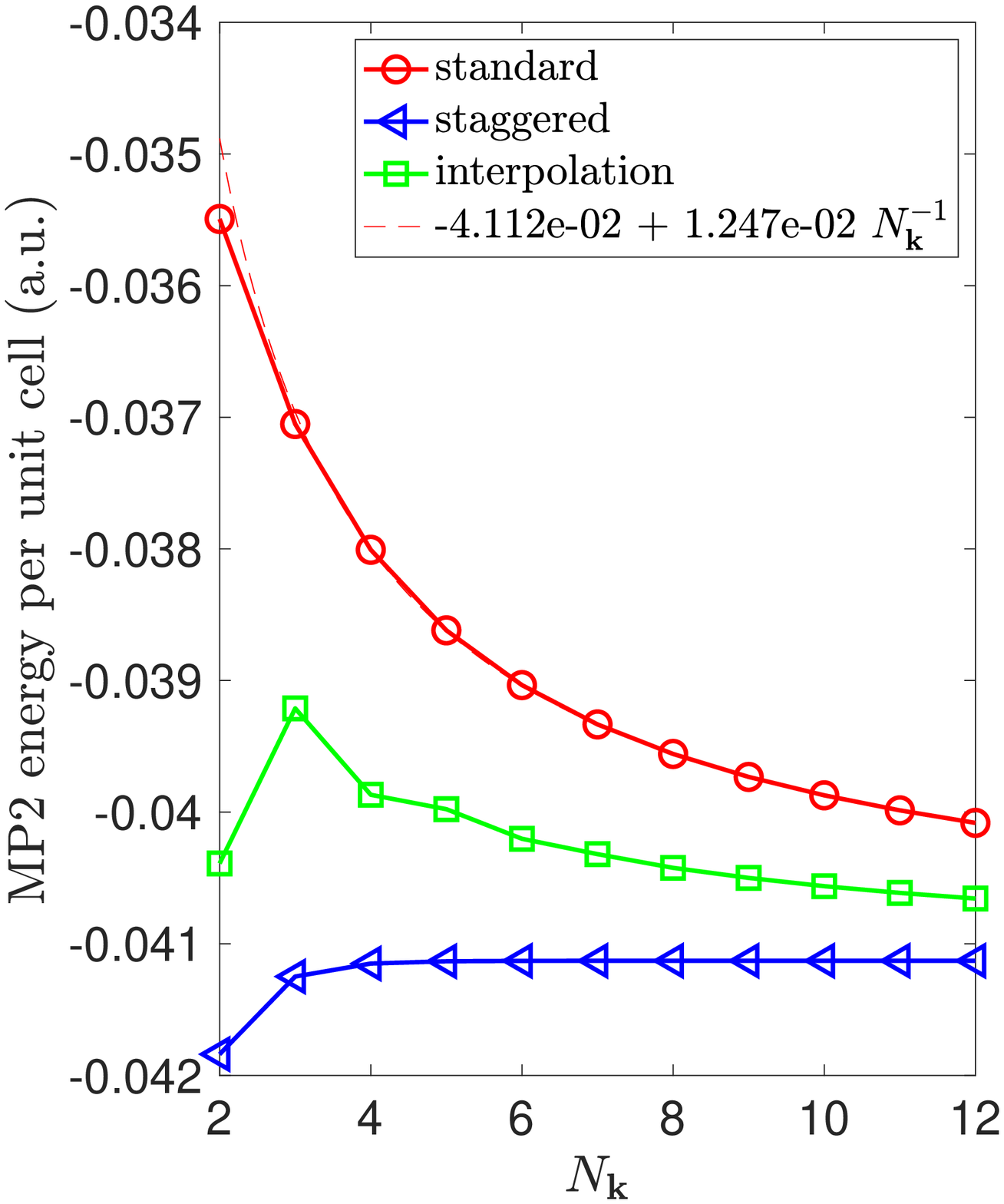}
        }
        \subfloat[Quasi-2D, anisotropic, direct]{
                \includegraphics[width=0.33\textwidth]{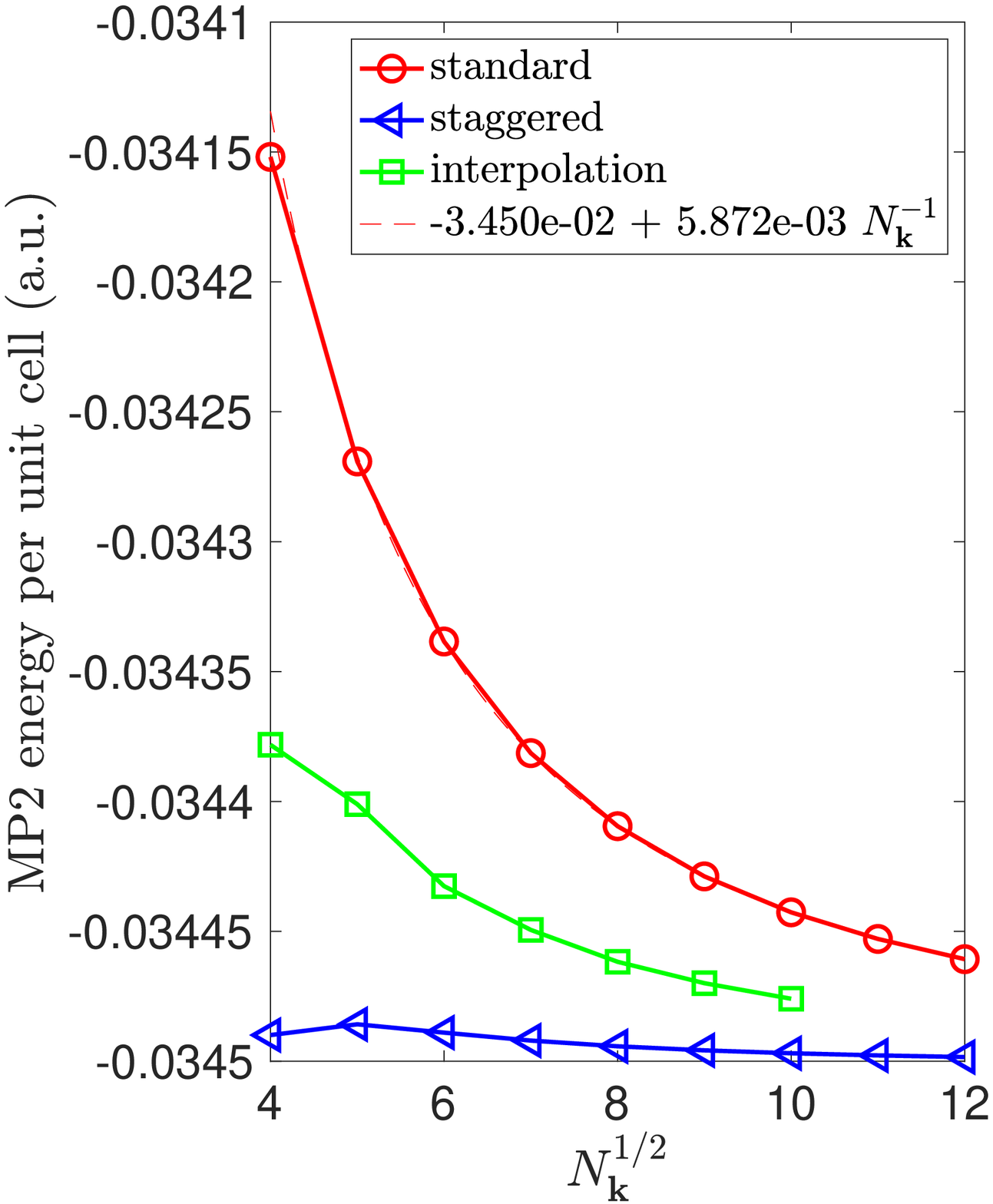}
        }
        \subfloat[3D, isotropic, direct]{
                \includegraphics[width=0.33\textwidth]{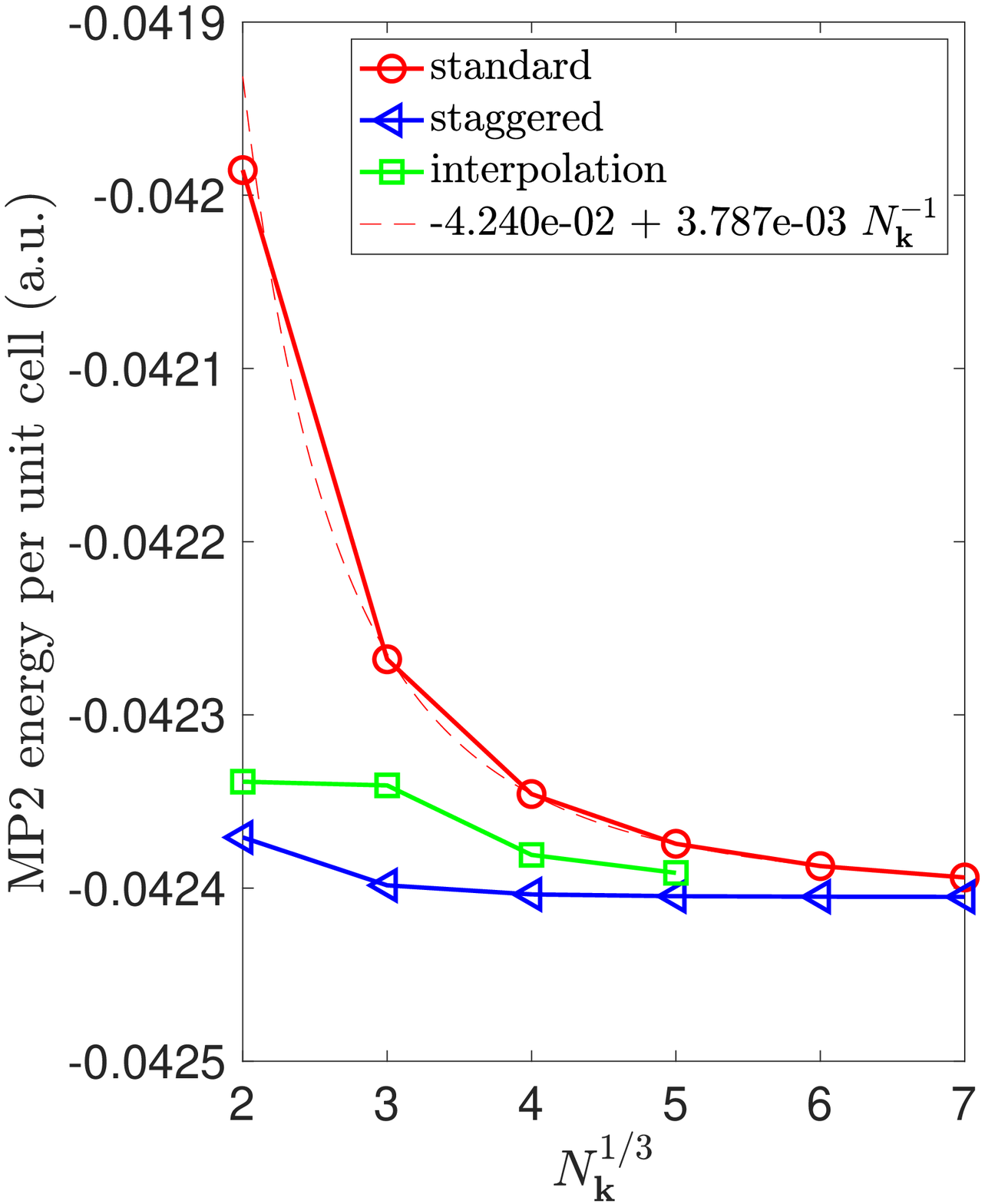}
        }
        
        \subfloat[Quasi-1D, anisotropic, exchange]{
                \includegraphics[width=0.33\textwidth]{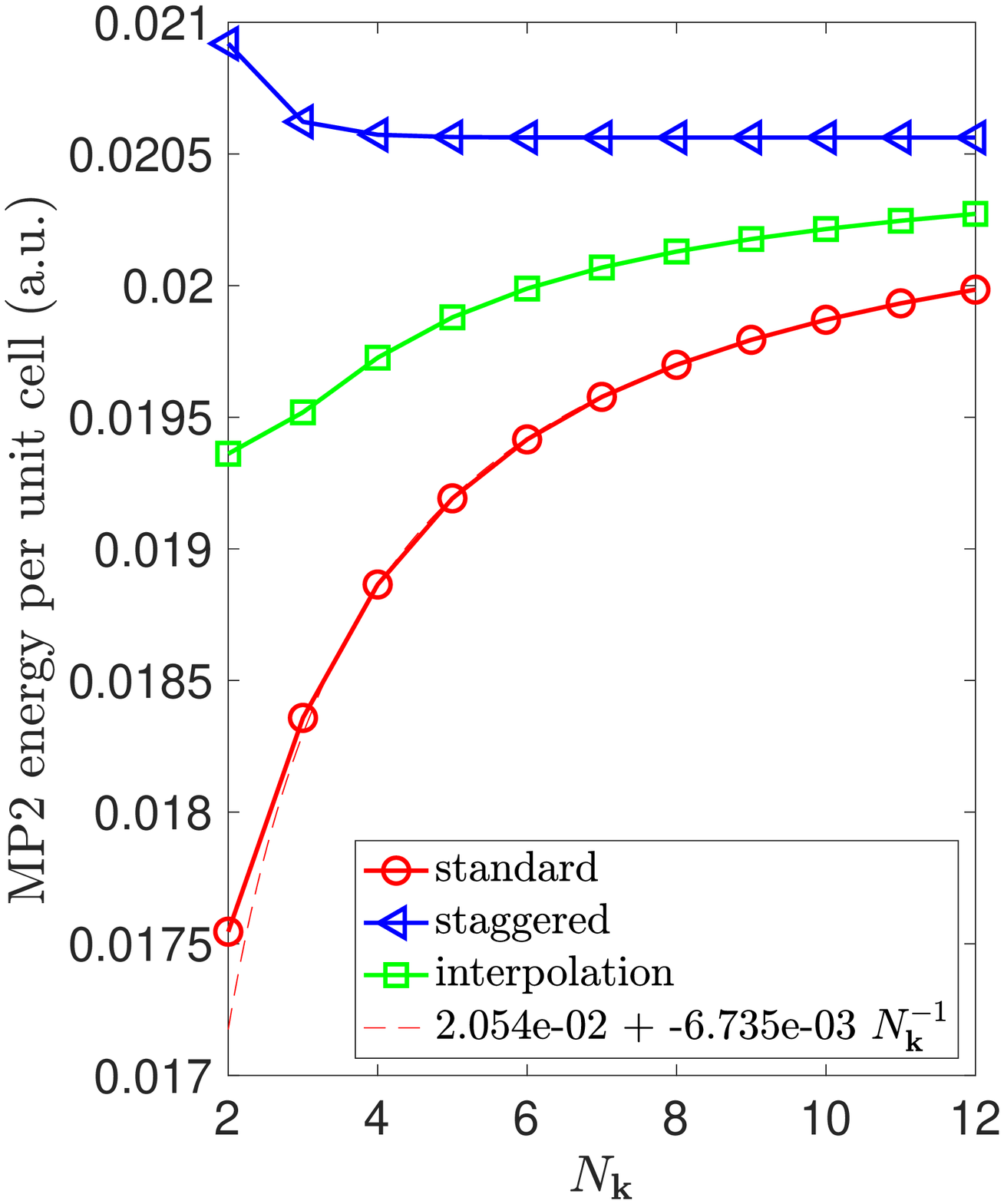}
        }
        \subfloat[Quasi-2D, anisotropic, exchange]{
                \includegraphics[width=0.33\textwidth]{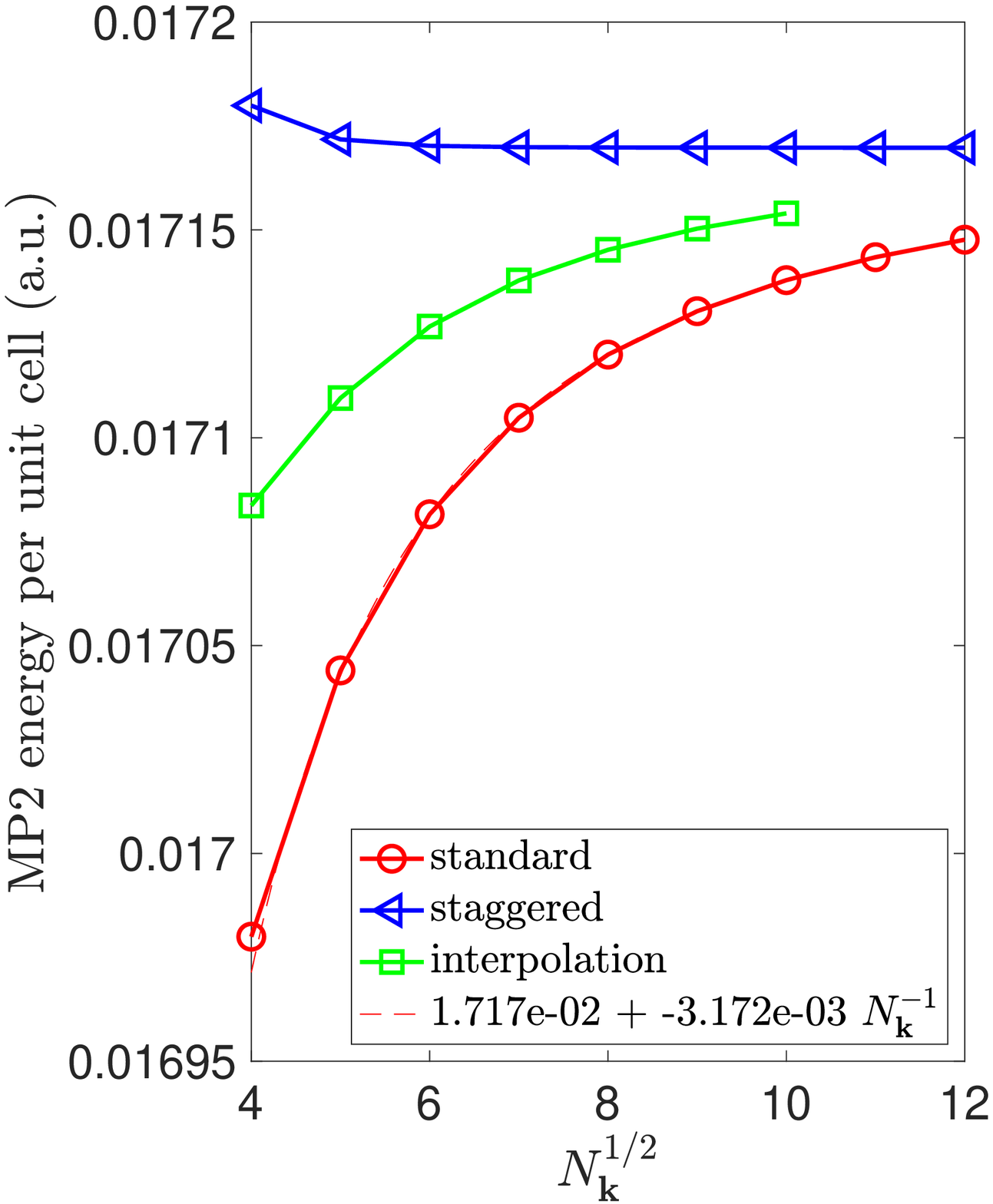}
        }
        \subfloat[3D, isotropic, exchange]{
                \includegraphics[width=0.33\textwidth]{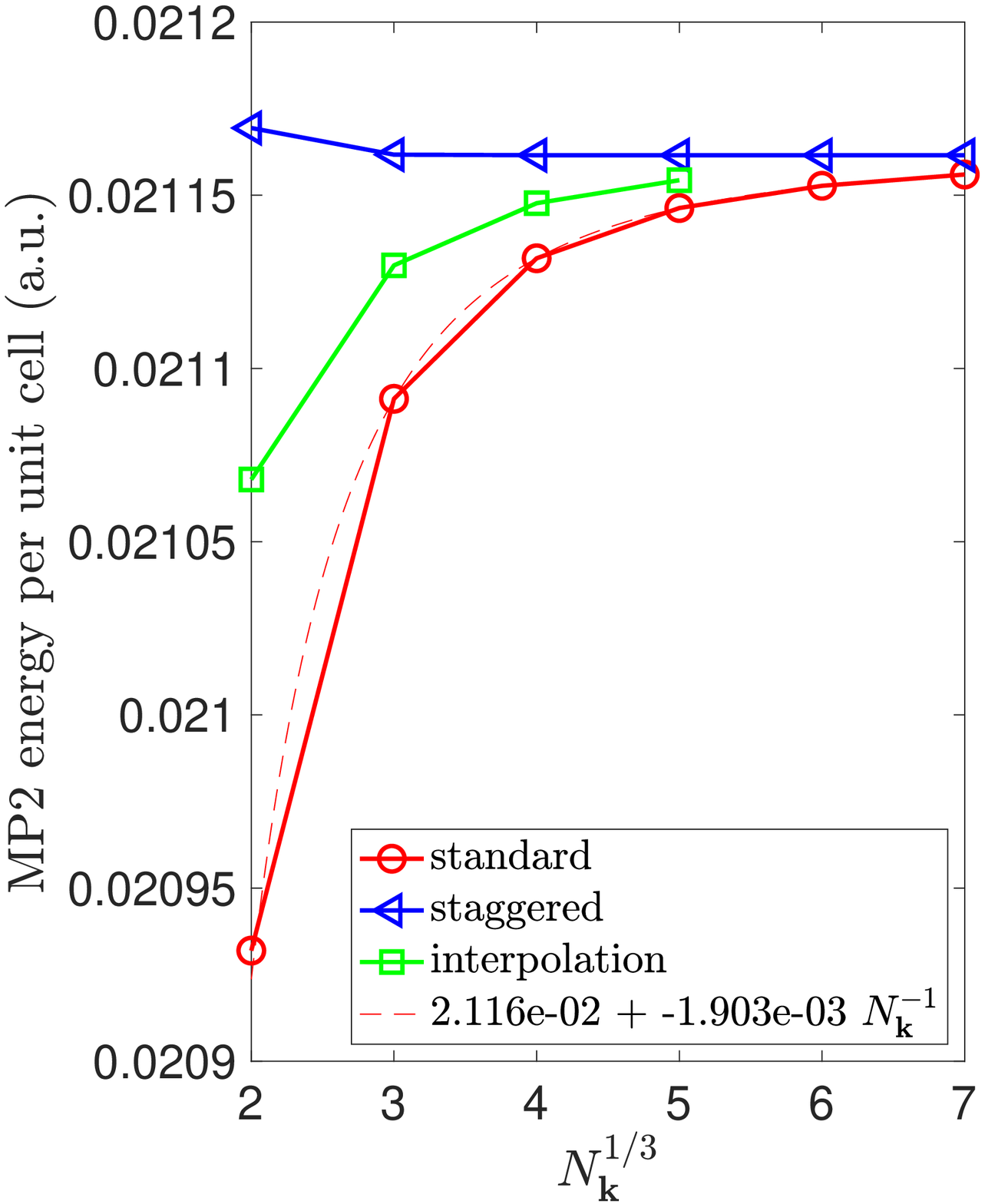}
        }
        \caption{Direct and exchange parts of the MP2 energy per unit cell computed by the standard method,  the staggered mesh method, and the structure factor interpolation method for anisotropic quasi-1D, anisotropic quasi-2D, and isotropic 3D model systems. 
        \label{fig:direct_exchange_mp2}
        }
\end{figure}

\begin{figure}[htbp]
        \centering
        \subfloat[Quasi-1D H2, direct]{
                \includegraphics[width=0.33\textwidth]{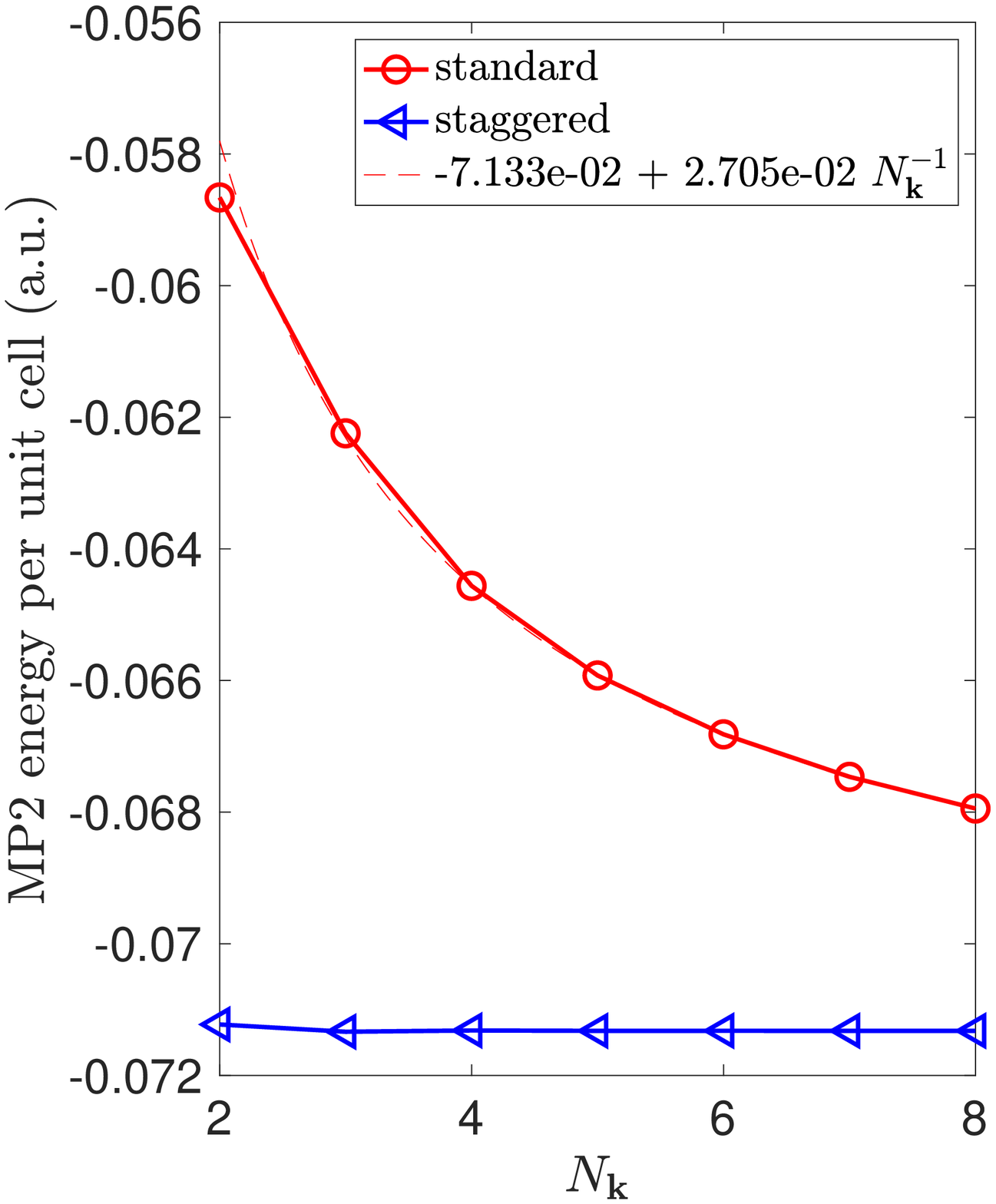}
        }
        \subfloat[Quasi-2D silicon, direct]{
                \includegraphics[width=0.33\textwidth]{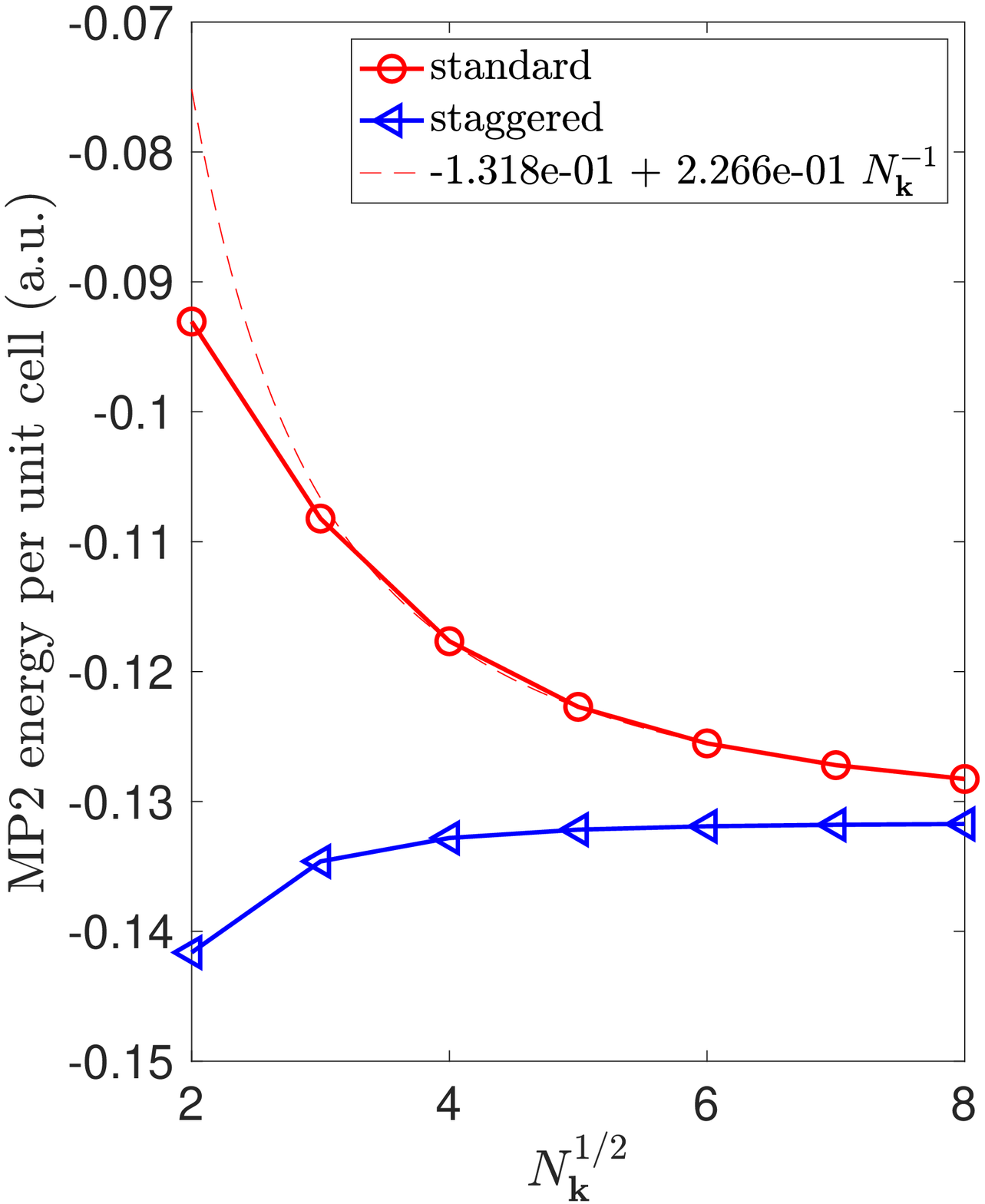}
        }
        \subfloat[3D diamond, direct]{
                \includegraphics[width=0.33\textwidth]{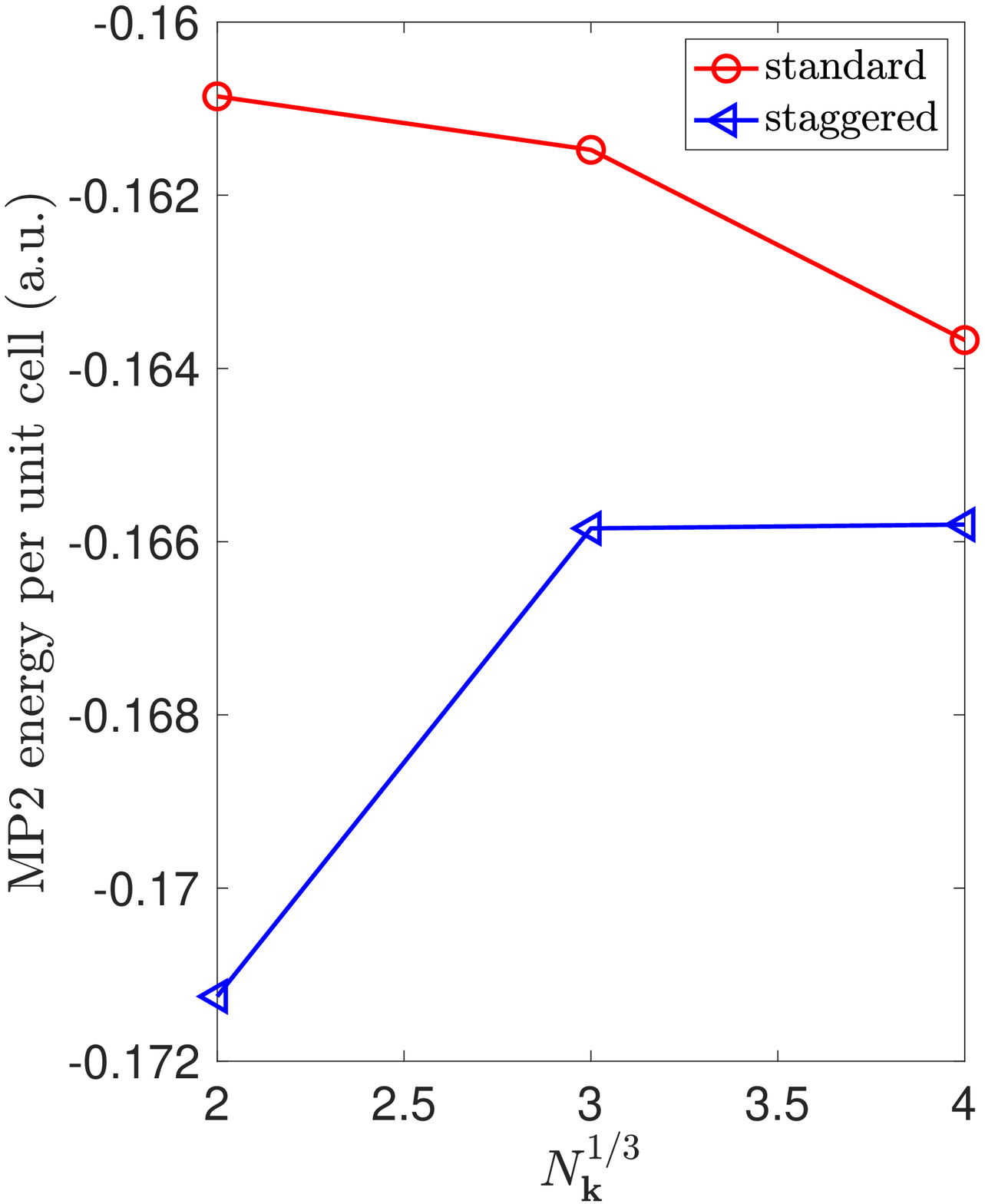}
        }
        
        \subfloat[Quasi-1D H2, exchange]{
                \includegraphics[width=0.33\textwidth]{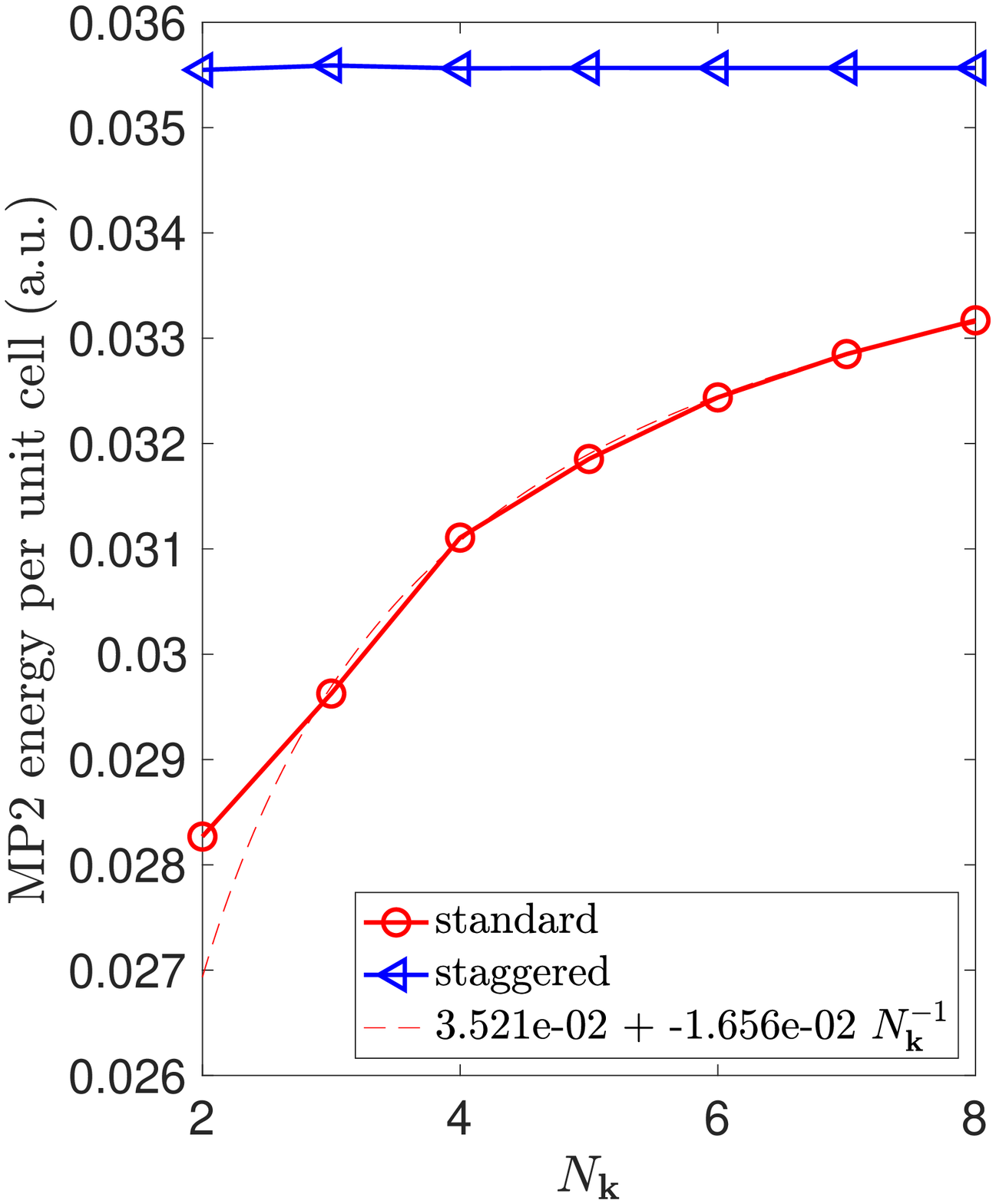}
        }
        \subfloat[Quasi-2D silicon, exchange]{
                \includegraphics[width=0.33\textwidth]{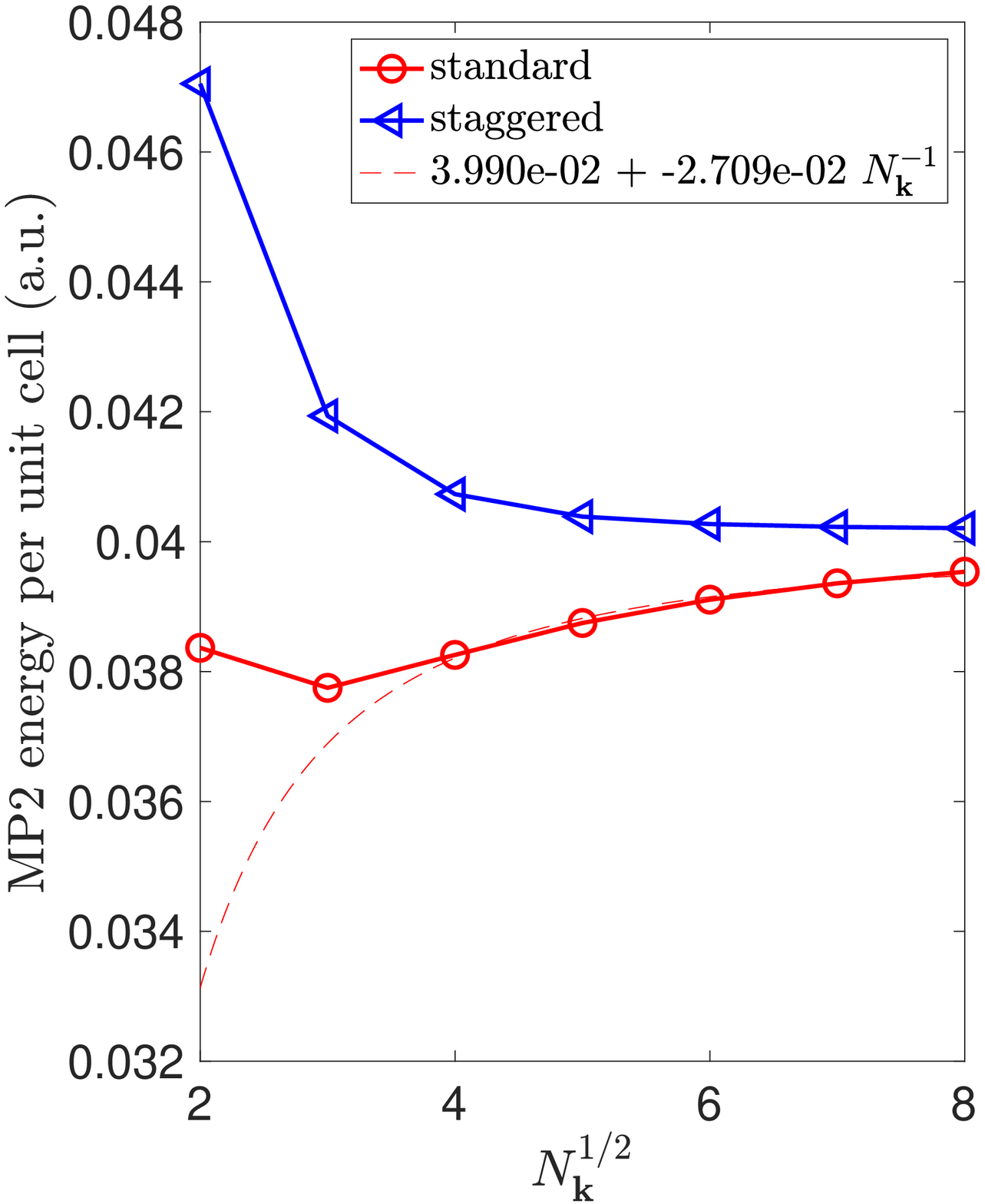}
        }
        \subfloat[3D diamond, exchange]{
                \includegraphics[width=0.33\textwidth]{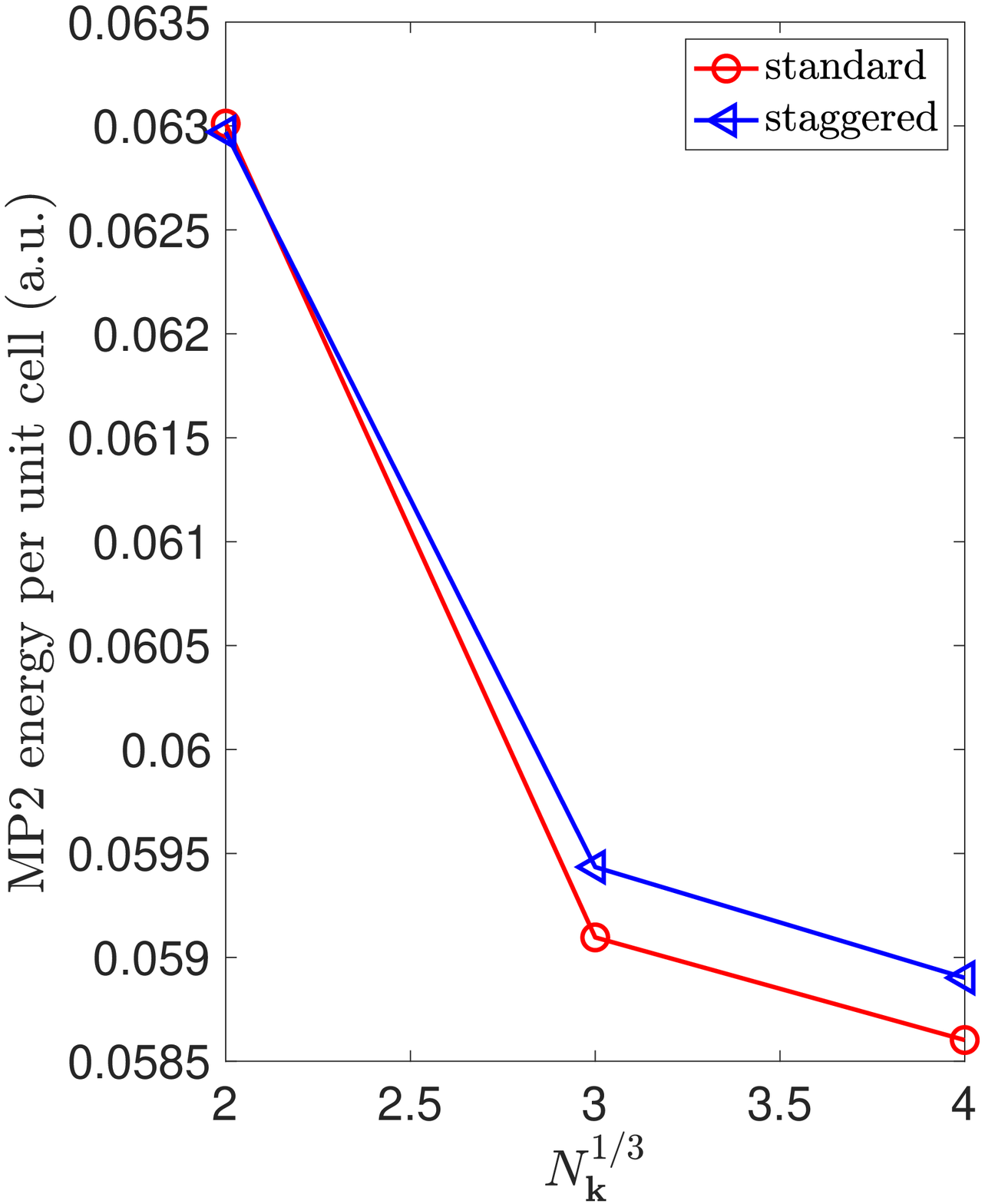}
        }
        \caption{Direct and exchange parts of the MP2 energy per unit cell computed by the standard method and  the staggered mesh method for quasi-1D hydrogen dimer, quasi-2D silicon, and 3D diamond systems. 
                \label{fig:direct_exchange_mp2_2}
        }
\end{figure}

\begin{figure}[htbp]
        \centering
        \subfloat[Quasi-1D H2]{
                \includegraphics[width=0.33\textwidth]{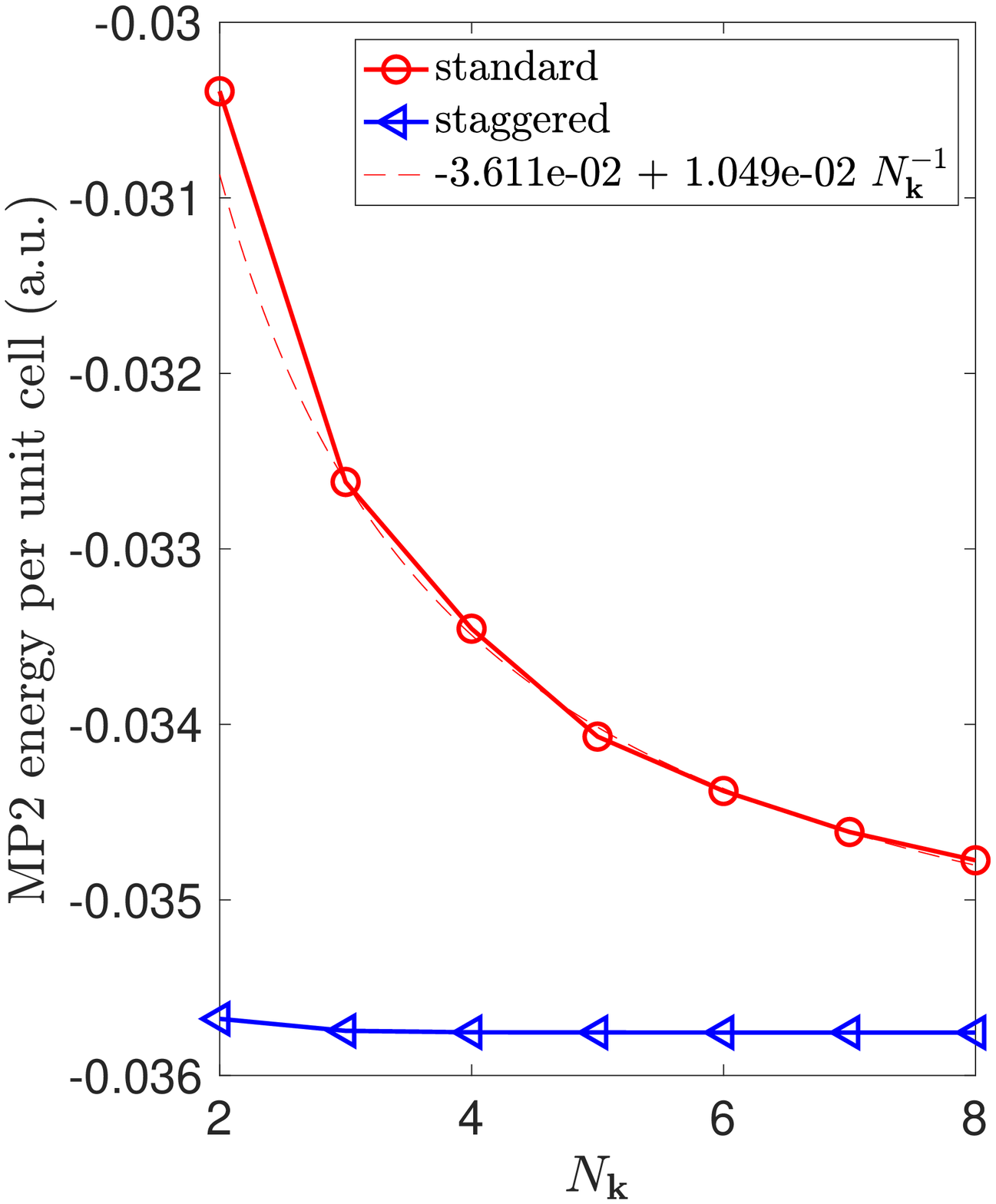}
        }
        \subfloat[Quasi-1D LiH]{
                \includegraphics[width=0.33\textwidth]{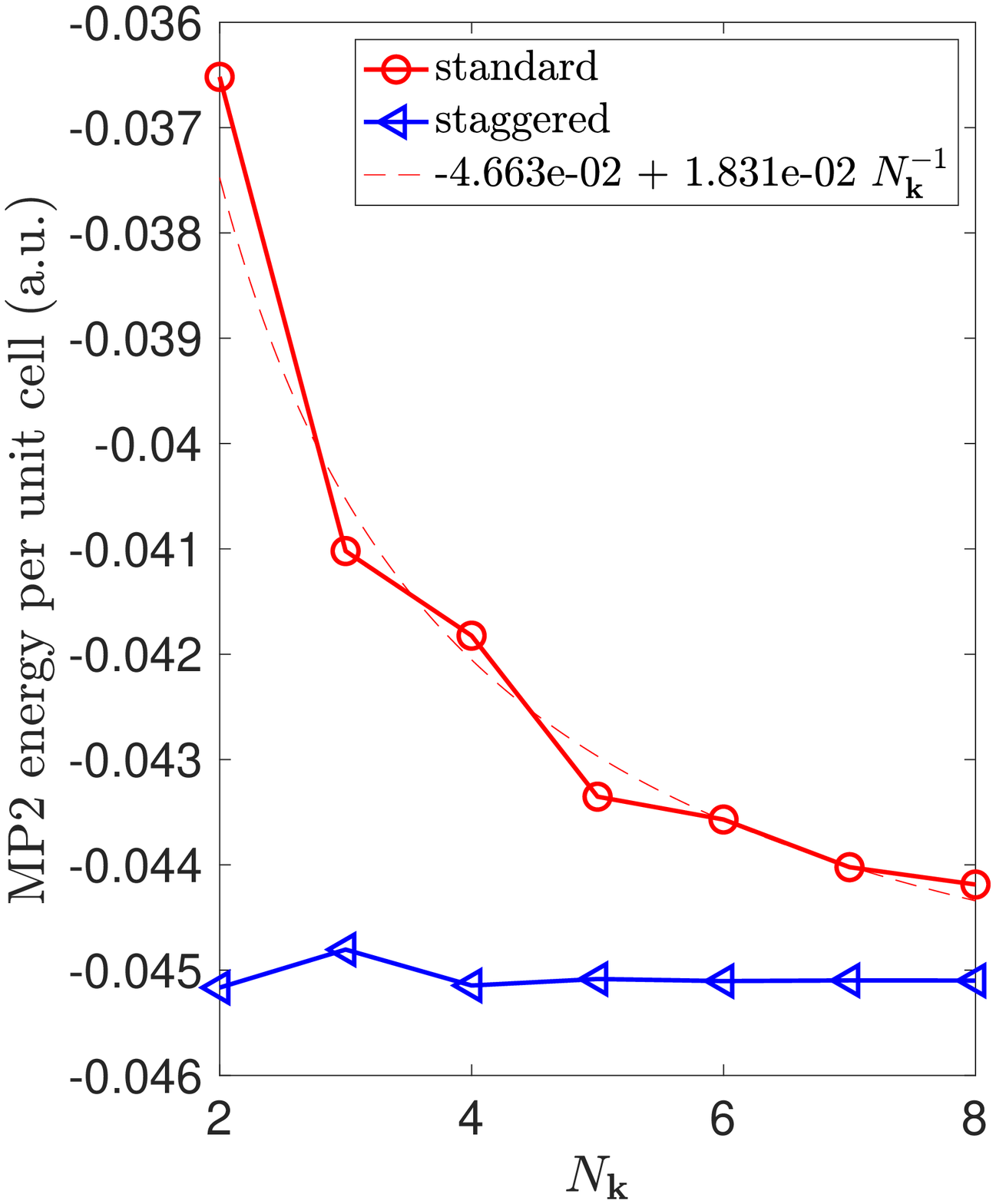}
        }
        \subfloat[Quasi-1D silicon]{
                \includegraphics[width=0.33\textwidth]{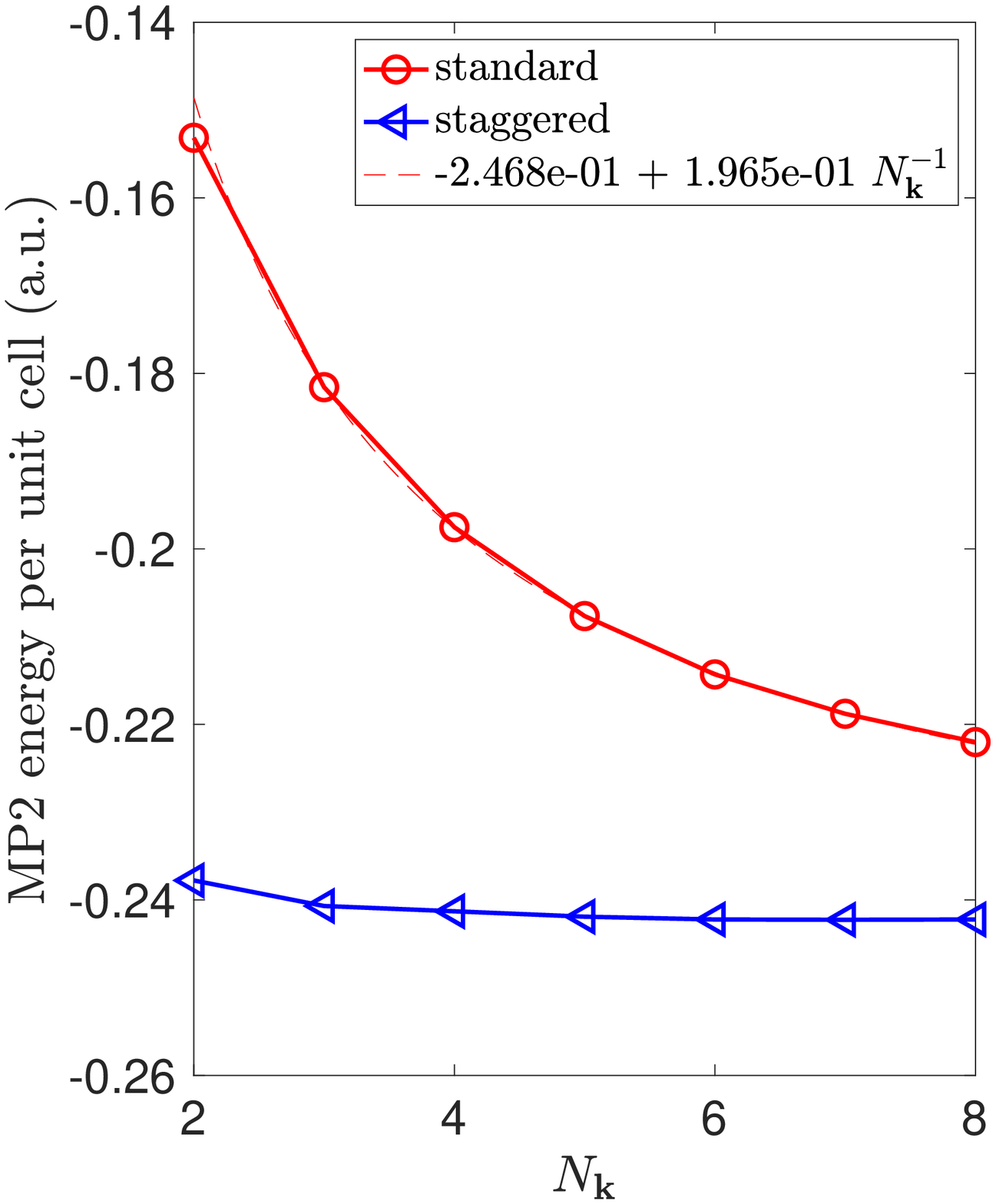}
        }
        
        \subfloat[Quasi-2D H2]{
                \includegraphics[width=0.33\textwidth]{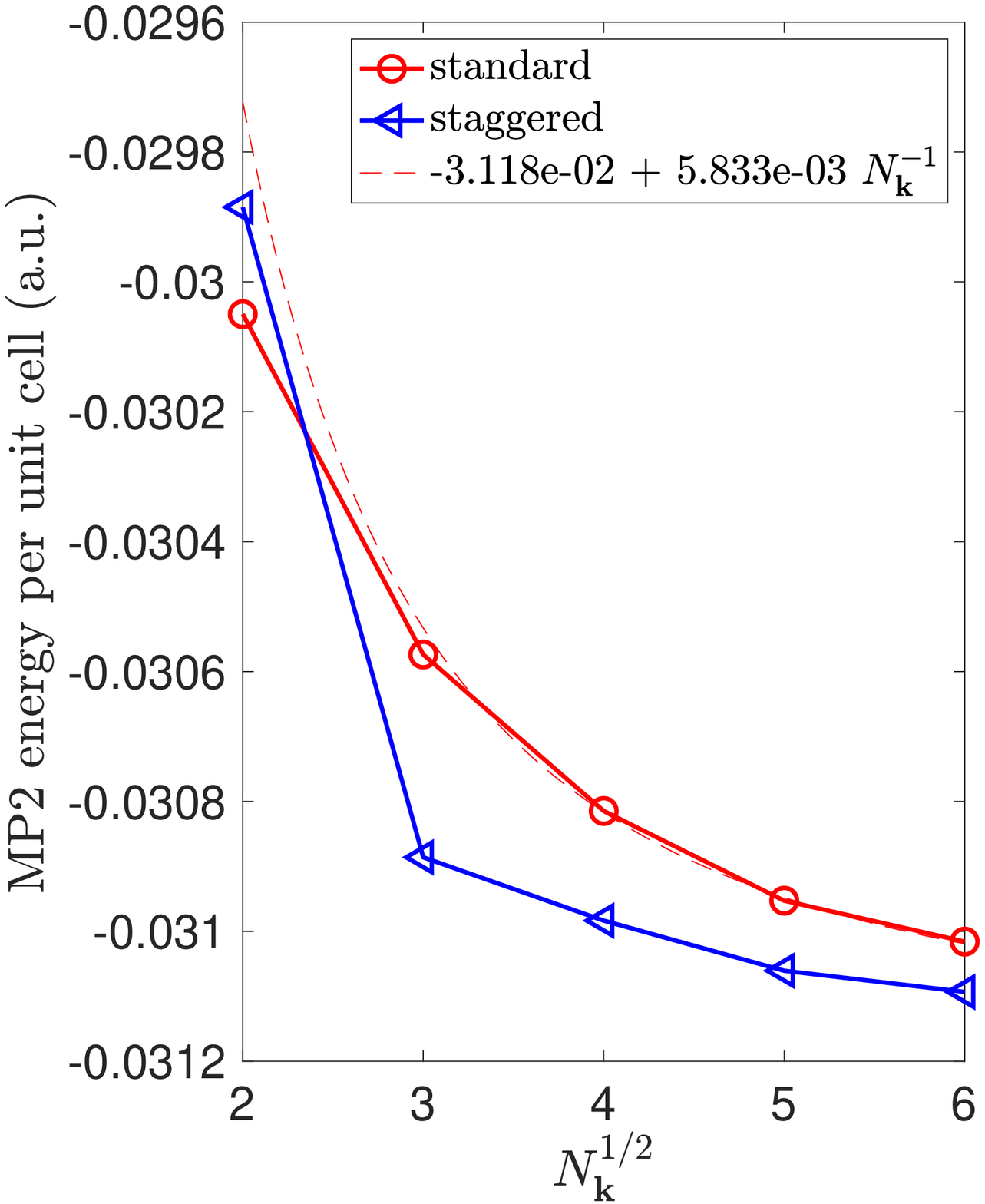}
        }
        \subfloat[Quasi-2D LiH]{
                \includegraphics[width=0.33\textwidth]{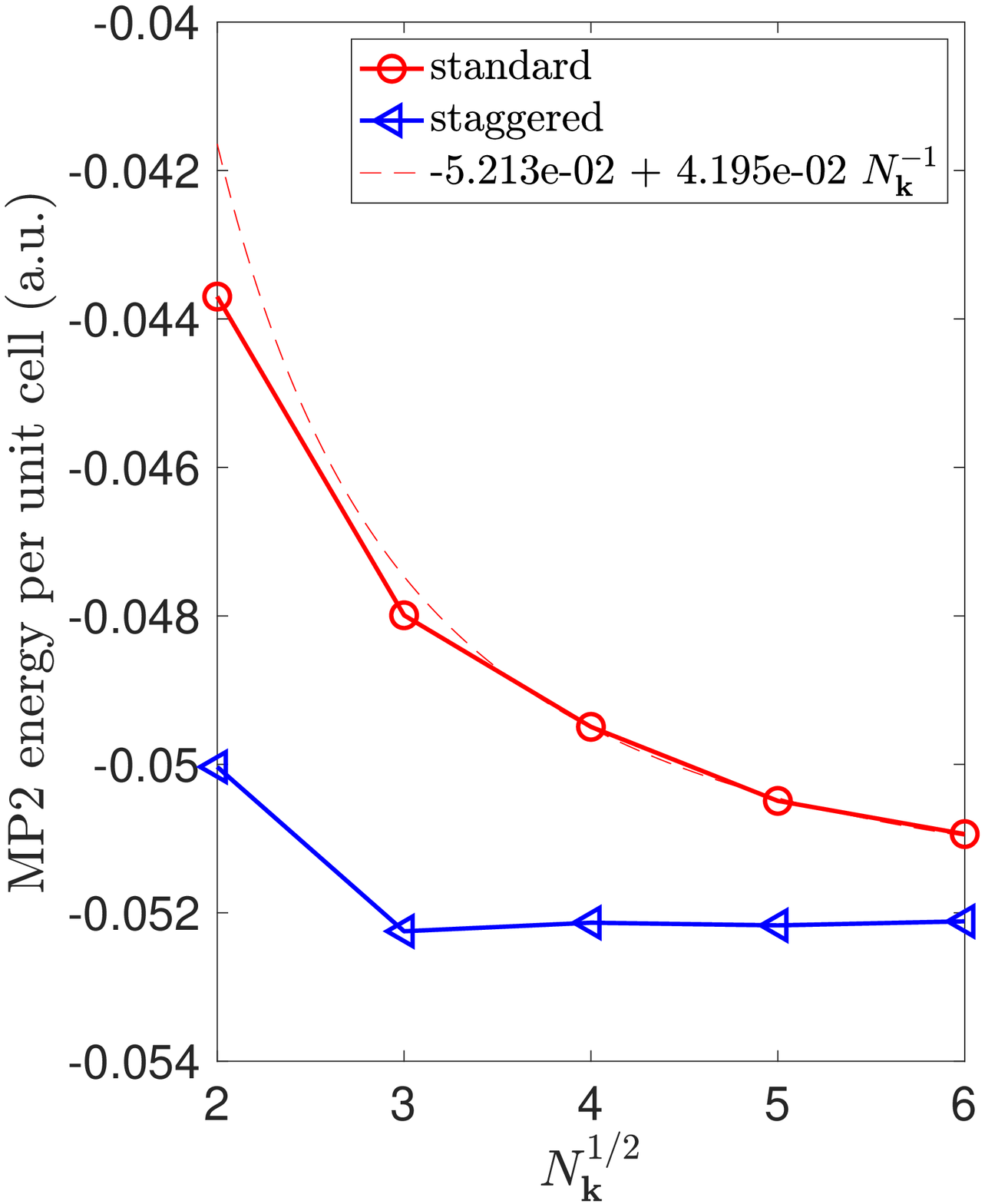}
        }
        \subfloat[Quasi-2D silicon]{
                \includegraphics[width=0.33\textwidth]{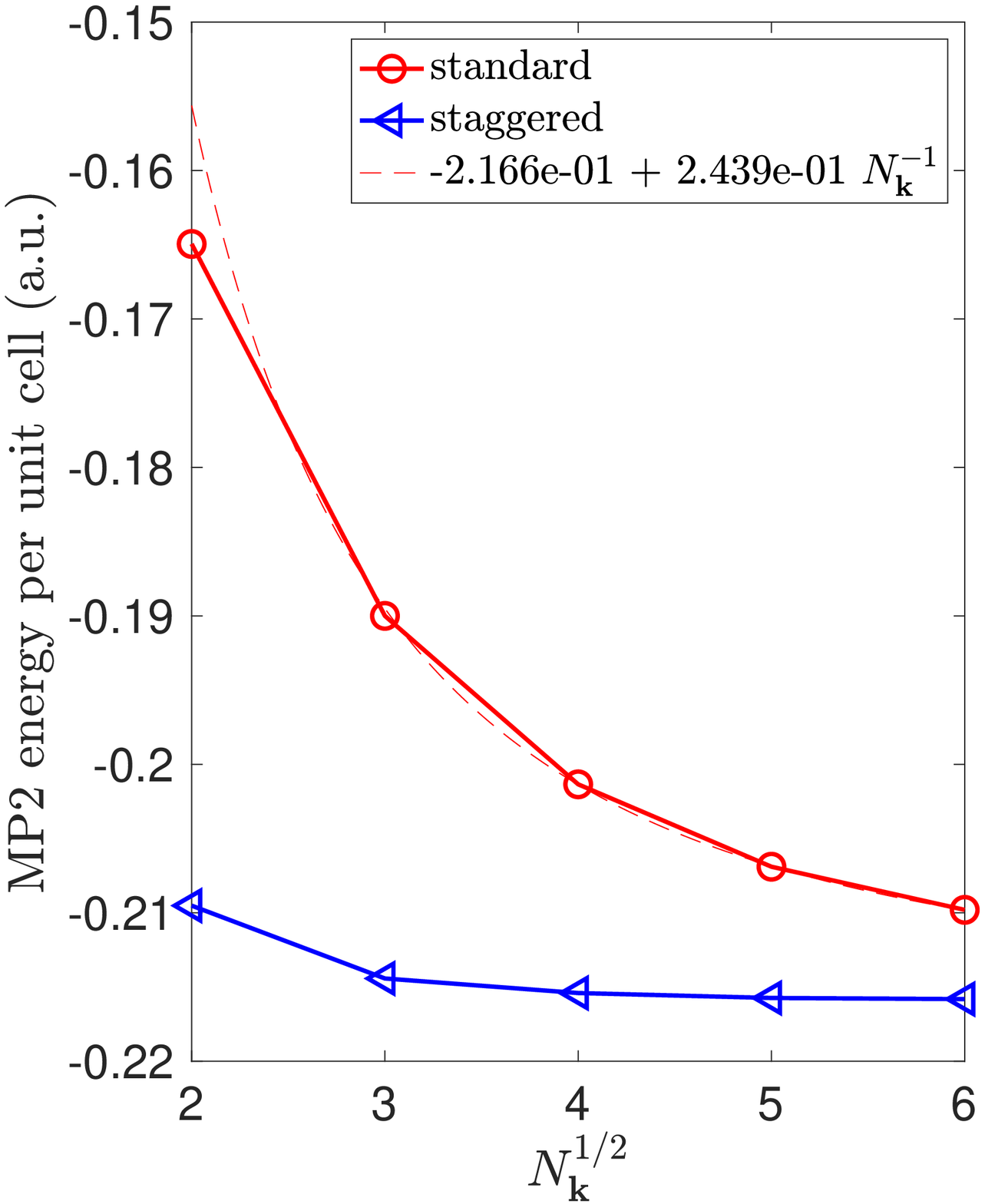}
        }
        \caption{MP2 energy per unit cell computed by the standard and the staggered mesh method for periodic hydrogen dimer, lithium hydride, and silicon systems with the gth-dzvp basis set. 
                \label{fig:dzvp_mp2}
        }
\end{figure}

\REV{\Cref{fig:direct_exchange_mp2} and \Cref{fig:direct_exchange_mp2_2} plot the direct and the exchange parts of MP2 energy per unit cell for several model and real systems in \cref{sec:numer_model} and \cref{sec:numer_pyscf}.
\Cref{fig:dzvp_mp2} plots the MP2 energy results for quasi-1D and quasi-2D  hydrogen dimer, lithium hydride, and silicon systems using the gth-dzvp basis set.}

\section*{Acknowledgement}

This work was partially supported by the Air Force Office of Scientific Research under award number FA9550-18-1-0095 (X.X.,L.L.), by the Department of Energy under Grant No. DE-SC0017867, and by the National Science Foundation under Grant No. DMS-1652330 (L.L.), and by the China Scholarship Council under File No. 201906040071. (X.L).  We thank Timothy Berkelbach and Garnet Chan for insightful discussions on the finite-size effects, \REV{and the anonymous referees for helpful suggestions.}

\begin{figure}
\centering
\includegraphics[width = 0.6\textwidth]{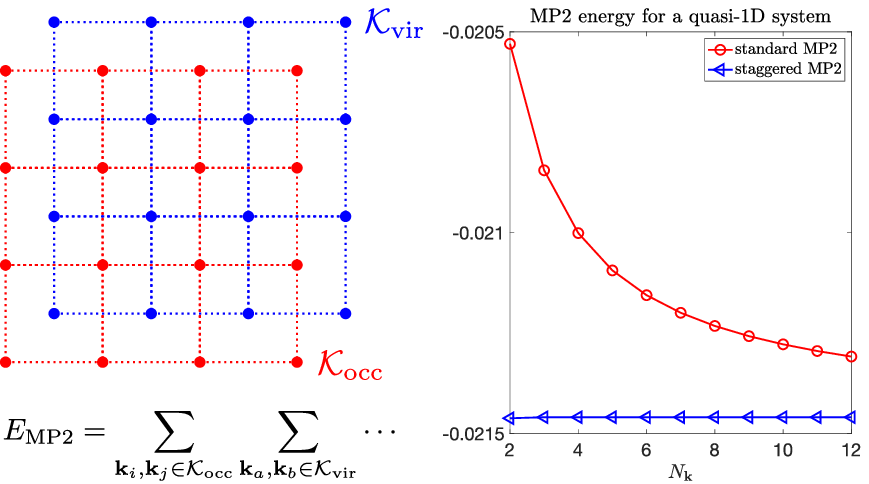}
\caption{TOC} 
\end{figure}

\bibliography{mp2}
\newpage

\end{document}